\documentstyle[preprint,aps,epsf]{revtex}

\def\fpj{\hspace{-.7cm}}

\def\thalf{{\textstyle{\frac{1}{2}}}}

\def\oneth{{\textstyle{\frac{1}{3}}}}
\def\thrhalf{{\textstyle{\frac{3}{2}}}}
\def\onefive{{\textstyle{\frac{1}{5}}}}
\def\twofive{{\textstyle{\frac{2}{5}}}}

\def\twothr{{\textstyle{\frac{2}{3}}}}

\def\fourthr{{\textstyle{\frac{4}{3}}}}

\begin{document}

\draft
\preprint{ SUNY-NTG-97-35 }
\title{Neutrino interactions in hot and dense matter}
\author {Sanjay Reddy, Madappa Prakash, and James M. Lattimer }
\address{Department of Physics \& Astronomy, SUNY at Stony Brook,
Stony Brook, New York 11794-3800 \\}
\date{\today}
\maketitle
\begin{abstract}
We study the charged and neutral current weak interaction rates relevant for
the determination of neutrino opacities in dense matter found in supernovae and
neutron stars.  We establish an efficient formalism for calculating
differential cross sections and mean free paths for interacting, asymmetric
nuclear matter  at arbitrary degeneracy.  The formalism is valid for both
charged and neutral current reactions.  Strong interaction corrections are
incorporated through the in-medium single particle energies  at the relevant
density and temperature.  The effects of strong interactions on the weak
interaction rates are investigated using both potential and effective
field-theoretical models of matter.  We investigate the relative importance of
charged and neutral currents for different astrophysical situations, and also
examine the influence of strangeness-bearing hyperons. Our findings show that
the mean free paths are significantly altered by the effects of strong
interactions and the multi-component nature of dense matter. The opacities are
then discussed in the context of the evolution of the core of a protoneutron
star.
\end{abstract}

\vspace*{0.2in} 
\narrowtext
\newpage

\section{INTRODUCTION}

The transport of neutrinos is an essential aspect of simulations of
gravitational collapse, supernovae, protoneutron stars and binary mergers of
compact objects. The neutrinos of all flavors emitted from newly formed neutron
stars in supernova explosions~\cite{SN87A} are the only direct probe of the
mechanism of supernovae and the structure of protoneutron stars.  The most
important ingredient of neutrino transport calculations in these simulations
is the neutrino opacity at supra-nuclear density~\cite{B1,MB,BL,WM,SuS,K,KJ}.
Yet, to date, calculations of neutrino opacities in dense matter have received
relatively little attention compared to other physical inputs such as the
equation of state (EOS).

Both charged current absorption and neutral current scattering reactions are
important sources of opacity. Neutral current processes involve all flavors
of neutrinos scattering on baryons and leptons.  While scattering from
electrons is important for energy and momentum transfer in the
process of thermalizing neutrinos, for both energy and lepton number transport,
neutrino-baryon scattering and absorption are the dominant processes.

Earlier work on neutrino interactions in matter at supra-nuclear densities that
have shaped our discussion in this paper are in Refs.~[9-32].
Various approximations have
almost always been made in the calculations of neutrino cross sections.  These
approximations concern  the degrees of degeneracy or relativity, the
composition, or the effects of the baryon-baryon interactions. The cross
section for neutrino-nucleon interactions has only been calculated exactly for
noninteracting gases by Schinder~\cite{PS} and for neutrino-electron
scattering by Mezzacappa and Bruenn~\cite{MB} .  Other workers have developed
limiting expressions for noninteracting gases for the purely elastic case, the
completely degenerate case, or the completely nondegenerate case. In fact,
most numerical simulations of supernovae, protoneutron star evolution, and
binary neutron star coalescence, have employed limiting expressions, derived
from those for noninteracting nucleonic matter, in which interactions were
included by simple scaling factors~\cite{B1,BL,K,KJ}.  The results may be
inconsistent with the underlying nuclear matter EOS.

Only a few attempts exist~\cite{S1,IP,GP,HW,RP1} in which the effects of
strong interactions on neutrino opacities were considered.   But these studies
were not performed for the intermediate degeneracies that are often encountered
in astrophysical environments.  In the case of neutral current opacities, the
effects of stong interactions were investigated for nondegenerate
nuclear matter by Sawyer~\cite{S1} and for degenerate neutron matter
by Iwamoto \& Pethick~\cite{IP}. Both of these efforts treated nucleons in the
nonrelativistic limit and predicted increases in the mean free paths by
factors of $\sim 2-3$, for densities in the range of $2-4$ times the nuclear
saturation density ($n_0=0.16$ fm$^{-3}$).   Subsequently, relativistic
calculations based on effective Lagrangian models for hot, but neutrino-poor
neutron star matter were performed by Horowitz and
Wehrberger~\cite{HW}. The neutral current differential cross sections
were calculated using linear response theory and mean free paths were
found to be factors of $1.5-2$ times that for noninteracting nucleons.
In the case of charged current interactions, the effects of interactions have
been considered by  Sawyer~\cite{S1} for nondegenerate matter and by
Goodwin and Pethick~\cite{GP} for degenerate matter.

Schinder's exact results for noninteracting nucleons are expressed in terms of
a lengthy series of Fermi integrals.   We have found, however, a simplification
to Schinder's results that not only expresses them in numerically simpler
fashion as Polylogarithmic functions, but are also easily generalized to the
case of interacting matter.  In contrast to previous work,  we calculate both
neutral and charged current opacities including effects of interactions arising
from the underlying EOS.

Another aspect of opacities which recent work has emphasized concerns the
importance of including the multi-component nature of dense matter on
neutrino opacities.  Prakash {\em et. al.}~\cite{PPLP}, and Reddy and
Prakash~\cite{RP1,RP2} have identified neutrino-hyperon absorption and
scattering reactions as being important new sources of opacity.  These include
absorption involving the $\Lambda$ and $\Sigma^-$ hyperons and
scattering involving the  $\Sigma^-$ hyperon.  These could play important roles
in calculations of the neutrino
signature of an evolving protoneutron star with hyperons~\cite{KJ}.

In this paper, we perform neutrino opacity calculations  for interacting matter
of arbitrary degeneracy and composition at supra-nuclear densities.
Interactions between the leptons are negligible and will be ignored.  Strong
interactions between the baryons, however, significantly alter the energy
spectra from their noninteracting forms, especially at high density. One of
our objectives is to explore the extent to which interactions among the baryons
affect the neutrino cross sections.  We will separately consider potential
models that are inherently nonrelativistic and relativistic field-theoretical
models.  Appendices A and B consider the formulation of these models in some
detail. We also include effects due to the multicomponent nature of interacting
matter and the possible presence of strangeness in the form of hyperons.  Where
possible, we provide analytical expressions for both charged and neutral
current differential scattering cross sections for a given EOS at a fixed
density, temperature and lepton concentration for given incoming neutrino
energy and scattering angle. The most detailed transport codes~\cite{B1}, which
solve the full Boltzmann transport equation, require differential cross
sections.  However, simpler transport methods~\cite{BL,WM,SuS,K,KJ} only need
angle and/or energy averaged opacities, which can be usually expressed in the
form of mean free paths.   In this paper, to facilitate comparisons between our
results and those involving various approximations,  we present numerical
results for mean free paths, leaving for subsequent publications detailed
results for the general opacities.  In particular, we present results for mean
free paths for thermodynamic conditions relevant to the evolution of
protoneutron stars.

The paper is organized as follows.  In Sec. II, we summarize the basic
relations needed to calculate both absorption and scattering opacities in hot
and dense matter.  We will consider in Sec. III the idealized situation in
which the baryons are treated as nonrelativistic, the leptons are assumed to be
massless, and the baryon-baryon interactions are neglected.  In this case, the
baryon energy spectra has the form $E=p^2/2M$.  Our results lead naturally
to previously obtained limiting cases such as the degenerate, nondegenerate,
and elastic approximations.  We will next consider, in Sec. IV, the
nonrelativistic Skyrme-like potential model~\cite{VB}, in which the spectrum
is given by $E = p^2/2M^* + U_{NR}$, where $U_{NR}$ denotes the
density-dependent potential contribution and $M^*$ is the Landau effective
mass, which is also generally density dependent.  Next, in Sec. V, we will
consider the effective field-theoretical Walecka-type model~\cite{SW} at the
mean-field level in which the spectrum is $E = {\sqrt{p^2+M^{*^2}}} + U_R$,
where $M^*$, the Dirac effective mass, and $U_R$, which accounts for
interactions of the fields, are both generally density dependent.  The kinetic
parts in both nonrelativistic and relativistic approaches contain the effects
of interactions insofar as they depend on density-dependent effective masses;
further, in both cases, the momentum dependence of the kinetic energy is
formally identical to their noninteracting counterparts.  The evaluation of
the cross section is therefore similar to the case of noninteracting baryons
for these particular models.  The numerical results, however, are sensitive to
the presence of interactions.  In Sec. VI, we show how additional baryonic
components, such as hyperons, affect the neutrino opacities.   In Sec. VII, we
compare our results with those of previous workers. In Sec. VIII, we
apply our results to the neutrino opacities encountered in a particular
astrophysical event:  that of a deleptonizing and cooling protoneutron star.
Finally, we summarize and provide an outlook in Sec. IX. Appendices~A and B
contain supporting material about the potential and field-theoretical models of
dense matter.

\section{NEUTRINO CROSS SECTIONS }

The total opacity of dense matter to neutrinos has contributions from both
neutral and charged weak currents.  Neutral currents are involved in
neutrino-baryon scattering while charged currents are involved in
neutrino-baryon absorption reactions.  We have considered neutral currents in
Ref.~\cite{RP2}, in which convenient expressions for the differential and total
scattering cross  sections were established. Here, we will
concentrate upon deriving similar relations for the charged current absorption
reactions. Further, the formalism
developed here may be easily generalized to include neutral current reactions
so that the present work in effect extends and replaces Ref.~\cite{RP2}.

The neutrino energies of interest to us are less than a few
hundred MeV; we may thus write the relevant interaction Lagrangian from
Wienberg--Salam theory~\cite{W,AS,SG} in terms of a
current--current interaction:
\begin{eqnarray}
{\cal L}_{int}^{cc} &=& \frac{G_F C}{\sqrt{2}} ~~l_\mu
j_W^\mu\, \qquad {\rm ~for} \qquad \nu_l + B_2 \rightarrow  l + B_4 \\
{\cal L}_{int}^{nc} &=& \frac{G_F }{\sqrt{2}} ~~l_\mu^{\nu}
j_Z^\mu\, \qquad {\rm ~for} \qquad \nu_l + B_2 \rightarrow  \nu_l + B_4 \,,
\label{gsw}
\end{eqnarray}
where $G_F\simeq 1.436\times 10^{-49}~{\rm erg~cm}^{-3}$ is the Fermi weak
coupling constant and the Cabibbo factor $C= \cos \theta_c$ for change of
strangeness $\Delta S=0$ and $C=\sin \theta_c$ for $\Delta S=1$ .  The lepton
and baryon weak charged currents are:
\begin{eqnarray}
l_\mu = {\overline \psi}_l \gamma_\mu
\left( 1 - \gamma_5 \right) \psi_\nu \,, \quad
j_W^\mu = {\overline \psi}_4 \gamma^\mu
\left( g_{V} - g_{A} \gamma_5 \right) \psi_2 \,.
\label{ccurrents}
\end{eqnarray}
Similarly, the baryon neutral currents are given by
\begin{eqnarray}
l_\mu^{\nu} = {\overline \psi}_{\nu} \gamma_\mu
\left( 1 - \gamma_5 \right) \psi_\nu \,, \quad
j_Z^\mu = \frac 12 {\overline \psi}_4 \gamma^\mu
\left( c_{V} - c_{A} \gamma_5 \right) \psi_2 \,,
\label{ncurrents}
\end{eqnarray}
where $2$ and $4 $ are the baryon initial state and final state labels,
respectively (these are identical for neutral current reactions). Other
particle labels and four-momenta $P_i$ are as shown in Fig.~1(a) for the
charged current reaction and Fig.~1(b) for the neutral current reaction.  The
vector and axial-vector coupling constants  $g_{V}$ and $g_{A}$  are  listed
in Table~I for the various charged current reactions of interest.  Similarly,
the couplings $c_V$ and $c_A$ for the neutral current reactions are listed in
Table~II.
Generally, the $\mu$ and $\tau$ neutrino charged current reactions are
kinematically suppressed; $\mu$ and $\tau$ neutrinos are thermally produced so
that their energies are of order $T<< m_{\mu}$.  On the other hand, neutral
current reactions are common to all neutrino species and the neutrino-baryon
coupling is independent of neutrino flavor. Neutrino coupling to leptons in
the same family is modified since the scattering may proceed due to both $W$
and $Z$  exchange; the couplings shown in Tables~I and II reflect this fact.
Numerical values of the  parameters that best fit data on charged current
semi-leptonic decays of hyperons are~\cite{JGGS}:  D=0.756, F=0.477,
$\sin^2\theta_W$=0.23 and $\sin\theta_c = 0.231$. Note that the amplitude for
the strangeness changing charged current is suppressed by the factor $\sin
\theta_c$. These couplings follow from SU(3) flavor symmetry for the baryons
and the quark model.  Corrections arising due to explicit $SU(3)$ breaking
terms have been recently investigated~\cite{MSJW} and in some cases are
about 10-30\%.

The cross section per unit volume of matter (or equivalently the
inverse collision mean free path) may be derived from Fermi's golden
rule and is given by
\begin{eqnarray}
\frac{\sigma(E_1)}{V}= 2\int\frac{d^3p_2}{(2\pi)^3} \int\frac{d^3p_3}{(2\pi)^3}
\int\frac{d^3p_4}{(2\pi)^3}
~(2\pi)^4\delta^4(P_1+P_2-P_3-P_4)~W_{fi} \nonumber \\
\quad~~\times f_2(E_2)(1-f_3(E_3))(1-f_4(E_4)) \,,
\label{cross}
\end{eqnarray}
where $P_i=(E_i,{\vec p}_i)$ denotes the four-momentum of particle $i$
(particle labels are as shown in Fig.~1(a)) and
the transition rate $W_{fi}$ is
\begin{equation}
W_{fi}=\frac{<|{\cal M}|^2>}{2^4E_1E_2E_3E_4} \,.
\end{equation}
Above, $|{\cal M}|^2$ is the squared matrix element and the symbol $<\cdot >$
denotes a sum over final spins and an average over the initial spins.
A common expression for both scattering and absorption may be written:
\begin{eqnarray}
W_{fi}=G_F^2
&&\bigg[({\cal V}+{\cal A})^2(1-v_2 \cos{\theta_{12}})(1-v_4\cos{\theta_{34}})
\nonumber\\
&+&~({\cal V}-{\cal A})^2(1-v_2\cos{\theta_{23}})(1-v_4\cos{\theta_{14}})
\nonumber\\
&-&~  ({\cal V}^2-{\cal A}^2)\frac{M^2}{E_2E_4}(1-\cos{\theta_{13}})\bigg] \,,
\label{rate}
\end{eqnarray}
where the vector and axial couplings ${\cal V}$ and ${\cal A}$, in the case of
absorption, stand for  $Cg_V$ and $Cg_A$, respectively.  For the reactions of
interest, $g_v$ and $g_A$ are listed in Table~I. Similarly, for the scattering
reactions of interest, ${\cal V}$ and ${\cal A}$ stand for $c_V/2$ and $c_A/2$,
respectively, which are listed in Table~II.   The particle velocities are
denoted by $v_i = p_i/E_i$, and the angle between the momentum vectors
$\vec{p_i}$ and $\vec{p_j}$ is denoted by $\theta_{ij}$. Further, $M$ is the
bare nucleon mass.  The functions $f_i(E_i)$ in Eq.~(\ref{cross}) denote the
particle distribution functions, which in thermal equilibrium are given by the
Fermi-Dirac functions
\begin{eqnarray}
f_i(E_i) = \left[ 1 + \exp \left(\frac{E_i-\mu_i}{T}\right)\right]^{-1} \,,
\end{eqnarray}
where $E_i$ are the single particle energies, $\mu_i$ are the corresponding
chemical potentials, and $T$ is the temperature.

In general, the single particle energies and chemical potentials depend on the
ambient matter conditions, i.e., the density and temperature, and also on the
interactions among the various particles.  The various chemical potentials are
determined by the conditions of charge neutrality and, in all but the most
extremely dynamical situations, chemical equilibrium.  In
some astrophysical situations, such as in the late stages of core collapse and
during the early stages of the evolution of a protoneutron star, neutrinos are
trapped on dynamical times within the matter~\cite{KS,TM} and chemical
equilibrium is established among the baryons and leptons. In this case, the
chemical potentials satisfy the relation
\begin{eqnarray}
\mu_{B_2}-\mu_{B_4}=\mu_e-\mu_{\nu_e}.
\label{bequil}
\end{eqnarray}
These situations are characterized by a trapped lepton fraction $Y_L=Y_e +
Y_{\nu_e}$, where $Y_e=(n_e-n_{e^+})/n_B$ and $Y_{\nu_e}=(n_{\nu_e} -
n_{\overline \nu_e})/n_B$ are the net electron and neutrino fractions,
respectively.  The evolution of a protoneutron star begins from a
neutrino-trapped situation with $Y_L\approx0.4$ to one in which the net
neutrino fraction vanishes and chemical equilibrium without neutrinos is
established.  In this case, the chemical equilibrium is modified by setting
$\mu_{\nu_e}=0.$  In all cases, the condition of charge neutrality requires
that
\begin{eqnarray} \sum_i \left(n_{B_i}^{(+)} + n_{\ell_i}^{(+)} \right) =
\sum_i \left(n_{B_i}^{(-)} + n_{\ell_i}^{(-)} \right)  \,,
\label{cneut}
\end{eqnarray}
where the superscript's $(\pm )$ on the number densities $n$
signify positive or negative charge.

Although neutrino opacities are required for a wide range of densities,
temperatures, and compositions, for the most part we will display results for
two limiting situations, namely beta equilibrium matter with either $Y_L=0.4$
or $Y_\nu=0.$  These are situations encountered in the evolution of a
protoneutron star~\cite{BBAL}, as discussed further in Sec. VII.

\section{NONRELATIVISTIC NONINTERACTING BARYONS}

For baryon densities $n_B \leq 5n_0$, where $n_0=0.16~{\rm fm}^{-3}$ is the
empirical nuclear equilibrium density, and in the absence of interactions
which could significantly alter their effective masses, baryons may be
considered as nonrelativistic.  The expression for $W_{fi}$ in
Eq.~(\ref{rate}) then  simplifies considerably, since  the baryon velocities
$v_i\ll 1$.   In this case, the terms involving the baryon velocities
may be safely neglected.  However, the term involving the angle between the
initial and final leptons remains.  For reactions involving nucleons,
this  term gives a small contribution, since it is proportional to ${\cal V}^2
- {\cal A}^2$.  For simplicity, and to make an apposite comparison with earlier
results in which this term was also neglected, we drop this term in this
section, but will return to a more complete analysis in the succeeding
sections.

Under these conditions, the transition rate $W_{fi}$ becomes a constant,
\begin{equation}
W_{fi}=G_F^2({\cal V}^2+3 {\cal A}^2),
\end{equation}
independent of the momenta of the participating particles, and the differential
cross section is given by
\begin{eqnarray}
\frac{1}{V}\frac{d^3\sigma}{d^2\Omega~dE_3}=\frac{G_F^2}{2\pi}
({\cal V}^2+3{\cal A}^2)(1-f_3(E_3)) S(q_0,q),
\end{eqnarray}
where the three-momentum transfer $\vec{q}=\vec{p_1}-\vec{p_3}$, so that
$q=|\vec{q}|$, and the energy transfer $q_0=E_1-E_3$. The function $S(q_0,q)$,
the so-called dynamic form factor or structure function, characterizes the
isospin response of the (nonrelativistic) system. It is simply the total phase
space available to  transfer energy $q_0$ and momentum $q$ to the
baryons.  We note that the differential cross section is needed in multi-energy
group neutrino transport codes.  However, more approximate neutrino transport
algorithms often only require the total cross section as a function of the
neutrino energy.  The cross section per unit volume given in Eq.~(\ref{cross})
then simplifies to
\begin{eqnarray}
\frac{\sigma(E_1)}{V} &=& G_F^2({\cal V}^2+3{\cal A}^2)
\int\frac{d^3p_3}{(2\pi)^3}~(1-f_3(E_3)) S(q_0,q) \label{nrfsig}, \\
S(q_0,q) &=& 2\int\frac{d^3p_2}{(2\pi)^3}
\int\frac{d^3p_4}{(2\pi)^3}~(2\pi)^4
\delta^4(P_1+P_2-P_3-P_4)f_2(E_2)(1-f_4(E_4)) \,,
\label{nrfstruc}
\end{eqnarray}
The total cross section given by Eq.~(\ref{nrfsig}) can be recast as
a double integral in $(q_0,q)$ space using $d^3p_3 =2\pi q(E_3/E_1)~ dq_0
dq$.  Since $E_3$ ranges between $0$ and $\infty$, the limits of $q_0$
are $-\infty$ and $E_1$.  The limits of $q$ are obtained by inspecting
the relation $q^2=E_1^2+E_3^2-2E_1E_3\cos\theta_{13}$ for
$\cos\theta_{13}=\pm 1$. Thus, $|q_0| < q < 2E_1-q_0$. One finds
\begin{equation}
\frac{\sigma(E_1)}{V}= \frac{G_F^2}{4\pi^2}({\cal V}^2+3{\cal A}^2)
\int_{-\infty}^{E_1} dq_0~
\frac{E_3}{E_1} (1-f_3(E_3))
\int_{\left|q_0\right|}^{2E_1-q_0} dq~ q S(q_0,q) \,.
\label{nrfsig1}
\end{equation}
This, of course, applies to both scattering and absorption
with appropriate changes of particle labels and coupling constants.

The integrals in Eq.~(\ref{nrfstruc}) can be performed analytically
and the result expressed in closed form for the noninteracting case
and for certain models of interacting matter.  The integral over the
final state momentum $p_4$ in Eq.~(\ref{nrfstruc}) may be performed by
exploiting the momentum delta function to obtain
\begin{equation}
S(q_0,q)=\frac{1}{2\pi^2}\int d^3p_2~ \delta(q_0+E_2-E_4)f_2(E_2)
(1-f_4(E_4)) \,.
\label{a1}
\end{equation}
We note that $E_4 = (\vec {p}_2 + \vec {q}~)^2/2M$, and we ignore the
difference between $M_2$ and $M_4$ as it is small compared to other
energy scales at high density for noninteracting matter.  We may
rewrite the energy delta function in terms of the angle between
$\vec{p}_2$ and $\vec{q}$:
\begin{equation}
\delta(q_0+E_2-E_4)=\frac{M}{p_2q}\delta(\cos \theta-\cos \theta_0)
\Theta(p_2^2-p_-^2) \,,
\label{edelta}
\end{equation}
where
\begin{eqnarray}
\cos \theta_0 = \frac{M}{p_2q}\left(q_0-\frac{q^2}{2M}\right) \,, \quad
p_-^2 = \frac{M^2}{q^2}\left(q_0-\frac{q^2}{2M}\right)^2
\end{eqnarray}
and $\Theta(x) = 1$ for $x \geq 0$ and zero otherwise.
Substituting these results in Eq.~(\ref{a1}) and performing the angular
integrals we obtain
\begin{equation}
S(q_0,q)=\frac{M}{\pi q}
\int_{p_-}^{\infty} dp_2~ p_2~ f_2(E_2)~(1-f_4(E_4)) \,. \\
\end{equation}
The remaining $p_2$ integral is performed by using the relation
\begin{equation}
\int~\frac{dx}{1+\exp(x)}~\frac{1}{1+\exp(-x-z)}=
-~\frac{1}{1-\exp(-z)}~\ln \frac{1+\exp(x)}{1+\exp(x + z)} \,.
\label{int}
\end{equation}
Thus, the free gas isospin density response function is given by
\begin{equation}
S(q_0,q)= \frac{M^2T}{\pi q}~\left[\frac{z}{1-\exp(-z)}
\left(1+\frac{\xi_-}{z}\right)\right],
\label{nrfstruc1}
\end{equation}
where
\begin{eqnarray}
z &=& \frac{q_0+\hat{\mu}}{T} \,, \quad \hat{\mu} = \mu_2-\mu_4 \,,
\nonumber  \\
\xi_- &=& \ln\left [\frac{1+\exp((e_--\mu_2)/T)}{1+\exp((e_-+q_0-\mu_4)/T)}
\right]\,, \quad e_-= \frac{p_-^2}{2M}
= \frac 14 \frac {(q_0-q^2/2M)^2}{q^2/2M}\,.
\label{xi}
\end{eqnarray}
This result generalizes a result obtained earlier for noninteracting
symmetric nuclear matter~\cite{KG}, in which $\hat\mu=0$, to the case of
asymmetric nuclear matter for conditions of arbitrary degeneracy.  This result,
which we further generalize to include nuclear interactions in the next
section, proves to be the key to being able to efficiently calculate opacities.
These results are easily specialized to the case of scattering by
noting that particle labels $2$ and $4$ are identical; thus
$\mu_2=\mu_4$.  The integrals in Eq.~(\ref{nrfsig1}), even with an analytical
expression for $S(q_0,q)$,  require numerical
evaluation; closed form expressions for arbitrary degeneracy cannot be
obtained.  However, in some limiting cases these integrals become analytic and
correspond to results obtained earlier and which are often used in
astrophysical simulations. \\

\noindent {\bf{\em Highly Degenerate Baryons ($\mu_i/T \gg 1)$:}} In
this situation, the participating particles all lie close to their
respective Fermi surfaces.  In this case, the $q$ integration may be
performed trivially, since the factor $(1+\xi_-/z)$ may be replaced by
$\Theta(\mu_2-e_-)$ or, equivalently, $\Theta(q-(p_{F_2}-p_{F_4}))$,
where $\Theta(x)=1$ for $x\geq 0$ and zero otherwise.  The integral to
be performed is then
\begin{eqnarray}
I_q = \int_{\left|q_0\right|}^{2E_1-q_0} dq~
\Theta(q-(p_{F_2}-p_{F_4}))  \cong
\int_{\left|\hat\mu\right|}^{2E_1+\hat\mu} dq~
\Theta(q-(p_{F_2}-p_{F_4}))  \,,
\end{eqnarray}
where in writing the rightmost relation, we have set $q_0 =
-\hat\mu$, since the exponentials in the $q_0$ integral are highly peaked at
this value at low temperature.  Thus,
\begin{eqnarray}
I_q = \left\{ \begin{array}
{r@{\quad:\quad}l}
2E_1\Theta(\hat\mu-(p_{F_2}-p_{F_4}))& {\rm for~~} \hat\mu > p_{F_2}-p_{F_4} \\
(2E_1-\hat\mu -p_{F_2} + p_{F_4})
\Theta (2E_1-\hat\mu -p_{F_2} + p_{F_4}) &
{\rm for~~} \hat\mu \leq p_{F_2}-p_{F_4}
\end{array} \right.
\end{eqnarray}
The upper limit on the remaining $q_0$ integral can be replaced by $+\infty$
since the integrand vanishes exponentially for positive values of $q_0$
due to final state Pauli blocking of the electron degeneracy.
With this substitution, it is straighforward to perform the $q_0$ integral
by noting that
\begin{eqnarray}
\int_{-\infty}^{+\infty}dz~\frac{z}{1-\exp{(-z)}}~\frac{1}{1+\exp{(z+\eta)}}=
\frac{1}{2}\left[\frac{\pi^2+\eta^2}{1+\exp{(\eta)}}\right].
\end{eqnarray}
The final result for the cross section per unit volume in the
degenerate approximation is given by
\begin{eqnarray}
\frac {\sigma(E_1)}{V} &=& \frac {G_F^2}{4\pi^3}
({\cal V}^2 + 3{\cal A}^2)
M^2 T^2 \Xi~(E_1+\hat{\mu})
\left[ \pi^2+\left( \frac{E_1-\mu_1}{kT} \right)^2 \right]
\frac {1}{1+\exp\left ((\mu_1-E_1)/ T\right)}
 \nonumber\\
{\rm with} ~~ \Xi &=&
\Theta (p_{F_4} + p_{F_3} - p_{F_2} - p_{F_1}) \nonumber\\
&+& {\displaystyle\frac {p_{F_4}+p_{F_3} - p_{F_2} + p_{F_1}}{2E_1} }
~\Theta(p_{F_1} - |p_{F_4}+p_{F_3}-p_{F_2}|) \,.
\label{dcsig}
\end{eqnarray}
In the above expression all terms proportional to $T/\mu_i$ are neglected as
they are small. Further, if we were to
assume that the incoming neutrino energy $E_1$
were equal to the neutrino chemical potential, the factor
$(E_1+\hat{\mu})$ could
be replaced by $\mu_e$. With this substitution, the above result coincides
exactly with that derived earlier  by Sawyer and Soni~\cite{SS}.

In the case of scattering, this expression simplifies to
\begin{eqnarray}
\frac {\sigma(E_1)}{V} &=& \frac {G_F^2}{16\pi^3}
(c_{V}^2 + 3c_{A}^2)M^2 T^2 E_{1}
\left[ \pi^2+\left(\frac{E_1-\mu_1}{kT} \right)^2 \right]
\frac {1}{1+\exp((\mu_1-E_1)/T)}\,.
\label{dnsig}
\end{eqnarray}
An analogous result was derived earlier by Iwamoto and Pethick ~\cite{IP}.

Note that the energy $e_-$ arises due to the kinematical condition
that ensures three-momentum conservation.  For $e_-\gg \mu_2$, in the
degenerate limit, the phase space rapidly vanishes.  Thus, at low
temperatures, this leads to the condition
\begin{equation}
q \ge |p_{F_2}-p_{F_4}| \,,
\label{urca}
\end{equation}
which is the threshold condition for the so-called direct Urca
process~\cite{LPPH}. Note that the maximum possible momentum transer
is $q=E_1+p_{F_3}$.  For a free gas in beta equilibrium with zero
neutrino chemical potential, the condition in Eq.~(\ref{urca}) is
usually not satisfied, since the neutron Fermi momenta are usually
much larger than those of protons and electrons. In contrast, in the
case when neutrinos are trapped, significantly larger proton fractions
are favored in beta equilibrium condition. This enables the threshold
condition to be easily fulfilled.  Similarly, strong interactions
also tend to increase the proton fraction, which in some cases allows
Eq.~(\ref{urca}) to be satisfied even in the vicinity of nuclear
densities~\cite{LPPH}.  The factor $1+\xi_-/z$ naturally
accounts for the threshold-like behavior with decreasing temperature.\\

\noindent {\bf{\em Nondegenerate matter}} ($\mu_i/T \ll -1$):
In the nondegenerate limit, one has
\begin{equation}
z\left(1+\frac{\xi_-}{z}\right)=z+\xi_-
\cong \exp\left({\mu_2-e_-\over T}\right)~[1-\exp(-z)] \,
\end{equation}
after expanding the logarithmic terms in Eq.~(\ref{xi}) to leading
order.  Thus, $S(q_0,q)\simeq(M^2T/\pi q)\exp((\mu_2-e_-)/T)$.
Assuming further that the effects due to final state Pauli blocking
may be neglected and the relevant energy transfer is small, $q_0 \ll
E_1$, one finds
\begin{equation}
\frac {\sigma(E_1)}{V}=
\frac{G_F^2({\cal V}^2+3{\cal A}^2)M^2T}{4\pi^3}
\exp\left( \frac{\mu_2}{T}\right) \int_{-\infty}^{E_1} dq_0~
\int_{0}^{2E_1} dq~\exp\left(-{e_-\over T}\right)
\end{equation}
In this case, the $q_0$ integration may be performed first.  To leading order
in $q$ and $T/M$ the result is $q(2\pi T/M)^{1/2}$.
The remaining $q$ integration is elementary and we obtain
\begin{equation}
\frac {\sigma(E_1)}{V}=\frac {G_F^2}{\pi} ({\cal V}^2 + 3{\cal A}^2)
E_1^2 ~n_2 \,,
\label{ndcsig}
\end{equation}
where $n_2 = 2(MT/2\pi)^{3/2}~\exp(\mu_2/T)$ is the neutron number
density in the nondegenerate limit.  In this limiting situation,  the total
cross section is simply that on a single baryon times the baryon number
density~\cite{TS}.\\

\noindent {\bf{\em The Elastic Approximation:}} We can also
derive the result of the so-called elastic  approximation~\cite{B1}, in
which it is assumed there is no energy or momentum transfer to the nucleons.
The effects of final state Pauli blocking and nucleon degeneracy are still
fully included.  In this case, the cross section may be obtained by considering
the response function for $q_0 \rightarrow 0$ and $q \rightarrow 0$:
\begin{eqnarray}
S(q_0\rightarrow 0,q \rightarrow 0) &=&
2\pi \delta (E_1-E_3) \int \frac {d^3p_2}{(2\pi)^3}~ f_2(E_2) (1-f_4(E_2))
\nonumber \\
&=& 2\pi \delta (q_0)
\frac{n_2-n_4}{1-\exp((\mu_4-\mu_2)/T)} \,,
\label{elstruc}
\end{eqnarray}
where $n_2$ and $n_4$ are the number densities of species 2 and 4,
respectively.  In obtaining this result, the identity
\begin{eqnarray}
f_2(E_2) (1-f_4(E_2)) =
\frac{f_2(E_2)-f_4(E_2)}{1-\exp((\mu_4-\mu_2)/T)} \,
\end{eqnarray}
was used.
One then finds
\begin{eqnarray}
\frac {\sigma(E_1)}{V} =
\frac {G_F^2}{\pi} ({\cal V}^2 + 3{\cal A}^2) ~~
\frac{n_2-n_4}{1-\exp((\mu_4-\mu_2)/T)} ~
E_1^2~(1-f_3(E_1)) \,.
\label{elcsig}
\end{eqnarray}
To specialize this result to the case of scattering, we need to take the limit
$(\mu_2-\mu_4)\rightarrow 0$:
\begin{eqnarray}
\frac {\sigma(E_1)}{V} =
\frac {G_F^2}{4\pi} (c_{V}^2 + 3c_{A}^2)~{\partial
n_2\over \partial(\mu_2/T)}\biggr|_T ~E_1^2~(1-f_{3}(E_1)) \,.
\label{elnsig}
\end{eqnarray}

We now present some results for noninteracting matter containing nucleons,
electrons, and trapped neutrinos  for $n_B$ and $T$ chosen to lie in the range
$(0.25-5)n_0$ and $(5-60)$ MeV, respectively.  For the beginning of the
evolution of a nascent neutron star, it is appropriate to consider the neutrino
trapped case in which the electron lepton fraction $Y_L = Y_e + Y_{\nu_e}$ is
held fixed at 0.4. At later times the neutrino chemical potential vanishes and
the matter contains thermally produced pairs; under these circumstances,
$Y_{\nu_e}$ and $\mu_{\nu_e}$ vanish and $Y_{L} = Y_e$ is determined by charge
and chemical equilibrium.

Fig.~2 shows the composition and the chemical potentials versus the density (in
units of the nuclear equilibrium density $n_0$) for $T=5,30,$ and 60 MeV for
the neutrino-trapped case.  The major effect of trapping is to keep the
electron concentration high so that matter is more proton rich in comparison to
the case in which neutrinos do not carry net lepton number; thus the threshold
conditon for the charged current rate is easily satisfied for all relevant
densities and temperatures of interest.  There is little variation in the
individual concentrations with either density or temperature. The individual
chemical potentials and ${\hat \mu}=\mu_n-\mu_p$ increase significantly with
density, but are relatively insensitive to variations in temperature as leptons
are degenerate except at the lowest densities.

In the neutrino-trapped case, in which the neutrinos are degenerate,
Pauli blocking ensures that the transport processes involve only neutrinos
close to the  Fermi surface.  Thus it is appropriate to calculate the  neutrino
mean free path  for a neutrino energy given by $E_1=\mu_{\nu_e}$.  The neutrino
absorption mean free path, $\lambda = (\sigma/V)^{-1}$, is shown in Fig.~3, as
solid lines for temperatures $T=5,30,$ and 60 MeV as a function
of density.  As expected,
the mean free path decreases with increasing density since $\mu_e$ increases
with density.  For  comparison, we also show the results of the three limiting
expressions for the extremely degenerate (Eq.~(\ref{dcsig}), dashed
curve),  nondegenerate (Eq.~(\ref{ndcsig}), dot-dashed curve), and
elastic (Eq.~(\ref{elcsig}), long dashed curve)  approximations.   It is
evident that, for the densities and temperatures of interest to us, it is
essential to include Pauli blocking of the final state electrons to obtain
reliable results. This is chiefly due to the fact that  electrons are
degenerate.  When all participating particles are extremely degenerate,
Eq.~(\ref{dcsig}) provides an excellent approximation to the exact result.
The failure of the elastic approximation in this regime is related to the fact
that the favored energy transfer is not zero as is assumed, but
$q_0=-\hat\mu$, which is generally large in magnitude. At high temperatures and
relatively low density, in which the nucleons are not extremely degenerate, the
cross section in Eq.~(\ref{elcsig}), based on the elastic approximation,
reasonably reproduces the exact results.  As expected, the nondegenerate
approximation is only realistic for extremely low densities.

Similarly, we have calculated the mean free path due to scattering reactions
assuming $E_1=\mu_{\nu_e}$, and these are compared with the charged current
results in Fig.~4 for $T=10$ MeV as a function of density.  The
total scattering opacity, which has contributions from neutrons (dashed
curve), protons (dot-dashed curve), and
electrons (dot-long dashed curve), is shown as the long-dashed curve labelled
$\lambda_S$. The results for electron scattering are actually taken from the
relativistic formulation presented in Sec. V; they cannot be evaluated from the
nonrelativistic formulation in this section because the electrons are highly
relativistic.
In the right panel, the ratio of scattering mean free path to that
of the absorption mean free path is plotted. The solid curve is the ratio of
the total scattering mean free path to that of the absorption mean free path,
the dashed curve is the ratio of the mean free due to the reaction
$\nu_e+n\rightarrow \nu_e+n$ to that of absorption.
$\nu_e+n\rightarrow e^-+p$. Clearly, the scattering mean free path is
dominated by neutrons since the electron and proton fractions are small.
However, the net electron-neutrino mean free path in the trapped
regime is dominated by the absorption reaction, for two reasons. First, the
charged current matrix element is four times larger and second, the
rate is proportional to $\mu_e$ while the scattering rate is
proportional to $\mu_{\nu_e}$.

In the neutrino-free case ($Y_{\nu_e}=0$), neutrinos still exist, but
as pairs and do not carry net
lepton number.  Thus the reference to this phase as neutrino-free is somewhat
misleading.  Fig.~5 shows the composition and the chemical potentials.
Compared to the neutrino-trapped case, Fig.~2, matter
contains a smaller proton fraction. Consequently, the
difference between the neutron Fermi momentum and proton (and electron) Fermi
momenta are large, making simultaneous energy and momentum conservation
impossibile. Thus, at low temperatures for all densities shown,
neutrino absorption due the reaction $\nu +n \rightarrow e^-+p$ is
kinematically suppressed and is unimportant; the dominant absorption
rates are those due to the modified Urca
reaction $\nu+n+X\rightarrow p+X+e^-$ where $X$
is a bystander particle.  The modified Urca reaction, however,
gives a small contribution to the total
opacity when compared with the total scattering opacity and can be
neglected. Although the direct reaction ($\nu+n\rightarrow e^-+p$) is
kinematically forbidden at zero temperature, it exhibits a strong temperature
dependance.  We also note, as can be seen in Fig. 5,
that beta-equilibrium and charge
neutrality favors a larger proton fraction at higher temperatures, which
enhances the importance of the direct process. In this sense, finite
temperature effects on the neutrino absorption reaction are similar to those
encountered due to the presence of a  finite neutrino chemical potential. It
increases both the neutrino energy and the proton fraction so as to
simultaneously fulfill momentum and energy conservation.

In Fig.~6, we plot the neutrino mean free path for neutrino-free
matter with $Y_{\nu_e}=0$. Because the neutrinos are thermal, we choose
$E_1=3T$ as the relevant energy for calculating the mean free path.
At low temperatures, the neutrino absorption mean free paths are very
large and increase with density as matter becomes increasingly degenerate.  For
higher temperatures, this behaviour is not seen and the larger neutrino energy
and a higher proton fraction both ensure that momentum and energy conservation
for the direct reaction are easily satisfied.  The extent to which the elastic
and nondegenerate approximations reproduce the exact numerical results may be
gauged from Fig.~6.  Note that the degenerate approximation is not an
appropriate limit because when the theta functions in Eq.~(\ref{dcsig}) are
zero causing the mean free paths to be infinite; thus the degenerate limit is
not plotted in this  case.

From Fig.~6 we see that the absorption rate shows a  strong temperature
dependance. In Fig.~7, we show the temperature dependance of the absorption
and scattering mean free paths at nuclear saturation density.  The individual
scattering contributions are shown in the left panel together with the total
scattering mean free path, which is shown as the
long dashed curve.  For this density, the scattering opacity is dominated by
neutron scattering.  The right panels show the ratio of the total scattering
mean free path to that of the absorption reaction $\nu+n\rightarrow e^-+p$.
For low temperatures, where the absorption is kinematically suppressed,
scattering dominates the net opacity.  However, with increasing temperature,
absorption increases and eventually dominates for $T\ge10$ MeV.

\section{NONRELATIVISTIC INTERACTING BARYONS }

We turn now to the effects of strong interactions on the charged and neutral
current neutrino cross sections.   To begin, we will continue to use the
nonrelativistic approximation for baryons.  Nonrelativistic potential
model descriptions of dense matter are based on a two-body potential fitted to
nucleon-nucleon scattering and a three-body term whose form is suggested by
theory and whose parameters are determined by the binding of few-body nuclei
and the saturation properties of nuclear matter (see for example,
Ref.~\cite{WFF}).  However, microscopic calculations of this type which
encompass asymmetric matter at an entropy/baryon in the range $s=1-2$ are not
yet available.  We therefore use a schematic potential model~\cite{PIPELR},
which is designed to reproduce the results of more microscopic calculations of
both symmetric and neutron matter at zero temperature and which can be extended
to asymmetric matter at finite temperature.   In addition to employing local
contact interactions, momentum dependent interactions arising from finite range
exchange forces can also be considered in such an approach.   The relevant
details are given in Appendix~A.

The calculation of the neutrino opacities is greatly simplified if the finite
range interactions are approximated by effective local interactions.  A notable
feature of this approximation is that the particle effective masses generally
drop more rapidly with increasing density compared to the case in which the
full nonlocality is retained. In this case, one retains only a quadratic
momentum dependence in the single particle spectrum, so that it takes a form
({\em cf.} Appendix~A) closely resembling that of a free gas:
\begin{equation}
E_i(p_i)=\frac{p_i^2}{2M_i^*}+U_i \,, \quad i=n,p \,.
\label{nrspec}
\end{equation}
Both the single particle potentials $U_i$ and the (Landau) effective
masses $M_i^*$ are density dependent.  Because the functional
dependence of the spectra on the momenta is identical to that of the
noninteracting case, the methods employed in the previous section are
amenable to the incorporation of the effects of strong interactions.
However, it is important now to retain the distinction between the
masses of the nucleons, in particular the effective masses, $M_2^*$
and $M_4^*$.

The dynamic form factor in Eq.~(\ref{nrfstruc}), with the single particle
spectrum in Eq.~(\ref{nrspec}), may be expressed in
essentially the same functional form as in Eq.~(\ref{nrfstruc1}).
The energy delta function can be recast in terms of the angle between
$\vec{p_2}$ and $\vec{q}$:
\begin{equation}
\delta(q_0+E_2-E_4)=\frac{M_4^*}{p_2q}\delta(\cos \theta-\cos \theta_0)
~\Theta(E_2 - e_-)\Theta(e_+ - E_2) \,,
\end{equation}
where
\begin{eqnarray}
\cos \theta_0 &=&
\frac{M_4^*}{p_2q}\left( c - \frac{\chi p_2^2}{2M_4^*}\right) \,, \quad
p_{\pm}^2 =\frac{2q^2}{\chi^2}\left[\left(1+\frac{\chi M_4^*c}{q^2}\right) \pm
\sqrt{1+\frac{2\chi M_4^*c}{q^2}}~\right]\,, \nonumber \\
E_2 &=& \frac{p_2^2}{2M_2^*}\,, \quad e_\pm=\frac{p_\pm^2}{2M_2^*}\,,
\label{p2pm}
\end{eqnarray}
where the new variables appearing above are defined by the relations
\begin{eqnarray}
\chi = 1- \frac{M_4^*}{M_2^*} \,, \quad
c =  q_0 + U_2 - U_4 - \frac{q^2}{2M_4^*} \,.
\end{eqnarray}
The factor $U_2-U_4$ is the potential energy gained in converting a
particle of species ``2'' to a particle of species ``4''. When the
intial and final state baryon masses are different, the $p_2$
integration is restricted to the interval $ p_+^2 \ge p_2^2 \ge
p_-^2$. Note also that the effective chemical potential $\mu_2-U_2$
replaces $\mu_2$.  With these changes, the response function
incorporates the effects of strong interactions at the mean field
level.  We can now generalize the definition of $\xi_-$, which appears in the
noninteracting response function, with $\xi_\pm$:
\begin{equation}
\xi_\pm = \ln\left[\frac{1+\exp((e_\pm-\mu_2+U_2)/T)}
{1+\exp((e_\pm+q_0-\mu_4+U_2)/T)}\right] \,.
\end{equation}
Collecting these
modifications together, and using the relation in Eq.~(\ref{int}), we have
\begin{eqnarray}
S(q_0,q)&=& \frac{M_2^*M_4^*T}{\pi q}~\frac{\xi_--\xi_+}{1-\exp(-z)} \,.
\label{nricstruc}
\end{eqnarray}
For the charged current, modifications due to interactions are
twofold.  First, the difference in the neutron and proton single
particle potentials appears in the response function and also in
$\hat\mu$.  Second, the response depends upon the nucleon effective
mass.  Since both $U_i$ and $M_i^*$ are strongly density dependent,
the opacities are significantly altered from those for the
noninteracting case at high density.

Finally, note that we can make the nondegenerate, elastic, and degenerate
approximations in analogy to those found for the noninteracting case.  The
expression for the degenerate approximation is nearly indentical to that for
the noninteracting case, because the condition $e_-\geq \mu_2-U_2$ is
equivalent to $q\geq |p_{F_2}-p_{F_4}|$.  We must only replace $M^2$ in
Eq.~(\ref{dcsig}) by $M^*_2M^*_4$.  The expression for the nondegenerate
approximation is identical to that for the noninteracting case
(Eq.~({\ref{ndcsig})), because for nondegenerate nucleons, the number density
is proportional to $(M^*_2T)^{3/2}\exp((\mu_2-U_2)/T)$.  Finally, in the
elastic case, we need only replace $\mu_4-\mu_2$ in Eq.~(\ref{elcsig}) with
$\mu_4-\mu_2+U_2-U_4$.

For the neutral currents, since the initial and final state particle
labels are identical for both leptons and baryons, we have the
simplifications
\begin{eqnarray}
z &=& \frac{q_0}{T}\,, \quad \mu_2 = \mu_4 \,, \quad
e_-=\frac{M_2^*}{2q^2}\left(q_0-\frac{q^2}{2M_2^*}\right)^2 \,,
\quad e_+=\infty\,.
\end{eqnarray}
Thus, $\xi_+=-z$, and one finds a result formally similar to
the noninteracting expression:
\begin{equation}
S(q_0,q)= \frac{M_2^{*2}T}{\pi q}
~\left[\frac{z}{1-\exp(-z)}\left(1+\frac{\xi_-}{z}\right)\right] \,.
\label{nrinstruc}
\end{equation}

Fig.~8 shows the composition and the chemical potentials versus the density for
charge-neutral matter containing interacting nucleons, electrons, and trapped
neutrinos in beta equilibrium.  While these results are qualitatively similar
to the case of noninteracting nucleons, interactions lead to lower values of
$Y_{\nu_e}$ and $\mu_{\nu_e}$, and larger values of $\hat\mu$, especially at
high densities.  The influence of these changes on the neutrino mean free paths
are depicted in Fig.~9, using $E_1=\mu_{\nu_e}$ as before for the
trapped-neutrino case, and the dynamic form factor Eq.~(\ref{nricstruc}).
Shown for comparison are the three limiting behaviors for the degenerate,
nondegenerate, and elastic approximations discussed in connection with the
noninteracting case.

The counter-intuitive result, that the mean free paths increase at high
density, is chiefly due to the use of nonrelativistic kinematics coupled with
the behavior of the Landau effective mass of the nucleons, which are displayed
in Fig.~10.  The cross section is proportional to $M^*_2M^*_4$, and the
dropping effective masses more than compensate for the higher neutrino chemical
potential at high densities.  However, the net effect of a decreasing
cross section at high density may be an anomalous result of using a
nonrelativistic theory in a situation in which the nucleons are at least
partially relativistic.  This also explains why the elastic approximation,
which is applicable in the case when the mass of the absorber is large,
provides a much poorer representation of the exact results than in the
noninteracting situation.
Related to this is the fact that  the high density
behavior of the nonrelativistic Skyrme-like EOS becomes dominated by the
$u^{8/3}$ density-dependence in the energy density.  This  leads to eventual
acausality, i.e., the speed of sound exceeds the speed of light.  While it is
possible to avoid this behavior in nonrelativistic  potential model approaches
by including more general forms of momentum dependence (see for example,
Ref.~\cite{PAL}),  a nonrelativistic description of matter at very high
densities is difficult to justify at a fundamental level.  In addition, the
calculation of the opacity is then greatly complicated by the fact that the
energy spectrum no longer resembles that of a free gas.  In the next section,
we consider field-theoretical models with relativistic kinematics which do not
have the above problems.

In the neutrino-free case, interactions change the composition significantly.
The proton fraction is larger when compared with the neutrino-free
noninteracting case.
Quantitatively, the proton fraction is sensitive to the
density dependence of the symmetry energy, which for the Skyrme interaction
we chose is nearly linearly increasing with density.
The composition and chemical potentials are shown in Fig.~11.
We shall first focus on the charged current reaction. At zero temperature,
the direct Urca reaction $\nu+n\rightarrow e^-+p$ is kinematically possible
for low energy neutrinos at and above a threshold density
when the proton fraction exceeds $1/9$~
\cite{LPPH}.  In the presence of muons, the proton fraction at threshold
density is slightly larger.
The threshold density for the model chosen here occurs at
$n=1.5n_0$ (for a free gas, by comparison,
the threshold density is  $n=73n_0$!).

In Fig.~12 the neutrino mean free paths are calculated for the neutrino-free
case. At low temperature, the absorption mean free path increases sharply below
the threshold density ($1.5n_0$). The mean free paths show similar behavior to
that of the neutrino-trapped case shown in Fig. 9, although in general they are
about 3 times smaller.  This factor can largely be understood by examining the
last three terms in the noninteracting expression Eq.~(\ref{dcsig}).  In the
neutrino-trapped case in which $E_1\approx\mu_\nu$ these terms yield
$\mu_e(Y_L=0.4)\pi^2/2$.  In the neutrino-free case in which $\mu_\nu=0$ and
$E_1=3T$, these terms yield $\mu_e(Y_\nu=0)(\pi^2+9)$.  The ratio of these two
cases is thus approximately $\mu_e(Y_L=0.4)/4.5\mu_e(Y_\nu=0)\simeq0.3$ using
values for $\mu_e$ from Figs.~8 and 11.

\section{RELATIVISTIC INTERACTING BARYONS}

At several times the nuclear equilibrium density $n_0$, the Fermi momentum and
effective nucleon mass are both expected to be comparable. Thus, a relativistic
description may be more appropriate.  Relativistic local quantum
field-theoretical models of finite nuclei and infinite nuclear matter have had
some success~\cite{SW}, albeit with rather more schematic interactions and with
less sophisticated approximations than their nonrelativistic counterparts.
Appendix~B contains a description of this approach in which the set of baryons
$B$ has been augmented to include strangeness-bearing hyperons.

The usual starting point for relativistic field theory calculations is the
mean-field approximation.  In this approximation, the presence of baryons
generates nonzero average values of the meson fields.  Baryons move
independently in these self-consistently generated mean fields.  In a model of
baryons Yukawa coupled to vector ($\omega_\mu$), isovector ($b_\mu$) and
scalar ($\sigma$) mesons, the latter simulating correlated pion-exchange, the
single particle energies are
\begin{eqnarray}
E_B = \sqrt {{\bf p}^2 + M_B^{*2} } + g_{\omega B} \omega_0
+ g_{\rho B} t_{3B} b_0 \equiv \sqrt {{\bf p}^2 + M_B^{*2}}+U\,,
\label{rspec}
\end{eqnarray}
where $M_B^{*} = M_B - g_{\sigma B}\sigma$ are the Dirac effective masses,
$\sigma$, $\omega_0$, and $b_0$ are the average values of the meson fields, and
$g_{\sigma B}$, $g_{\omega B}$, and $g_{\rho B}$ are the strong interaction
couplings of the different meson fields to baryons, and $t_{3B}$ is the third
component of isospin for the baryons.  This equation defines the effective
potential $U$.  For more details, see Appendix~B.

We turn now  to calculate the cross section for absorption of electron
neutrinos in a multi-component system described by a Walecka-type
effective field-theoretical model at the mean field level. In this model,
the nucleons become increasingly relativistic due to the rapidly dropping
nucleon effective masses.  The nonrelativistic approximation
for the hadronic current discussed thus far cannot be justified, since, at high
density, terms of
order $p/M^*$ are of order unity and must be retained.  It is convenient to
express the angles appearing in Eq.~(\ref{rate}) in terms of energy and
momentum transfers $(q_0,q)$, and to define the response functions
in terms of appropriate current-current correlation or polarization
functions~\cite{HW}. To see how this is accomplished, we start with
a general expression for the differential cross section~\cite{FW,DS}
\begin{eqnarray}
\frac {1}{V} \frac {d^3\sigma}{d^2\Omega_3 dE_3} =  -\frac {G_F^2}{128\pi^2}
\frac{E_3}{E_1}~
\left[1-\exp{\left(\frac{-q_0-(\mu_2-\mu_4)}{T}\right)}\right]^{-1}~
(1-f_3(E_3))~{\rm Im}~~(L^{\alpha\beta}\Pi^R_{\alpha\beta}) \,,
\label{dcross}
\end{eqnarray}
where the incoming neutrino energy is $E_{1}$ and the outgoing electron energy
is $E_{3}$. The factor $[1-\exp((-q_0-\mu_2+\mu_4)/T)]^{-1}$ arises
due to the fluctuation-dissipation theorem, since particles labeled `2' and '4'
are in thermal equilibrium at temperature $T$ and in chemical equilibrium with
chemical potentials $\mu_2$ and $\mu_4$, respectively. The final state blocking
of the outgoing lepton is accounted for by the Pauli blocking factor
$(1-f_3(E_3))$. The lepton tensor $L_{\alpha\beta}$ is given by
\begin{equation}
L^{\alpha\beta}= 8[2k^{\alpha}k^{\beta}+(k\cdot q)g^{\alpha\beta}
-(k^{\alpha}q^{\beta}+q^{\alpha}k^{\beta})\mp i\epsilon^{\alpha\beta\mu\nu}
k^{\mu}q^{\nu}]
\end{equation}
The target particle retarded polarization tensor is
\begin{equation}
{\rm Im} \Pi^R_{\alpha\beta} =
\tanh{\left(\frac{q_0+(\mu_2-\mu_4)}{2T}\right)} {\rm Im}~\Pi_{\alpha\beta}
\,,\\
\end{equation}
where $\Pi_{\alpha\beta}$ is the time ordered or causal polarization and is
given by
\begin{equation}
\Pi_{\alpha\beta}=-i \int
\frac{d^4p}{(2\pi)^4} {\rm Tr}~[T(G_2(p)J_{\alpha} G_4(p+q)J_{\beta})]\,.
\end{equation}
Above, $k_{\mu}$ is the incoming neutrino four-momentum and $q_{\mu}$ is  the
four-momentum  transfer. In writting the lepton tensor, we have neglected the
electron mass term, since typical electron energies are of the order of a few
hundred MeV.  The Greens' functions $G_i(p)$ (the index $i$
labels particle species) describe the propagation of baryons at finite
density and temperature.  The current operator $J_{\mu}$ is $\gamma_{\mu}$
for the vector current and $\gamma_{\mu}\gamma_5$ for the axial current. Given
the structure of the particle currents, we have
\begin{eqnarray}
\Pi_{\alpha\beta}
&=&{\cal V}^2\Pi_{\alpha\beta}^{V}+{\cal A}^2\Pi_{\alpha\beta}^{A}
-2{\cal V}{\cal A}\Pi_{\alpha\beta}^{VA} \,.
\end{eqnarray}
For the vector polarization,
$\{J_\alpha,J_\beta\}::\{\gamma_\alpha,\gamma_\beta\}$,
for the axial polarization,  $\{J_\alpha,J_\beta\} ::
\{\gamma_\alpha\gamma_5,\gamma_\beta\gamma_5\}$ and for the mixed
part,  $\{J_\alpha,J_\beta\} ::\{\gamma_\alpha\gamma_5,
\gamma_\beta\}$. Using vector current conservation and translational
invariance, $\Pi_{\alpha\beta}^V$ may be written in terms of two independent
components. In a frame where $q_{\mu}=(q_0,|q|,0,0)$, we have
\begin{eqnarray*}
\Pi_T = \Pi^V_{22} \qquad {\rm and} \qquad
\Pi_L = -\frac{q_{\mu}^2}{|q|^2}\Pi^V_{00} \,.
\end{eqnarray*}
The axial current-current correlation function can be written as a vector
piece plus a correction term:
\begin{eqnarray}
\Pi_{\mu \nu}^A=\Pi^V_{\mu \nu}+g_{\mu \nu}\Pi^A \,.
\end{eqnarray}
The mixed, axial current-vector current correlation function is
\begin{eqnarray}
\Pi_{\mu \nu}^{VA}= i\epsilon_{\mu, \nu,\alpha,0}q^{\alpha}\Pi^{VA}\,.
\end{eqnarray}
The above mean field or Hartree polarizations, which characterize the
medium response to the neutrino, have been explicitly evaluated in previous
work~\cite{LH,SMS} for the case of
neutrino scattering.  Since the structure of the charged current is
similar to that of the neutral current, we may write
\begin{eqnarray}
\frac{1}{V}\frac{d^3\sigma}{d^2\Omega dE_{\nu}} &=& -\frac{G_F^2}{16
\pi^3}~ \frac {E_3}{E_1} q_{\mu}^2 ~(1-f_3(E_3))~
\left[1-\exp{\frac{-q_0-(\mu_2-\mu_4)}{T}}\right]^{-1}[AR_1+R_2+BR_3] \,,
\nonumber \\
A&=&\frac{4E_1E_3+q_{\mu}^2}{2q^2} \,, \quad B= E_1+E_3 \,.
\label{dcross2}
\end{eqnarray}
From Eq.~(\ref{dcross2}), we see that the response of a relativistic system to
the charged or neutral current probe may be written in terms of three
response functions
$R_1$, $R_2$, and $R_3$.  (In contrast, the response in a nonrelativistic
system, Eq.~(\ref{nrfstruc}),  is characterized by a single isospin density
response function. This is due to the fact that  when the baryon velocity $v_i
\sim 1$, the angular terms in Eq.~(\ref{rate}) cannot be dropped.).  The
various response functions required have been studied in earlier works for
neutral current reactions~\cite{HW,RP2}.  Their generalization to the case of
charged current reactions are:
\begin{eqnarray}
R_1&=&({\cal V}^2+{\cal A}^2) ~[{\rm Im}~ \Pi^R_L(q_0,q)+{\rm Im}~ \Pi^R_T(q_0,q)]\\
R_2&=&({\cal V}^2 + {\cal A}^2)~{\rm Im}~ \Pi^R_T(q_0,q)
- {\cal A}^2~{\rm Im}~ \Pi^R_A(q_0,q)\\
R_3&=&2{\cal V}{\cal A} ~{\rm Im}~ \Pi^R_{VA}(q_0,q)  \,.
\end{eqnarray}
These response functions have been written in terms of the
imaginary part of the polarization functions, whose causal components
are given in Ref.~\cite{LH,SMS} for symmetric nuclear matter.  We present
extensions of these results to
asymmetric matter, and, in particular, to unlike particle-hole
excitations. For space like excitations, $q_{\mu}^2\le0$, they are
given by
\begin{eqnarray}
{\rm Im}~ \Pi_L(q_0,\vec{q})
&=&2 \pi\int\frac{d^3p}{(2\pi)^3}
~~\frac{E_p^{*2}-|p|^2\cos^2\theta}{E_p^*E_{p+q}^*}~~\Theta\\
{\rm Im}~ \Pi_T(q_0,\vec{q})
&=&\pi\int\frac{d^3p}{(2\pi)^3}
~~\frac{q_{\mu}^2/2-|p|^2(1-\cos^2\theta)}{E_p^*E_{p+q}^*}~~\Theta\\
{\rm Im}~ \Pi_A(q_0,\vec{q})
&=&2\pi\int\frac{d^3p}{(2\pi)^3}
~~\frac{M_2^{*^2}}{E_p^*E_{p+q}^*}~~\Theta\\
{\rm Im}~ \Pi_{VA}(q_0,\vec{q})
&=&2\pi\int\frac{d^3p}{(2\pi)^3}
~~\frac{q_\mu^2M_2^*}{|q^2|E_p^*E_{p+q}^*}~~\Theta \,.
\end{eqnarray}
In the above,
\begin{eqnarray}
\Theta&=&F(E_p^*,E_{p+q}^*)[\delta(q_0-(E_{p+q}-E_p))+
\delta(q_0-(E_p-E_{p+q})]\\
F(E_p^*,E_{p+q}^*)&=&f_2(E^*_p)(1-f_4(E_{p+q}^*))\\
E_{p}^*&=&\sqrt{|p|^2+M_2^{*2}}\,,  ~~\quad
E_{p} = E_{p}^*+U
\end{eqnarray}
The particle distribution functions $f_i(E)$ are given by the
Fermi-Dirac distribution functions
\begin{eqnarray}
f_i(E_p^*)=\frac{1}{1+\exp((E_p^*-\nu_i)/kT)} \,,
\label{FD}
\end{eqnarray}
where $\nu $ is the effective chemical potential defined by
\begin{equation}
\nu_i=\mu_i - U_i =\mu_i- (g_{\omega B_i}\omega_0 + t_{3 B_i}g_{\rho B_i}b_0)\,,
\end{equation}
The single particle spectrum in Eq.~(\ref{rspec}) is discussed in Appendix B.
The angular integrals are performed by exploiting the delta functions.  The
three dimensional integrals can be reduced to the following one dimensional
integrals:
\begin{eqnarray}
{\rm Im}~ \Pi_L(q_0,{q})
&=& \frac{q_{\mu}^2}{2\pi |q|^3}
\int_{e_-}^{\infty}dE~~[(E+q_0/2)^2-|q|^2/4]\nonumber\\
&\times&[F(E,E+q_0)+F(E+q_0,E)]\\
{\rm Im}~ \Pi_T(q_0,{q})
&=& \frac{q_{\mu}^2}{4\pi |q|^3}
\int_{e_-}^{\infty}dE~~[(E^*+q_0/2)^2+|q|^2/4+|q|^2M_2^{*^2}/q_{\mu}^2)]
\nonumber\\&\times & [F(E,E+q_0)+F(E+q_0,E)]\\
{\rm Im}~ \Pi_A(q_0,{q})
&=&\frac{M_2^{*^2}}{2\pi|q|}\int_{e_-}^{\infty}dE~~[F(E,E+q_0)+F(E+q_0,E)]\\
{\rm Im}~ \Pi_{VA}(q_0,{q})
&=&\frac{q_\mu^2}{8\pi |q|^3}
\int_{e_-}^{\infty}dE~~[2E+q_0][F(E,E+q_0)+F(E+q_0,E)]\ \,.
\end{eqnarray}
The lower cut-off $e_-$ arises due to kinematical restrictions
and is given by
\begin{eqnarray}
e_-&=&-\beta \frac{\tilde{q_0}}{2}
+\frac{q}{2}\sqrt{\beta^2-4\frac{M_2^{*2}}{q^2-\tilde{q_0^2}}}\,,
\end{eqnarray}
where
\begin{equation}
\tilde{q_0}=q_0+U_2-U_4 \,,\quad
\beta = 1+\frac{M^{*^2}_4-M^{*^2}_2} {q^2-\tilde{q_0^2}}\,.
\end{equation}
It is convenient to reexpress the polarization functions as follows:
\begin{eqnarray}
{\rm Im}~ \Pi^R_L(q_0,q)&=&\frac{q_{\mu}^2}{2\pi |q|^3}
\left[I_2 + q_0I_1+\frac{q_{\mu}^2}{4}I_0\right] \\
{\rm Im}~ \Pi^R_T(q_0,q)&=&\frac{q_{\mu}^2}{4\pi |q|^3}
\left[I_2 + q_0I_1+
\left(\frac{q_{\mu}^2}{4} +
\frac{q^2}{2}+M_2^{*^2}\frac{q^2}{q_{\mu}^2}\right) I_0\right] \\
{\rm Im}~ \Pi^R_A(q_0,q)&=&\frac{M_2^{*^2}}{2\pi |q|}I_0\\
{\rm Im}~ \Pi^R_{VA}(q_0,q)&=&\frac{q_{\mu}^2}{8\pi |q|^3}[q_0I_0+2I_1] \,,
\label{polar}
\end{eqnarray}
where we used the one-dimensional integrals
\begin{equation}
I_n=\tanh{\left(\frac{q_0+(\mu_2-\mu_4)}{2T}\right )}\int_{e_-}^{\infty}dE~E^n~[F(E,E+q_0)+F(E+q_0,E)]\,.
\end{equation}
These integrals may be explicitly expressed in terms of the Polylogarithmic
functions
\begin{equation}
Li_n(z) = \int_0^z \frac {Li_{n-1}(x)}{x} \,dx \,,
\qquad Li_1(x) = \ln (1-x)
\end{equation}
which are defined to conform to the definitions of Lewin~\cite{LL}.
This Polylogarithm representation is particularly useful and compact:
\begin{eqnarray}
I_0 &=& T~z\left(1+\frac{\xi_1}{z}\right)\,, \\
\label{iis0}
I_1 &=& T^2~z\left(\frac{\mu_2-U_2}{T}  -
\frac{z}{2}+\frac{\xi_2}{z}+\frac{e_-\xi_1}{zT}\right) \,,\\
\label{iis1}
I_2 &=& T^3~z\left(\frac{(\mu_2-U_2)^2}{T^2} -
z\frac{\mu_2-U_2}{T} +
\frac{\pi^2}{3}+\frac{z^2}{3}-2\frac{\xi_3}{z}+2\frac{e_-\xi_2}{Tz}
+\frac{e_-^2\xi_1}{T^2z}\right) \,,
\label{iis2}
\end{eqnarray}
where $z=(q_0+(\mu_2-\mu_4))/T$ and the factors $\xi_n$ are given by
\begin{equation}
\xi_n = Li_n(-\alpha_1) - Li_n(-\alpha_2)\,,
\label{dlis}
\end{equation}
with
\begin{equation}
\alpha_1 = \exp\left((e_--\mu_2+U_2)/T\right)\,,\quad
\alpha_2 = \exp\left((e_-+q_0-\mu_4+U_2)/T\right) \,.
\end{equation}

\noindent We note that the nonrelativistic structure function for
neutral current scattering, Eq.~(\ref{nrinstruc}) is, aside from the
factor $M_2^{*^2}T/\pi q$, equal to $I_0$ since
$\xi_1\equiv\xi_-$.

The total cross section is the double integral in $(q_0,q)$ space:
\begin{eqnarray}
\frac{\sigma(E_1)}{V}=
\frac{G_F^2}{2\pi^2E_1^2} \int_{-\infty}^{E_1}
dq_0~ \frac{(1-f_3(E_3))}{1-\exp{\frac{-q_0-(\mu_2-\mu_4)}{T}}}
\int_{|q_0|}^{2E_1-q_0} dq~q~q_{\mu}^2~ [AR_1+R_2+BR_3]\,.
\label{master}
\end{eqnarray}
Eq.~(\ref{master}) allows us to calculate the cross section per unit volume,
or equivalently the inverse mean free path, consistently with the relativistic
field-theoretical model in the mean field approximation. This naturally
incorporates the effects of strong interactions, Pauli blocking of final state
particles and the contribution of relativistic terms to the baryon currents.
In the case of neutral currents, some of the terms in the above are slightly
simplified:
\begin{eqnarray}
z = \frac{q_0}{T} \,, \quad \mu_2 = \mu_4\,, \quad
e_- = -\frac{q_0}{2}
+\frac{q}{2}\sqrt{1-4\frac{M_2^{*2}} {q_{\mu}^2}}
\end{eqnarray}

We note that other expressions for the cross sections, based on a
similar formalism, have been derived earlier in the literature~
\cite{HW,RP2}. However, negative values of $q_0$ were not taken into account in
Ref.~\cite{HW} and the response functions used in Refs.~\cite{HW,RP2}
inadvertantly omitted the factor $(1-\exp(-z))$ in the denominator of
Eqs.~(\ref{dcross2}).  While the qualitative results in
Refs.~\cite{HW,RP2} remain unchanged, they are quantitatively affected.

In Fig.~13, the  individual particle fractions and the relevant chemical
potentials are shown  for a
lepton fraction $Y_L=0.4$. As in the case of the potential  model, interactions
in this model lead to  larger proton fractions compared to the noninteracting
case.  These results are almost indistinguishable from those of the potential
model.
The absorption (upper panels) and scattering (lower panels) mean free paths,
calculated using Eq.~(\ref{master}), are shown in Fig.~14 for three different
temperatures, $T=5,30,$ and $60$ MeV, as a function of baryon density. At
low temperatures, the neutrino mean free path has relatively little
variation with density due to the dropping baryon effective masses at high
density.  Nevertheless, the mean free paths eventually decrease with
increasing density, unlike in the potential model.  We attribute this to the
inclusion of relativistic kinematics.

The results for the neutrino-free case are shown in Figs.~15 and 16.
Qualitatively, the results
have similar density and temperature behaviour as in both the neutrino-trapped
case and in the nonrelativistic
case.  The upper panels show
the absorption mean free paths and the lower panels show the scattering mean
free path. The threshold density for the charged current reaction to be
kinematically allowed at zero temperature in this model occurs at $1.7 n_0$.
This accounts for the sudden increase in absorption mean free path below this
density at low temperatures. At higher temperatures, this threshold-like
behaviour is much less pronounced, neutrino absorption mean free path dominate
the total opacity for electron neutrinos. From the leftmost panels in Fig.~16,
we see that even at $T=5$ MeV, despite kinematical restrictions at low density
the absorption reaction always dominates over the scattering reaction.

\section{COMPARISON WITH OTHER WORKS}

Calculations of protoneutron star evolution have been performed by several
groups~\cite{BL,WM,SuS,K,KJ}.  At supra-nuclear densities, these
groups used for the most part neutrino cross sections as originally
described in Burrows and Lattimer~\cite{BL}.  Those cross sections
were generated by
interpolation among limiting formulae for degenerate and nondegenerate matter
and neutrinos analogous to the limiting formulae given in Sec. III for
noninteracting, nonrelativistic baryonic matter.   A direct comparision of
those results with the results obtained here for arbitrary degeneracy
and including
interactions is therefore not straightforward. First, the limiting expressions
should include the effects of interactions. Second, the results will be
sensitive to the interpolation algorithm, especially the interpolation
parameter.

In addition, relativistic effects become increasingly important with density
in a field-theoretical description, since the nucleon effective mass decreases
with increasing density.  Quantitatively, the effects arising due to the
relativistic structure of the baryon currents is small and is typically of
order $E{\nu}/M^*$.  However, relativistic kinematics introduces important
corrections (of order $p_f/M^*$) in the degenerate limit. To facilitate an
illustrative comparision of our results with other results often used in the
literature, we study the neutrino mean free path in the trapped regime.  In
this case,  the dominant neutrino cross sections are those due to absorption
(on neutrons) and scattering off neutrons and protons. The appropriate limiting
expressions for the absorption and scattering cross sections
for the degenerate case become
\begin{eqnarray}
\frac {\sigma^A_{D,n}(E_{\nu})}{V} &=& \frac {G_F^2}{4\pi^3}
(g_{V}^2 + 3g_{A}^2)E^{*2}_{F_n} T^2 \mu_{e}
\left[ \pi^2+\left(\frac{E_\nu-\mu_{\nu}}{kT} \right)^2 \right]
\frac {1}{1+\exp((\mu_{\nu}-E_{\nu})/T)}\,, \\
\frac {\sigma^S_{D,i}(E_{\nu})}{V} &=& \frac {G_F^2}{16\pi^3}
(c_{V}^2 + 3c_{A}^2)E^{*2}_{F_i} T^2 E_\nu
\left[ \pi^2+\left(\frac{E_\nu-\mu_{\nu}}{kT} \right)^2 \right]
\frac {1}{1+\exp((\mu_{\nu}-E_{\nu})/T)}\,,
\end{eqnarray}
where $E_{F_i}^2=p_{Fi}^2+M_i^{*2}$ replaces $M_i^2$ appearing in the formulae
for nonrelativistic, noninteracting nucleons Eq.~(\ref{dcsig})
($i=n,p$). For scattering off neutrons and protons, $c_V$ and $c_A$
are appropriately chosen.
The nondegenerate cross sections $\sigma^A_{ND,n}$ and $\sigma^S_{ND,i}$ are
unchanged from Eq.~(\ref{ndcsig}).

We compare the results for the total cross section
obtained by interpolation as opposed to the exact
integrations ($\sigma_{rpl}$) for the case of neutrino-trapped matter
with $Y_L=0.4$.  For this
exercise, we chose a field-theoretical model and therefore used the limiting
formulas for the $\sigma_D$'s and $\sigma_{ND}$'s just described. The
total cross section is the sum of those from absorption on neutrons and
scattering off neutrons and protons.
As an illustration,
we first employ the particular interpolation scheme of Keil and
Janka~\cite{KJ}:
\begin{equation}
\sigma_1= \sum_{i=n,p} \left(
\sigma_{D,i}^S~\frac{X_i}{1+X_i} + \sigma_{ND,i}^S~\frac{1}{1+X_i}
\right) +
\sigma_{D,n}^A~\frac{X_n}{1+X_n} + \sigma_{ND,n}^A~\frac{1}{1+X_n} \,,
\label{int1}
\end{equation}
where $X_i={\rm max}[0,\eta_i]$ where $\eta_i$ is the degeneracy parameter.
For a relativistic model, the appropriate degeneracy parameter is
$\eta_i=(\nu_i-M_i^*)/T$ which can be seen by reference to Eq.~(\ref{FD}).
Fig.~17 (lower left panel) shows the ratio of our exact integration compared
with $\sigma_1$ as a function of density and temperature and assuming
$E_\nu=\mu_\nu$.  This interpolation gives reasonable cross sections in the
nondegenerate limit but is poor in the degenerate limit.  This behavior is
easily understood by examining the ratio $\sigma_{ND}^A/\sigma_D^A$, which for
$E_\nu=\mu_\nu$ is proportional to $nE_\nu^2/(E_{F_n}^{*2}T^2\mu_e)$.  We can
neglect the relatively weak density dependence of $E_{F_n}^*$, since the drop
in $M_n^*$ is compensated somewhat by the increase of $k_{F_n}$, and note that
in the neutrino-trapped case both $E_\nu=\mu_\nu$ and $\mu_e$ scale as
$n^{1/3}$.  Thus, $\sigma_{ND}^A/\sigma_D^A\sim
n^{4/3}/T^2\sim(E_{Fn}/T)^2=\eta^2$ where $E_{F_n}$ is the nonrelativistic
neutron Fermi energy.  In the degenerate limit, therefore, the $\sigma_{ND}^A$
contribution can still dominate that from $\sigma_D^A$. This may also be
inferred by comparing with the ratio $\sigma_{rpl}/\sigma_{ND}$ in the
degenerate regions as shown in the upper left panel of Fig.~17, where
$\sigma_{ND} = \sigma_{ND}^A + \sigma_{ND}^S$.

These results suggest that a better interpolation formula is
\begin{equation}
\sigma_3= \sum_{i=n,p} \left(
\sigma_{D,i}^S~\frac{X_i^3}{1+X_i^3} + \sigma_{ND,i}^S~\frac{1}{1+X_i^3}
\right) +
\sigma_{D,n}^A~\frac{X_n^3}{1+X_n^3} + \sigma_{ND,n}^A~\frac{1}{1+X_n^3} \,.
\label{int2}
\end{equation}
The ratio of the exact to these interpolated results are shown Fig.~17.
The interpolation now correctly goes to the required limits; for
reference the exact result is compared directly with the two limiting
forms in the upper left (nondegenerate) and upper right (degenerate)
panels.  Nevertheless, significant errors for intermediate degeneracies
still exist, which underscores the importance of the relatively simple
exact formulae we have
found.  There is no significant need to use interpolations any longer.

\section{MULTI-COMPONENT ENVIRONMENT}

Strangeness-bearing components can appear in dense matter either in the form of
hyperons, kaon condensate, or quarks.   In this section, we investigate the
effects of multi-components on the neutrino mean free paths by concentrating on
the possible presence of hyperons. We present specific results for a
relativistic field-theoretical model in which the baryons, $B$, interact via
the exchange of $\sigma,\omega,$ and $\rho$ mesons~\cite{SW}. The relevant
details are  presented in Appendix~B.

For a typical case, the
particle fractions in matter containing hyperons are shown in the upper
panels of Fig.~18  for different temperatures in the neutrino-trapped case,
$Y_L=0.4$.   The lower  panels show the behaviour of the relevant chemical
potentials.  Note that the neutrino  chemical potential increases rapidly with
the onset of hyperons in matter (compare with the results in Fig.~12) due to
the increase in the positive charge in the system.  This is typical for matter
in which strangeness, due to any source, appears.
Consequently, in matter with hyperons neutrino energies are somewhat larger
than those in nucleons-only matter.

The absorption mean free paths in matter containing hyperons are shown in
Fig.~19. The various  reactions that contribute to the total absorption opacity
are calculated using Eq.~(\ref{master}) by using the appropriate chemical
potentials  and neutrino coupling constants.  Strangeness-changing reactions
are Cabibbo suppressed and hence their contribution to the total opacity is
negligible. The relative importance of the various  reactions are shown in the
upper  panels. Pauli blocking and kinematic restrictions account for the
threshold-like structure seen at low temperatures, particularly for the reaction
$\nu_e+\Sigma^-\rightarrow\Lambda+e^-$.
This is analogous to the threshold behavior seen for the reaction
$\nu_e+n \rightarrow e^-+p$ in hyperon-free matter. The kinematical
restrictions on reactions involving baryons require that the initial and final
state baryon Fermi momenta not be vastly different. When too large a difference
exists, the phase space is highly suppressed. At higher temperatures, these
kinematical restrictions are relaxed.  The higher neutrino energy and the
lower degeneracy of the baryons account for the qualitative trends of the
results at the higher temperatures $T=30$ MeV and $T=60$ MeV.
The solid lines in the  lower panels
show the net mean free path from all contributions from absorption reactions.
For comparison, the
net mean free path in a  model  without hyperons is shown by the dashed
curves.  In general, the presence of hyperons has the effect of decreasing the
neutrino mean free path.

In  Fig.~20, the scattering mean free paths are shown.  The upper panels show
the individual contributions to the total scattering mean free path due to the
various reactions of interest.   In the lower panels, the net mean free paths
in hyperonic matter (solid curves) are compared with those in nucleons-only
matter (dashed curves).   The appearance of hyperons again decreases the
neutrino
mean free path.  The overall scattering mean free path is less than 50\% larger
than the absorption mean free path, so that scattering provides an important
contribution.

In summary, the decrease in mean free paths in a multi-component environment
may be understood by noting that: (1) A larger
neutrino  chemical potential results in larger neutrino energies, (2) hyperons
decrease the degeneracy of baryons, and (3) hyperons  provide additional
channels of scattering reactions to occur.

When neutrinos carry no net lepton number, they do not play a role in
determining the composition of matter. In neutrino-free matter, the hyperons
appear in larger numbers and at threshold densities which are lower than in
neutrino-rich matter. In Fig.~21, particle fractions (upper panels)  and the
electron chemical potentials (lower panels) in  neutrino-free matter with
hyperons are shown.  Since the neutrinos in this case are thermal,
we present results for  absorption and scattering mean free paths
calculated for $E_{\nu}=3T$.

In  Fig.~22 are shown the relative contributions of the dominant charged
current reactions
(upper panels)  and the total neutrino absorption mean free paths (lower
panels).  At low temperatures, the reactions show the expected threshold-like
behavior.  The total
absorption opacity due to all possible reactions (solid curves) is shown in the
lower panels.  For comparison, results in matter without hyperons is also shown
by the dashed curves. As in the neutrinos-trapped case,  the neutrino mean free
paths are smaller in matter with hyperons, albeit by a smaller amount.
Scattering off hyperons, shown in Fig. 23, clearly leads to important
modifications to the neutrino mean free path in the neutrino-free case.

\section{NEUTRINO TRANSPORT IN AN EVOLVING PROTONEUTRON STAR}

\subsection{Introduction}

One of the most important applications of dense matter opacity calculations is
the environment of a newly formed neutron star.  A protoneutron star is formed
subsequent to the core bounce of a massive star in a gravitational collapse
supernova.  Its early evolution has been investigated in
Refs.~\cite{BL,WM,SuS,K,KJ}.
Detailed studies of the dynamics of core collapse and supernova indicate that
within milliseconds of the shock wave formation the formerly collapsing stellar
core settles into nearly hydrostatic equilibrium, with a relatively low entropy
and large lepton content.

The entropy per baryon, $s$, is about one or less (measured in units of
Boltzmann's constant), which corresponds to a temperature of about 10-20 MeV.
The electron lepton fraction $Y_{L}=Y_{e}+Y_{\nu_e}$ at bounce in the interior
is estimated to be about $0.4$.  The $\nu_e$'s formed and trapped in the core
during collapse are degenerate with a chemical potential of about 300 MeV.  In
addition, because no $\mu$- or $\tau$- leptons were present when neutrino
trapping occurred, the net numbers of either $\mu$ or $\tau$ leptons is
zero: e.g., $Y_{\nu_\mu}=-Y_\mu\approx0$.  The $\mu$ and $\nu_\mu$  chemical
potentials in beta equilibrium are related by
$\mu_\mu-\mu_{\nu_\mu}=\mu_e-\mu_{\nu_e}$, each side of which has a value of
order 100 MeV.  Unless $\mu_\mu > m_\mu c^2$, however, the net number of
$\mu$'s or $\nu_\mu$'s present is zero.

The temperature and entropy increase beyond about 0.5 M$_\odot$ from the center
because of shock heating~\cite{BL}. During the early deleptonization or
neutrino loss phase, the stellar interior gains entropy because of resistive
neutrino diffusion, and the regions near the protoneutron star surface lose
entropy because of neutrino losses.  Eventually, this reverses the positive
temperature and entropy gradients in the interior. On time scales of about
10-15 s, the center reaches a maximum entropy of about 2 and a maximum
temperature of 40--60 MeV.  This time coincides with the loss of virtually all
the net trapped neutrino fraction $Y_{\nu_e}$ in the interior, although there
are still considerable numbers of neutrino pairs of all flavors present in
thermal equilibrium.  This time, therefore, marks both the end of the
deleptonization phase and the onset of the cooling phase of the protoneutron
star.

The baryonic composition of dense matter is greatly affected by the degree to
which neutrinos are trapped.  Thus there exists an unambiguous compositional
difference between the initial deleptonization (neutrino-trapped) and cooling
(neutrino-free) phases~\cite{PIPELR}.  This difference could be enhanced if
strangeness, in the form of hyperons, kaons or quark matter, is considered.
Neutrino trapping delays the appearance of the strange matter components to
higher baryon densities.  This implies that during the early deleptonization
phase, matter may consist mostly of nonstrange baryons, except possibly at the
very center of the star.  But the cooling phase may be characterized by the
presence of a substantial amount of strange matter, due to both the decreasing
threshold density for their appearance and to the increasing central density of
the star.  It is therefore of considerable interest to examine the behavior of
neutrino opacities in matter characterized by these different  thermodynamic
conditions.

\subsection{Semi-analytic treatment of neutrino transport}

The neutrino transport in the bulk of the interior may be treated in the
diffusion approximation, since neutrinos are very nearly in thermal equilibrium
due to the large weak-interaction rates and the typical mean free path is very
small compared to the stellar radius until after about a minute when the mean
neutrino energy becomes very small. In the diffusion approximation, neglecting
the effects of general relativity, the rate of
change of electron lepton number is related to the electron
neutrino number gradient by
\begin{eqnarray}
n\frac{\partial Y_L}{\partial t} =
\frac{1}{r^2}\frac{\partial}{\partial r}
\left[r^2\int\frac{c}{3}\left(\lambda_\nu(E_\nu)
\frac{\partial n_\nu(E_\nu)}{\partial r}-\lambda_{\bar\nu}(E_\nu)
\frac{\partial n_{\bar\nu}(E_\nu)}{\partial r}\right)\right] dE_\nu\,,
\label{ndyl}
\end{eqnarray}
where $n$ is the baryon number density,
\begin{eqnarray}
n_\nu(E_\nu)=\frac{E_\nu^2}{2\pi^2(\hbar c)^3} f_\nu(E_\nu),\qquad
n_{\bar\nu}(E_\nu)=\frac{E_\nu^2}{2\pi^2(\hbar c)^3} f_{\bar\nu}(E_\nu)
\end{eqnarray}
are the electron neutrino and antineutrino number densities, respectively,
 at energy $E_\nu$,
$f_\nu=(1+e^{(E_\nu-\mu_\nu)/T})^{-1}$ and $f_{\bar\nu}=
(1+e^{(E_\nu+\mu_\nu)/T})^{-1}$ are the neutrino and anti-neutrino
distribution functions,
and $Y_L=Y_\nu+Y_e$ is the total number of leptons per baryon.  The net
neutrino mean free path is due to both absorption and scattering:
\begin{eqnarray}
\lambda_\nu^{-1}(E_\nu)=\frac{1}{1-f_\nu(E_\nu)}\sum_{i=a,s}
\frac {\sigma_i(E_\nu)}{V}\,,
\end{eqnarray}
where $\sigma_a$ and $\sigma_s$ are the absorption and scattering cross
sections, respectively, and the factor $1-f_\nu(E_\nu)$ accounts for the
inverse process~\cite{IP}.  A similar relation exists for the antineutrino mean
free path, but uses $f_{\bar\nu}$ and the appropriate absorption and scattering
cross sections ($\bar\sigma_a, \bar\sigma_s$) for $\bar\nu_e$s.
Eq.~(\ref{ndyl}) can be rewritten in terms of the
diffusion coefficients
\begin{eqnarray}
D_n = \int_0^{\infty}dE_{\nu}~E_{\nu}^n~\lambda_\nu(E_{\nu})
~f_\nu(E_{\nu})(1-f_\nu(E_{\nu}))
\label{dn}
\end{eqnarray}
and a similarly defined $D_{\bar n}$ as follows:
\begin{eqnarray}
n\frac{\partial Y_L}{\partial t} =
\frac{c}{6\pi^2(\hbar c)^3}\frac{1}{r^2}\frac{\partial}{\partial r}
\left[r^2\left((D_2+D_{\bar 2})\frac{\partial(\mu_\nu/T)}{\partial r}
- (D_3-D_{\bar 3})\frac{\partial (1/T)}{\partial r}\right)\right].
\label{dyl}
\end{eqnarray}
The diffusion constants are related to the conventional
Rosseland mean free paths by
\begin{eqnarray}
\lambda_n = \frac {D_n}
{\int_0^{\infty}dE_{\nu}~E_{\nu}^n~f(E_{\nu})(1-f(E_{\nu}))}.
\label{lambda2}
\end{eqnarray}
The diffusion constants $D_2, D_3, D_{\bar 2}$ and $D_{\bar 3}$ only
contain contributions from electron-type neutrinos, since only these are
associated with changes in electron-neutrino number.

During the deleptonization phase, the neutrinos are mostly degenerate.
Thus, the diffusion of antineutrinos can be essentially ignored and
the neutrino properties are essentially only a function of $\mu_\nu$.
For nearly degenerate matter in beta equilibrium, one can show that
\begin{eqnarray}
\frac{\partial Y_L}{\partial Y_\nu} \simeq
\left(\frac{\partial Y_L}{\partial Y_\nu}\right)_o
\frac{\mu_{\nu,o}}{\mu_\nu}
\end{eqnarray}
where the subscript $o$ indicates values at the beginning of deleptonization.
For example, for $\mu_{\nu,o}\sim200$ MeV,
$(\partial Y_L/\partial Y_\nu)_o\simeq3$.
It is instructive to explore the transport behavior for various assumptions
concerning the energy dependence of the mean free path.  Supposing that
$\lambda_\nu(E_\nu)=\lambda_o(n, T) (\mu_{\nu,o}/E_\nu)^{m}$,
one finds in the degenerate limit
\begin{eqnarray}
D_n=\lambda_o T \mu_{\nu,o}^m\mu_\nu^{n-m}\,. \qquad\qquad n\ge m
\end{eqnarray}
Therefore, assuming that during the deleptonization phase the remnant
is nearly hydrostatic, Eq.~(\ref{dyl}) can be written in terms of the
neutrino chemical potential as
\begin{eqnarray}
\left(\frac{\partial Y_L}{\partial Y_\nu}\right)_o
\frac{3\mu_{\nu,o}^{1-m}\mu_\nu}{c}
\frac{\partial\mu_\nu}{\partial t} =
\frac{1}{r^2}\frac{\partial}{\partial r}
\left[r^2\lambda_o\mu_\nu^{2-m}\frac{\partial\mu_\nu}{\partial r}\right] \,.
\label{dyl1}
\end{eqnarray}
Note that the terms involving the temperature gradient in
Eq.~(\ref{dyl}) vanish in the degenerate limit irrespective of the
value of $m$.  The temperature now only enters through $\lambda_o(n,
T)$.  Inasmuch as the temperature in the interior of the neutron star
varies with time only by a factor of about two during the
deleptonization, while $\mu_\nu$ varies by a much larger factor, a
simple understanding of deleptonization can be obtained by treating
$\lambda_o$ as a constant in both space and time.

As shown in Ref.~\cite{PIPELR}, approximate solutions of
Eq.~(\ref{dyl1}) can be found by separating the time and radial
dependences in $\mu_\nu$:
\begin{eqnarray}
\mu_\nu=\mu_{\nu,o}\phi(t)\psi(r)\,,
\end{eqnarray}
with $\phi(0)=1$ and $\psi(0)=1$. This separation is justifiable since the
remnant is essentially hydrostatic.  The functional forms of $\phi$
and $\psi$, and the eigenvalue of the solution, depend upon the value
of $m$.  In Ref.~\cite{PIPELR}, it was assumed that $m=2$, in which
case $\phi$ decreases linearly with time: $\phi=1-t/\tau$. If,
instead, $m=0$, then $\phi$ varies like $(1+t/\tau)^{-1}$.  In the
case of $m=1$, $\phi$ varies like $\exp{-(t/\tau)}$. In each case,
$\tau$ is a deleptonization time and is proportional to
$R^2/c\lambda_o$ where $R$ is the stellar radius. In detail, one finds
\begin{eqnarray}
\tau = \left(\frac{\partial Y_L}{\partial Y_\nu}\right)_o
\frac{3R^2}{c\lambda_o} \frac{(3-m)^{2/(3-m)}}{\xi_{n,1}^2} \,,
\end{eqnarray}
where $\xi_{n,1}$ is the Lane-Emden radial eigenvalue for index
$n=2/(3-m)$~\cite{Chandra}.  For the cases $m=0,1$ and 2, one finds $n=2/3, 1$,
and 2, and $\xi_{n,1}=2.871, \pi$, and 4.353, respectively.  Below, we examine
the behavior of $D_2$ with respect to $\mu_\nu$, for fixed temperature, in the
interior of a star at the onset of deleptonization, as it appears that the
details of deleptonization fundamentally depend upon this behavior.

As the net electron-neutrino fraction disappears, the number diffusion equation
becomes irrelevant and one should consider the energy diffusion
equation~\cite{PIPELR}
\begin{eqnarray}
nT{\partial s\over\partial t}={1\over r^2}{\partial\over\partial r}\left[
r^2\int{c\over3}\sum_\ell\lambda_{\nu,\ell}{\partial\epsilon_{\nu,\ell}
(E_\nu)\over\partial r} dE_\nu\right]-n\mu_\nu{\partial Y_L\over\partial t}.
\label{tevol1}
\end{eqnarray}
Here, $s$ is the entropy per baryon and $\epsilon_{\nu,\ell}=n_{\nu,\ell}
E_\nu$ are the energy densities, at energy $E_\nu$, of neutrinos (or
antineutrinos) of species $\ell$. Similarly, $\lambda_{\nu,\ell}$ represents
the net mean free path of each neutrino and antineutrino species.  This
equation assumes the matter to be in beta equilibrium.  The energy densities
and mean free paths, in contrast to the deleptonization situation, contain
important contributions from all three types of neutrinos.  The $\lambda$'s of
electron-type neutrinos are due to both absorption and scattering, but those of
tau-type neutrons have only scattering contributions.  Those pertaining to
muon-type neutrinos, besides having scattering contributions, may also have
absorption contributions if muons are present, which is generally the situation
only when $Y_\nu<0.02$ above nuclear densities.

In the neutrino-free case, setting $\mu_\nu=0$ in Eq.~(\ref{tevol1}) results in
\begin{equation} nT{\partial s\over\partial t}={c\over6\pi(\hbar c)^3}{1\over
r^2}{\partial \over\partial r}\left[r^2\sum_\ell{D_{4,\ell}\over T^2} {\partial
T\over\partial r}\right]. \label{tevol} \end{equation} As is the case with the
mean free paths, the diffusion coefficients $D_{4,\ell}$ have contributions
from all three types of neutrinos. Since the baryon matter, which dominates the
specific heat, remains fairly degenerate throughout the cooling epoch, the
entropy varies practically linearly with the temperature. In the nondegenerate
approximation for the baryons and the neutrinos, the relativistic cross
sections are expected to vary as $E_\nu^2 n$, so that $D_{4,\ell}\propto
T^3/n$.  However, the baryons are degenerate so we should expect the cross
sections to be modified by an extra degeneracy factor of $T/\mu_n$.  Therefore,
we can anticipate that the diffusion constants $D_{4,\ell}$ will behave as
$T^2\mu_n/n$.  A separation of Eq.~(\ref{tevol}) into temporal and spatial
variations reveals that the temperature in the star is then expected to be
roughly linearly decreasing with time. The predicted temperature profile will
depend somewhat upon the density profile of the star; assuming the density to
be approximately constant, one finds that the temperature profile is that of an
$n=2$ Lane-Emden polytrope. This is different from the predictions in
Ref.~\cite{PIPELR}, which, however, did not account for the degenerate nature
of baryonic matter during the cooling stage.  However, numerical simulations
have demonstrated the qualitative correctness of these remarks.

\subsection{Results }

Fig. 24 shows $D_2$ plotted versus density for beta equilibrium matter with
$Y_L=0.4$ for temperatures between 5 and 30 MeV for the field-theoretical model
adopted earlier.  Both nucleons-only matter and matter in which hyperons are
assumed to appear are shown.  One sees that  both nucleons-only matter and
matter containing hyperons have relatively small variations of $D_2$ with
density at fixed temperature for $n\geq1.5n_s$.  In the limit of degenerate
matter and neutrinos, this results from the asymptotic behavior of the cross
sections at the Fermi surface, which is  $\sigma\propto T^2\mu_e\mu_n\mu_p$ in
the relativistic case. (In the nonrelativistic case, see Eqs.~(\ref{dcsig}) and
(\ref{dnsig}).)  Thus, $D_2$ is expected to behave like
$\mu_\nu^2/(T\mu_e\mu_n\mu_p)$ in this case, and this has relatively little
density dependence at high densities since the increase in $\mu_e\mu_n\mu_p$ is
compensated by the increase in $\mu_\nu^2$.
This implies that the case $m=2$ could
approximate the deleptonization stage of a protoneutron star, which implies
that $\mu_\nu$ should decrease linearly with time.  Since
$D_2$ is roughly proportional to $1/T$, one has that $\lambda_o$ is roughly
proportional to $1/T^2$.  The rising temperature during deleptonization will
modify the linear decrease of $\mu_\nu$.  However, numerical simulations
show that the linear behavior is approximately correct.

The effect of hyperons appearing during the deleptonization phase at first
does not appear to be appreciable, especially if hyperons appear only
relatively late in the evolution.  In the presence of hyperons, the diffusion
constant $D_2$ decreases from its value for the nucleons-only case for a given
temperature. However, the appearance of hyperons is accompanied by a general
temperature decrease because of the addional degrees of freedom which increase
the heat capacity of the system.  This results in a very small net change for
deleptonization times.  More important effects from hyperons can be expected at
the late stages of deleptonization and  during the cooling epoch following
deleptonization. The total $D_4=\sum_\ell D_{4,\ell}$ is shown as a function of
density for several values of the temperature between 5 and 30 MeV in Fig. 25.
As suggested in the previous section, an overall $T^2$ behavior is observed.
The large (factor of 2) decrease in $D_4$ in hyperon-bearing matter relative to
nucleons-only matter for a given $T$ would, at first glance seems to imply
longer  diffusion and cooling times.  However, the fact that the temperatures
are smaller in the presence of hyperons somewhat compensates for this effect
and preliminary calculations~\cite{PRPL} show that there is an overall increase
in cooling times for models which contain hyperons.  Thus, there are important
feedbacks operating between the stellar structure, the EOS and the opacities
that may only be addressed in detailed simulations.

\section{SUMMARY AND OUTLOOK}

In summary, we have calculated both charged and neutral current neutrino cross
sections in dense matter at supra-nuclear densities.  We have identified
new sources of neutrino opacities involving strange particles and have computed
their weak interaction couplings.  The weak interaction cross sections are
greatly affected by the composition of matter which is chiefly determined by
the strong interactions between the baryons.  We have, therefore,  performed
baseline calculations by considering the effects strong interactions on the
in-medium single particle spectra, which also determines the composition
through the EOS. The formalism we have developed allows us to calculate the
cross sections efficiently for matter at arbitrary matter degeneracy.  From
our results, various limiting forms used earlier in the literature are also
easily derived. This, in addition to providing valuable  physical insights,
allows us to assess the extent to which the various approximations are valid in
both free and interacting matter.

To explore the influence of baryonic interactions on the neutrino cross
sections, we have examined both nonrelativistic potential and relativistic
field-theoretical models that are commonly used in the calculation of the EOS.
We sought to identify the common features shared by these models.   At the
mean-field level in both cases, a relatively simple structure for the
single particle spectrum leads to analytic expressions for the response
function, which in turn
facilitates a first study of the role of strong interactions
on the neutrino cross sections at temperatures of relevance in  astrophysical
applications. Specifically, when the effects of momentum dependent potential
interactions can be treated adequately in the effective mass approximation
({\em i.e.,} only the quadratic term in momentum is retained), analytic
expressions obtained for noninteracting matter are straighforwardly
extended to include interactions.  Effective field-theoretical models at the
mean-field level offer a similar opportunity since the single particle specturm
is that of a  free Dirac spectrum, but with a density dependent effective mass.
Note that in both cases, energy shifts in the spectrum arising through density
dependent interactions are naturally included. Investigating the effects of
more complicated momentum dependent interactions will be necessarily more
involved.  We have also calculated neutrino cross sections in matter containing
hyperons and assessed their influence on the total opacities.

We have examined the role played by neutrino opacities in determining the
deleptonization and cooling times of a newly born neutron star as it evolves in
time.  We showed analytically
how these times are related to the relevant diffusion
constants and, thus, the opacities.  Although the main physical issues
involved are clarified in such an approach, a quantitative assesment must await
calculations using a more complete protoneutron star evolution code. The
results of such a calculation will be reported elsewhere~\cite{PRPL}.

Some important aspects of our work are:
\begin{enumerate}
\item An exact and efficient calculation of the phase space valid for arbitrary
degeneracy of asymmetric matter for both nonrelativistic and relativistic
interacting baryons.  The simplicity of the resulting formulae may be
especially useful in the calculation of electron scattering, for example.
\item With the formalism presented here, it is no longer necessary or desirable
to employ ``generic'' ({\it i.e.,} EOS independent) neutrino
opacities in astrophysical simulations.  Rather, tables of opacities can be
constructed directly from quantities already known from the EOS determination.
For example, the needed quantities are the chemical
potentials, effective masses and single particle potentials of the baryon
components.  The first of these are generally tabulated in the EOS itself.  In
the nonrelativistic formalism, the effective masses and single particle
potentials are simple functions of the particle number densities and kinetic
energy densities. In the relativistic formalism, the
effective masses and single particle potentials are straightforwardly related
to the particle number densities, internal energy and pressure by the
minimization conditions.
\end{enumerate}

In the near future, we will prepare tables of both nonrelativistic potential
and relativistic mean-field models of the EOS combined with their opacities,
and make these generally available. Models including hyperons, kaon condensates
and/or quarks will be computed.

It is difficult to state categorically how the opacities reported here will
affect the current generation of supernova or protoneutron star calculations.
For example, although it seems clear that the addition of hyperons in matter
will decrease mean free paths for a given density and temperature, the
hyperon-bearing matter has a smaller temperature for a given entropy and
density.  Smaller temperatures tend to increase the mean free paths, and this
effect appears to more than cancel the original decrease.  The feedback between
opacities and EOS in a given astrophysical setting must be calculated
consistently.

It must be stressed that while we have included some effects of strong
interactions through density dependent single particle excitations, there
remain important collective effects from both density and spin/isospin
dependent excitations~\cite{S1,IP,HW} and from other density dependent
in-medium correlations~\cite{WR}. The magnitude of these effects has so far
only been assesed in some special cases such as nondegenerate symmetric matter
or pure degenerate neutron matter or neutrino-poor beta-equilibrated matter.
Furthermore, coupling to the $\Delta(1230)$ isobar~\cite{BRHO} and screening by
virtual particle-hole pairs created in the final state interactions~\cite{I}
could reduce the effective matrix elements.  The net impact of such effects
could be a reduction of some cross sections by as much as a factor of 2 at high
densities.  Work is in progress to calculate the role these effects on the
neutrino cross sections at all temperatures of
relevance and for all possible compositions, and will be reported subsequently.


\section*{ACKNOWLEDGMENTS}

We thank J. Pons for helpful discussions and for carefully reading the
manuscript. This work was supported in part by the U.S. Department of Energy
under contracts DOE/DE-FG02-88ER-40388 and DOE/DE-FG02-87ER-40317.

\newpage
\section*{APPENDIX A: POTENTIAL MODELS}

Here, we outline a  potential model  for a system of neutrons and
protons at finite temperature.  With suitable choices of
finite range interactions and parameters, this model reproduces the results of
the more microscopic calculations (for more details, see, for example,
Ref.~\cite{PIPELR}). We begin with the energy density
\begin{eqnarray}
\varepsilon = \varepsilon_n^{(kin)} + \varepsilon_p^{(kin)} + V(n_n,n_p,T) \,,
\label{expal1}
\end{eqnarray}
where $n_n$ ($n_p$) is the neutron (proton) density and the total density
$n=n_n+n_p$. The contributions arising from the kinetic parts are
\begin{eqnarray}
\varepsilon_n^{(kin)} + \varepsilon_p^{(kin)} =
2\int\frac {d^3k}{(2\pi)^3} \, \frac {\hbar^2k^2}{2m} \left( f_n + f_p\right)\;,
\label{kinetic}
\end{eqnarray}
where the factor 2 denotes the spin degeneracy and $f_i~$ for $i=n,p$ are the
usual Fermi-Dirac distribution functions and $m$ is the nucleon mass. It is
common to employ local contact interactions to model the nuclear potential.
Such forces lead to power law density-dependent terms in $V(n)$. Including the
effect of finite-range forces between nucleons, we parameterize the potential
contribution as
\begin{eqnarray}
V(n_n,n_p,T) &=&
 An_0\left[\thalf -\oneth\left(\thalf + x_0\right)(1-2x)^2\right]u^2
\nonumber \\
 && + {Bn_0\left[1 - \twothr \left(\thalf + x_3\right)
(1-2x)^2\right]u^{\sigma+1}  }
\nonumber \\
&& +\twofive u \sum_{i=1,2} \left\{ (2C_i+4Z_i)
2 \int \frac {d^3k}{(2\pi)^3} \, g(k,\Lambda_i) \left( f_n + f_p\right)
\right.\nonumber \\
&& +(C_i-8Z_i)
\left. 2 \int \frac {d^3k}{(2\pi)^3} \, g(k,\Lambda_i)
[ f_n (1-x) + f_p x] \right\} \;,
\label{expal2}
\end{eqnarray}
where $x=n_p/n$ and $u=n/n_0$, with $n_0$ denoting equilibrium nuclear
matter density.  The function $g(k,\Lambda_i)$ is suitably
chosen to simulate finite range effects.  The constants $A,~B,\sigma
,~C_1$, and $C_2$, which enter in the  description of symmetric
nuclear matter, and the additional constants $x_0,~x_3, ~Z_1$, and $Z_2$, which
determine the properties of asymmetric nuclear matter, are treated as
parameters that are constrained by empirical knowledge.

The single particle spectrum $e_i$ entering the Fermi-Dirac distribution
functions $f_i$ may be written as
\begin{eqnarray}
e_i(k) = \frac {\hbar^2 k^2}{2m_i} + U_i(k;n,x,T) \,,
\label{spectrum}
\end{eqnarray}
where the single particle potential $U_i(n,x,k;T)$, which is explicitly momentum
dependent, is obtained by a functional differentiation of the potential energy
density in Eq.~(\ref{expal2}), with respect to the distribution functions
$f_i$.  Explicitly,
\begin{eqnarray}
U_i(n,x,k;T)  &=& \onefive u \left[ \sum_{i=1,2} \left\{
5C_i\pm (C_i-8Z_i)(1-2x) \right\} \right] g(k,\Lambda_i)
\nonumber \\
&+& Au \left[ 1 \mp \twothr \left( \thalf +x_0 \right) (1-2x) \right]
\nonumber \\
&+& Bu^\sigma
\left[1 \mp \fourthr \frac {1}{\sigma + 1} (1-2x)
- \twothr \frac {(\sigma-1)}{(\sigma+1)} \left(\thalf
+ x_3\right) (1-2x)^2 \right]
\nonumber \\
&+& \twofive \frac{1}{n_0} \sum_{i=1,2} \left\{ (2C_i + 4Z_i) 2
\int \frac {d^3k}{(2\pi)^3} \, g(k,\Lambda_i) f_i(k) \right. \nonumber \\
&+&\left. (3C_i-4Z_i)2
\int \frac {d^3k}{(2\pi)^3} \, g(k,\Lambda_i) f_j(k) \right\} \,,
\label{upot}
\end{eqnarray}
where the upper (lower) sign in $\mp$ is for neutrons (protons) and $i \neq j$.

The Landau effective mass is defined through the relation
\begin{eqnarray}
\frac {m_i^*}{m} &=& \frac {k_{F_i}}{m_i}
\left[ \left.
\frac {\partial e_k}{\partial k}\right|_{k_{F_i}} \right]^{-1} \nonumber \\
&=& \left[ 1 + {1 \over 5} u \sum_{i=1,2}
\left\{ 5C_i \pm (C_i-8Z_i)(1-2x) g^{'}(k,\Lambda_i)|_{k_{F_i}} \right\}
\right]^{-1} \,,
\label{emass}
\end{eqnarray}
where the prime denotes a derivative with respect to momentum.
The finite range  interactions may be approximated by effective local
interactions by retaining only the quadratic momentum dependence: i.e.,
$g(k,\Lambda_i) = 1-(k/\Lambda_i)^2$. The energy density in Eq.~(\ref{expal2})
and the single particle potential in Eq.~(\ref{upot}) then take the forms
stemming from Skyrme's effective interactions~\cite{VB}.
Combining the finite range quadratic momentum term with the free
kinetic energy term,
the single particle spectrum in
Eq.~(\ref{spectrum}) may be written for Skyrme-like interactions as
\begin{eqnarray}
e_i(k) = \frac {\hbar^2 k^2}{2m^*_i} + {\tilde U}_i(n,x,;T) \,.
\label{sspectrum}
\end{eqnarray}
This resembles the free particle spectrum, but with a density dependent
effective mass $m^*_i$ given by Eq.~(\ref{emass}) with
$g^{'}(k,\Lambda_i)|_{k_{F_i}} = 1/(R_i^2E_F^{(0)})$, where $R_i =
\Lambda_i/(\hbar k_F^{(0)})$ and $E_F^{(0)} = (\hbar k_F^{(0)})^2/(2m)$ is the
Fermi energy of symmetric nuclear matter at the equilibrium density.
Other forms of $g(k,\Lambda_i)$, with more than a
quadratic momentum dependence, also offer a viable description of the
energy density and the single particle potential
(see for example, Ref.~(\cite{PIPELR})),
but do not lead to a spectrum resembling the free particle spectrum in
Eq.~(\ref{sspectrum}).  This simple form of the spectrum allows a direct
evaluation of the neutrino opacities including the effects of interactions
along the lines developed for
noninteracting baryons in Sec. III.
For more general momentum dependent interactions,
the evaluation of the neutrino
opacities requires more complicated techniques which take into account the
full momentum dependence in the four-momentum conserving delta functions.

The parameters $A,~B,\sigma ,~C_1$, and $C_2$,  are determined from constraints
provided by the empirical properties of symmetric nuclear matter  at the
equilibrium density $n_0=0.16$ fm$^{-3}$.  With appropriate choices of the
parameters, it is possible to parametrically vary the nuclear incompressibility
$K_0$ so that the dependence on the stiffness of the EOS may be explored.
In this work, we have chosen $K_0=180$ MeV for which
\begin{eqnarray}
A =159.47\,, ~~B= -109.04\,, ~~\sigma=0.844 \,, ~~C_1=-41.28\, \quad
{\rm and} \quad C_2=23 \,.
\label{palpar}
\end{eqnarray}
Except for the dimensionless $\sigma$, all quantities above
are in MeV. The finite-range parameters $\Lambda_1 = 1.5p_F^{(0)}$ and
$\Lambda_2=3p_F^{(0)}$.

In the same vein, by suitably choosing the parameters $x_0,~x_3,~Z_1$, and
$Z_2$, it is possible to obtain  different forms for the density dependence  of
the symmetry energy $S(n)$ defined by the relation
\begin{eqnarray}
E(n,x) = \varepsilon (n,x)/n = E(n,1/2) + S(n) (1-2x)^2 + \cdots \;,
\end{eqnarray}
where $E$ is the energy per particle, and $x=n_p/n$ is the proton fraction.
Inasmuch as the density dependent terms associated with powers higher than
$(1-2x)^2$ are generally small, even down to $x=0$, $S(n)$ adequately describes
the properties of asymmetric matter.   The need to explore different forms of
$S(n)$ stems from the uncertain behavior at high density and has been amply
detailed in earlier publications~\cite{LPPH,PAL}.  In this work,
we have chosen the potential part of the symmetry energy to vary
as $u$.  For this case,
\begin{eqnarray}
x_0=-0.410, ~~x_3 = -0.5\, ~~Z_1= -11.56~{\rm MeV}, \quad {\rm and} \quad
Z_2=-4.421~{\rm MeV}
\label{expalpar}
\end{eqnarray}

Since repulsive contributions that vary faster than linearly give rise to
acausal behavior at high densities, care must be taken to screen such repulsive
interactions~\cite{PAL}. This may be achieved by dividing the term
proportional to $u^{\sigma +1}$ (when $\sigma >1$) by the factor
\begin{eqnarray}
{1 + \twothr  B^\prime \left[\thrhalf - \left(\thalf + x_3\right)
(1-2x)^2\right]u^{\sigma-1}  } \,,
\end{eqnarray}
where $~B^\prime$ is a small parameter introduced to maintain causality.  Note
that the single particle potential in Eq.~(\ref{upot}) must then be accordingly
modified.  The appropriate terms may be obtained by using the relations
$(\partial/\partial n_n)|_{n_p} = \partial/\partial n - (x/n) \partial/\partial
x$ for neutrons and  $(\partial/\partial n_p)|_{n_n} =  \partial/\partial n +
[(1-x)/n] \partial/\partial x$ for protons.

For a fixed baryon density $n$, proton fraction $x$, and temperature $T$,  an
iterative procedure may now be employed to calculate  the density dependent
single particle potentials ${\tilde U}_n$ and ${\tilde U}_p$, and  the chemical
potentials $\mu_n$ and $\mu_p$.  The calculational procedure is detailed in
Ref.~\cite{PIPELR}.  These quantities, in conjunction with the requirements of
chemical equilibrium in Eq.~(\ref{bequil}) and charge neutrality in
Eq.~(\ref{cneut}) determines the composition of stellar matter at finite
temperature.   The specification of the spectrum in Eq.~(\ref{sspectrum}) then
allows for a  calculation of the neutrino opacities which are consistent with
the underlying EOS.

\newpage

\section*{APPENDIX B: EFFECTIVE FIELD-THEORETICAL  MODELS}

In a Walecka-type relativistic field-theoretical model the interactions
between baryons are mediated by the exchange of $\sigma,\omega$, and $\rho$
mesons.  The Lagrangian density is given by~\cite{SW},
\begin{eqnarray*}
L &=& L_H +L_\ell \nonumber \\
  &=& \sum_{B} \overline{B}(-i\gamma^{\mu}\partial_{\mu}-g_{\omega B}
\gamma^{\mu}\omega_\mu
-g_{\rho B}\gamma^{\mu}{\bf{b}}_{\mu}\cdot{\bf t}-M_B+g_{\sigma B}\sigma)B \\
&-& \frac{1}{4}W_{\mu\nu}W^{\mu\nu}+\frac{1}{2}m_{\omega}^2\omega_{\mu}\omega^
{\mu} - \frac{1}{4}{\bf B_{\mu\nu}}{\bf
B^{\mu\nu}}+\frac{1}{2}m_{\rho}^2 b_{\mu}b^{\mu}\\
&+& \frac{1}{2}\partial_{\mu}\sigma\partial^{\mu}\sigma -\frac{1}{2}
m_{\sigma}^2\sigma^2-U(\sigma)\\
&+& \sum_{l}\overline{l}(-i\gamma^{\mu}\partial_{\mu}-m_l)l \,.
\end{eqnarray*}
Here, $B$ are the Dirac spinors for baryons and $\bf t$ is the isospin
operator. The sums include baryons $B=n,p,\Lambda,\Sigma$, and $\Xi$, and
leptons, $l = e^-$ and $\mu^-$. The field strength tensors for the $\omega$
and
$\rho$ mesons are $W_{\mu\nu} = \partial_\mu\omega_\nu-\partial_\nu\omega_\mu$
and ${\bf B}_{\mu\nu} =  \partial_\mu{\bf b}_\nu-\partial_\nu{\bf b}_\mu$,
respectively.  The potential $U(\sigma)$ represents the self-interactions of
the scalar field and is taken to be of the form
\begin{eqnarray}
U(\sigma) =  \frac{1}{3}bM_n(g_{\sigma N}\sigma)^3 + \frac{1}{4}c(g_{\sigma
N}\sigma)^4\,.
\end{eqnarray}
Electrons and muons are included in the model as noninteracting particles,
since their interactions give small contributions compared to those of
their free Fermi gas parts.

In the mean field approximation, the partition function (denoted by $Z_H$) for
the hadronic degrees of freedom is given by
\begin{eqnarray}
\ln Z_H~ &=&~\beta V\left[\thalf m_{\omega}^2\omega_0^2+\thalf
m_{\rho}^2b_0^2 - \thalf m_{\sigma}^2\sigma^2-U(\sigma)\right]\nonumber\\
&&\qquad\qquad+ 2V\sum_B \int\frac{d^3k}{(2\pi)^3} \,\ln\left(1+{\rm e}
^{-\beta(E^*_B-\nu_B)}\right)\,,
\label{hyp2}
\end{eqnarray}
where $\beta = (kT)^{-1}$ and $V$ is the volume.
The contribution of antibaryons is
not significant for the thermodynamics of interest here, and is therefore
not included in Eq.~(\ref{hyp2}). Here, the effective baryon masses
$M_B^*=M_B-g_{\sigma B}\sigma$ and
$E^*_B=\sqrt{k^2+M^{*2}_B}$. The chemical potentials are given by
\begin{equation}
\mu_B = \nu_B+g_{\omega B}\omega_0+g_{\rho B}t_{3B}b_0\;,\label{hyp3}
\end{equation}
where $t_{3B}$ is the third component of isospin for the baryon. Note
that particles with $t_{3B}=0$, such as the $\Lambda$ and $\Sigma^0$
do not couple to the $\rho$.  The effective chemical potential
$\nu_B$ sets the scale of the temperature dependence of the thermodynamical
functions.

Using $Z_H$, the thermodynamic quantities can be obtained in the standard way.
The pressure $P_H=TV^{-1}\ln Z_H$, the number density for species $B$, and the
energy density $\varepsilon_H$ are given by
\begin{eqnarray}
n_B~&=&~2\int\frac{d^3k}{(2\pi)^3}
\left({\rm e}^{\beta(E^*_B-\nu_B)}+1\right)^{\!-1}\;,\nonumber\\
\varepsilon_H~&=&~\thalf m_{\sigma}^2\sigma^2+U(\sigma)+
\thalf m_{\omega}^2\omega_0^2+\thalf m_{\rho}^2 b_0^2
+2\sum_B
\int\frac{d^3k}{(2\pi)^3}
E_B^*\left({\rm e}^{\beta(E^*_B-\nu_B)}+1\right)^{\!-1}\;.\label{hyp4}
\end{eqnarray}
The entropy density is then given by
$s_H=\beta(\varepsilon_H+P_H-\sum_B\mu_Bn_B)$.

The meson fields are obtained by extremization of the partition function,
which yields the equations
\begin{eqnarray}
&&\fpj m_{\omega}^2\omega_0=\sum_Bg_{\omega B} n_B\quad;\quad
m_{\rho}^2b_0=\sum_Bg_{\rho B}t_{3B}n_B\;,\nonumber\\
&&\fpj m_{\sigma}^2\sigma=-\frac{dU(\sigma)}{d\sigma}
+\sum_B g_{\sigma B} ~~2\hspace{-1mm}\int\!\frac{d^3k}{(2\pi)^3}
\frac{M_B^*}{E_B^*}\left({\rm e}^{\beta(E^*_B-\nu_B)}+1\right)^{\!-1} \,.
\label{hyp5}
\end{eqnarray}
The total partition function $Z_{total}=Z_HZ_L$, where $Z_L$ is the standard
noninteracting partition function of the leptons.

The additional conditions needed to obtain a solution are provided by the
charge neutrality requirement, and, when neutrinos are not
trapped, the set of equilibrium chemical
potential relations required by the general condition
\begin{eqnarray}
\mu_i = b_i\mu_n - q_i\mu_l\,,
\label{beta}
\end{eqnarray}
where $b_i$ is the baryon number of particle $i$ and $q_i$ is its
charge.
For example, when $\ell=e^-$, this implies the equalities
\begin{eqnarray}
&&\fpj \mu_{\Lambda} = \mu_{\Sigma^0} = \mu_{\Xi^0} = \mu_n \,, \nonumber \\
&&\fpj \mu_{\Sigma^-} = \mu_{\Xi^-} = \mu_n+\mu_e \,, \nonumber \\
&&\fpj \mu_p = \mu_{\Sigma^+} = \mu_n - \mu_e \,.
\label{murel}
\end{eqnarray}
In the case that the neutrinos are trapped, Eq. (\ref{beta}) is replaced by
\begin{eqnarray}
\mu_i = b_i\mu_n - q_i(\mu_l-\mu_{\nu_\ell})\,.
\label{tbeta}
\end{eqnarray}
The new equalities are then obtained by the replacement
$\mu_e \rightarrow \mu_e - \mu_{\nu_e}$ in Eq.~(\ref{murel}).  The
introduction of additional variables, the neutrino  chemical potentials,
requires additional constraints, which we supply by fixing the lepton fractions,
$Y_{L\ell}$, appropriate for conditions prevailing in the evolution of the
protoneutron star.  The contribution to pressure from neutrinos of a given
species is $P_\nu=(1/24\pi^2)\mu_\nu^4$.

In the nucleon sector, the constants $ g_{\sigma N}, g_{\omega N},
g_{\rho N}, b$, and $c$ are determined by reproducing the nuclear matter
equilibrium density $n_0=0.16~{\rm fm}^{-3}$, and the binding energy per
nucleon ($\sim 16$ MeV),  the symmetry energy ($\sim 30-35$ MeV),
the compression modulus ($200~{\rm MeV} \leq K_0 \leq 300~{\rm MeV})$, and the
nucleon Dirac effective mass $M^*=(0.6-0.7)~\times 939$ MeV at $n_0$.
Numerical values of the coupling constants so chosen are:
\begin{eqnarray}
g_{\sigma N}/m_\sigma &=& 3.434~{\rm fm}, \quad
g_{\omega N}/m_\omega = 2.674~{\rm fm}, \quad
g_{\rho N}/m_\rho = 2.1~{\rm fm}, \nonumber \\
b &=& 0.00295, \quad {\rm and} \quad  c= -0.00107 \,.
\end{eqnarray}
These couplings yield a symmetry energy of  32.5 MeV, a
compression modulus of 300 MeV, and $M^*/M=0.7$. This particular choice of
model  parameters are from Glendenning and Moszkowski~\cite{GM} and will be
referred to as GM1 hereafter. The prevalent uncertainty in the  nuclear matter
compression modulus  and the effective mass $M^*$ does not allow for a unique
choice of these  coupling constants.   The high density behaviour of  the EOS
is sensitive to the strength of the meson coupling constants employed.  Lacking
definitive experimental and theoretical constraints, this choice of parameters
may be considered typical.

The hyperon coupling constants may be determined by reproducing the binding
energy of the $\Lambda$ hyperon in nuclear matter~\cite{GM}.
Parameterizing the hyperon-meson couplings in terms of nucleon-meson couplings
through
\begin{eqnarray}
x_{\sigma H}=g_{\sigma H}/g_{\sigma N},~~~
x_{\omega H}=g_{\omega H}/g_{\omega N}
,~~~x_{\rho H}=g_{\rho H}/g_{\rho N} \,,
\end{eqnarray}
the $\Lambda$ binding energy at nuclear density is given by
\begin{eqnarray}
(B/A)_\Lambda = -28 = x_{\omega \Lambda} g_{\omega N} \omega_0
- x_{\sigma \Lambda} g_{\sigma N} \sigma_0\,,
\end{eqnarray}
in units of MeV.  Thus, a particular choice of  $x_{\sigma
\Lambda}$ determines $x_{\omega \Lambda}$ uniquely.   To keep the number of
parameters small, the coupling constant
ratios for all the different hyperons are assumed to be the same.
That is
\begin{eqnarray}
x_\sigma = x_{\sigma\Lambda} = x_{\sigma\Sigma} = x_{\sigma\Xi} = 0.6\,,
\end{eqnarray}
and similarly for the $\omega$
\begin{eqnarray}
x_\omega = x_{\omega\Lambda} = x_{\omega\Sigma} = x_{\omega\Xi} = 0.653 \,.
\end{eqnarray}
The $\rho$-coupling is of
less consequence and is taken to be of similar order, i.e.
$x_\rho = x_\sigma$\,.

\newpage
\vskip 10pt
\noindent TABLE I: Charged current vector and axial couplings. Numerical values
are quoted using D=0.756 , F=0.477, $\sin^2\theta_W$=0.23 and $\sin^2\theta_c =
0.053$ (see Ref.~\cite{JGGS}).  The couplings for the same reactions involving
antineutrinos are identical, and $\ell$ stands for $e-$, $\mu-$ or $\tau-$ type
leptons.  For corrections arising due to explicit $SU(3)$ breaking terms, see
Ref.~\cite{MSJW}.
\vskip 5pt
\begin{center}
\begin{tabular}{lccc}
\hline \hline
{Reaction } & $g_V$  & $g_A $ & $\Delta S$  \\
\hline
$\nu_\ell + n \rightarrow \ell^- + p$ & $ 1 $ & $D+F=1.23$ & $0$\\
$\nu_\ell + \Sigma^-\rightarrow \ell^- +\Lambda$ & $0$ &
$\sqrt{2/3}D=0.62$ & $0$ \\
$\nu_\ell + \Sigma^-\rightarrow \ell^- +\Sigma^0$ &$\sqrt{2}$ &
$\sqrt{2}F=0.67$ & $0$  \\
$\nu_\ell + \Sigma^0\rightarrow \ell^- +\Sigma^+$&$-\sqrt{2}$
  &$-\sqrt{2}F=-0.67$& $0$ \\
$\nu_\ell + \Lambda\rightarrow \ell^- +\Sigma^+$& $0$ & $-\sqrt{2/3}D=-0.62$& $0$\\
$\nu_\ell + \Lambda\rightarrow \ell^- +p$ &$-\sqrt{3/2}$
& $-\sqrt{3/2}(F+D/3)=0.89$& $1$\\
$\nu_\ell + \Sigma^0\rightarrow \ell^- + p$ & $1$ & $\sqrt{1/2}D=0.54$ & $1$\\
$\nu_\ell + \Sigma^-\rightarrow \ell^- +n$ & $-1$ & $D-F=0.28$ & $1$ \\
$\nu_e+\mu^-\rightarrow\nu_\mu+e^-$&$1$&$1$&$0$\\
\hline \hline
\end{tabular}
\end{center}

\vskip .5in
\noindent TABLE II: Neutral current vector and axial couplings. Neutral current
couplings with baryons of all neutrino species, including antineutrinos, are
identical, and $\ell$ stands for $e-, \mu-,$ or $\tau-$ type neutrinos.
Neutrino interactions with leptons have the same matrix elements as those with
antineutrinos of the same flavor. For corrections arising due to explicit
$SU(3)$ breaking terms, see Ref.~\cite{MSJW}.
\vskip 5pt
\begin{tabular}{lcc}
\hline \hline
{Reaction } & $c_V$  & $c_A $  \\
\hline
$\nu_e+e^-\rightarrow\nu_e+e^-$&$1+4\sin^2\theta_W=1.92$&$1$\\
$\nu_{\mu}+\mu^-\rightarrow\nu_{\mu}+\mu^-$&$1+4\sin^2\theta_W=1.92$&$1$\\
$\nu_e+\mu^-\rightarrow\nu_e+\mu^-$&$-1+4\sin^2\theta_W=-0.08$&$-1$\\
$\nu_{\mu,\tau}+e^-\rightarrow\nu_{\mu,\tau}+e^-$&$-1+4\sin^2\theta_W=-0.08$
&$-1$\\
$\nu_\ell + n \rightarrow \nu_\ell + n$ & $-1$ & $-D-F=-1.23$ \\
$\nu_\ell + p \rightarrow \nu_\ell+ p$ & $ 1-4\sin^2\theta_W=0.08$ & $D+F=1.23$\\
$\nu_\ell + \Lambda\rightarrow \nu_\ell +\Lambda$ & $-1$ & $-F-D/3=-0.73 $ \\
$\nu_\ell + \Sigma^-\rightarrow \nu_\ell +\Sigma^-$ &$ -3+4\sin^2\theta_W=-2.08$ &
 $D-3F=-0.68$  \\
$\nu_\ell + \Sigma^+\rightarrow \nu_\ell +\Sigma^+$&$ 1-4\sin^2\theta_W=0.08$ & D+F
=1.23 \\
$\nu_\ell + \Sigma^0\rightarrow \nu_\ell +\Sigma^0$ & $-1$ & $D-F=0.28$\\
$\nu_\ell + \Xi^-\rightarrow \nu_\ell +\Xi^-$ & $ -3+4\sin^2\theta_W=-2.08 $ & $D-3F
=-0.68$ \\
$\nu_\ell + \Xi^0\rightarrow \nu_\ell +\Xi^0$ & $-1$ & $-D-F=-1.23$ \\
$\nu_\ell+\Sigma^0 \rightarrow \nu_\ell+\Lambda$ & $0$ & $2D/\sqrt 3=0.87$
 \\ \hline
\end{tabular}
\vskip 5pt

\newpage

\newcommand{\IJMPA}[3]{{ Int.~J.~Mod.~Phys.} {\bf A#1}, #3 (#2)}
\newcommand{\JPG}[3]{{ J.~Phys. G} {\bf {#1}}, #3 (#2)}
\newcommand{\AP}[3]{{ Ann.~Phys. (NY)} {\bf {#1}}, #3 (#2)}
\newcommand{\NPA}[3]{{ Nucl.~Phys.} {\bf A{#1}}, #3 (#2)}
\newcommand{\NPB}[3]{{ Nucl.~Phys.} {\bf B{#1}}, #3 (#2)}
\newcommand{\PLB}[3]{{ Phys.~Lett.} {\bf {#1}B}, #3 (#2)}
\newcommand{\PRv}[3]{{ Phys.~Rev.} {\bf {#1}}, #3 (#2)}
\newcommand{\PRC}[3]{{ Phys.~Rev. C} {\bf {#1}}, #3 (#2)}
\newcommand{\PRD}[3]{{ Phys.~Rev. D} {\bf {#1}}, #3 (#2)}
\newcommand{\PRL}[3]{{ Phys.~Rev.~Lett.} {\bf {#1}}, #3 (#2)}
\newcommand{\PR}[3]{{ Phys.~Rep.} {\bf {#1}}, #3 (#2)}
\newcommand{\ZPC}[3]{{ Z.~Phys. C} {\bf {#1}}, #3 (#2)}
\newcommand{\ZPA}[3]{{ Z.~Phys. A} {\bf {#1}}, #3 (#2)}
\newcommand{\JCP}[3]{{ J.~Comput.~Phys.} {\bf {#1}}, #3 (#2)}
\newcommand{\HIP}[3]{{ Heavy Ion Physics} {\bf {#1}}, #3 (#2)}

{}

\newpage
\section*{FIGURE CAPTIONS}

\noindent Fig. 1. Lowest order Feynman diagram for $\nu_l + B_2 \rightarrow l +
B_4$. The symbols $B_i$ and $l$ denote baryons and leptons, respectively.
$P_i$ are the particles' four-momenta and $q_\mu=(q_0,{\vec q})$ is the
four-momentum transfer. Fig.~(1a) is the absorption reaction and Fig.~(1b)
is the scattering reaction.

\noindent Fig. 2. Composition and chemical potentials for
noninteracting matter in beta equilibrium at $Y_L=0.4$ at different
temperatures.  Top panels: Individual concentrations $Y_i=n_i/n_B$,
where $i = n,~p,~e^-$ and $\nu_e$.  Bottom panels: The lepton chemical
potentials and $\mu = \mu_n - \mu_p = \mu_e - \mu_{\nu_e}$.
\bigskip

\noindent Fig. 3. Absorption mean free paths $\lambda =\sigma/V$ in
noninteracting matter in beta equilibrium at different temperatures.
The neutrino energy is taken to be the Fermi energy ($E_\nu=\mu_\nu$).
The solid curves show exact results from Eqs.~(\ref{nrfsig1}) and
(\ref{nrfstruc1}), the dashed curves are from the degenerate
approximation (Eq.~(\ref{dcsig})), the dot-dashed curves are the
nondegenerate results from Eq.~(\ref{ndcsig}), and the long dashed
curves represent the elastic approximation from Eq.~(\ref{elcsig}).
\bigskip

\noindent Fig. 4. Comparison between scattering and absorption
reactions at $T=10$ MeV for neutrino-trapped matter ($Y_L=0.4$),
assuming $E_\nu=\mu_\nu$. The neutrino energy was set equal to the
local neutrino chemical potential. In the left panels the contribution
to the neutrino mean free path due to individual reactions are
shown. In the right panel, the solid curve is the ratio of the
scattering to absorption mean free paths and the the dashed curve
is the neutron scattering/absorption ratio.
\bigskip

\noindent Fig. 5. Compositions and chemical potentials for
noninteracting matter in beta equilibrium with no trapped neutrinos
($Y_\nu=0$) at different temperatures. Top panels: Individual
concentrations $Y_i=n_i/n_B$, where $i = n,~p,$ and $e^-$. Bottom
panels: The electron chemical potential $\mu_e = \mu_n - \mu_p$.
\bigskip

\noindent Fig. 6. Absorption mean free paths $\lambda =\sigma/V$ in
noninteracting matter in beta equilibrium with no trapped neutrinos at
different temperatures. The neutrino energy is taken to be $E_1=3T$.
The leftmost panel clearly shows that at low temperatures the mean
free path is very large since the reaction is kinematically supressed.
\bigskip

\noindent Fig. 7. Temperature dependance of the neutrino mean free
path at nuclear saturation density for $Y_\nu=0$ and $E_\nu=3T$. Left
panel: Individual contributions due to scattering and absorption
reactions to the mean free path, the long-dashed curve is the total
scattering opacity. Right panel: Ratio of the total scattering opacity
to that of absorption.
\bigskip

\noindent Fig. 8. Particle fractions (upper panels) and lepton
chemical potentials (lower panels) in the nonrelativistic potential
Skyrme model for $Y_L=0.4$ at various temperatures. Note that the
proton fraction is larger than in the noninteracting case.
\bigskip

\noindent Fig. 9. Neutrino absorption mean free path in neutrino-trapped
matter with $Y_L=0.4$. Results for the limiting cases discussed
in the text are also shown, with the same notation as in Fig. 3.
\bigskip

\noindent Fig. 10. Neutron (solid curve) and proton (dashed curve) effective
masses as a function of baryon density in the Skyrme model. The curves
shown are for $T=30 MeV$.
\bigskip

\noindent Fig. 11. Compositions (upper panels) and electron chemical
potentials (lower panels) for neutrino-free matter at various
temperatures in the Skryme model.  Note that the proton fraction is
larger when compared with the noninteracting case and, at low
densities, is quite sensitive to the temperature.
\bigskip

\noindent Fig. 12. Upper panel: Neutrino absorption mean free path in
neutrino-free matter with the Skyrme model, assuming $E_\nu=3T$.  The
exact and limiting cases discussed in the text are shown, with the
same notation as in Fig. 3.  For $T=5$ MeV, the elastic approximation
is off-scale.  Lower panel: Comparison of absorption and scattering
mean free paths for this case.
\bigskip

\noindent Fig. 13. Compositions (upper panels) and chemical potentials
(lower panels) in field-theoretical model for nucleonic
matter in beta equilibrium with fixed lepton number $Y_L=0.4$.
\bigskip

\noindent Fig. 14. Absorption (upper panels) and scattering (lower
panels) mean free paths $\lambda =\sigma/V$ for neutrino energy
$E_{\nu}=\mu_{\nu_e}$ in beta equilibrium matter with $Y_L=0.4$,
for the field-theoretical model.
\bigskip

\noindent Fig. 15. Composition (upper panels) and chemical potentials
(lower panels) matter in beta equilibrium with no trapped neutrinos
($Y_\nu=0$) for the field-theoretical model.
\bigskip

\noindent Fig. 16. Absorption (upper panels) and scattering (lower
panels) mean free paths $\lambda =\sigma/V$ for neutrino energy
$E_{\nu}=3T$ in matter with no trapped neutrinos for the
field-theoretical model.
\bigskip

\noindent Fig. 17. Comparision with previous cross sections employed in
protoneutron star calculations. The upper panels show comparisions with
limiting forms valid for the nondegenerate (left) and degenerate (right)
matter. The lower panels show comparisions with interpolated cross sections;
the interpolation scheme in Eq.~(\ref{int1})is shown in the left panel and
that in Eq.~(\ref{int2}) is shown in the right panel.
\bigskip

\noindent Fig. 18. Compositions (upper panels) and chemical potentials
(lower panels) for neutrino-trapped matter ($Y_L=0.4$) including
hyperons in the field-theoretical model.
$\mu=\mu_n-\mu_p=\mu_e-\mu_{\nu_e}$.
\bigskip

\noindent Fig. 19. Absorption mean free paths $\lambda
=(\sigma/V)^{-1}$ for neutrino energy $E_{\nu}=\mu_{\nu_e}$ in matter
with fixed lepton number ($Y_L=0.4$) in the field-theoretical
model.  Upper panels show both the results for matter without hyperons
(dashed curves) and with hyperons (solid curves).  Lower panels show
the relative contributions to the absorption mean free path for matter
with hyperons.
\bigskip

\noindent Fig. 20. Scattering mean free paths for neutrino energy
$E_{\nu}=\mu_{\nu_e}$ in matter with fixed lepton number ($Y_L=0.4$)
in the field-theoretical model.  Upper panels show the results
for the toal scattering mean free path for matter without hyperons
(dashed curves) and with hyperons (solid curves).  Lower panels show
the relative contributions to the scattering mean free path for matter
with hyperons.
\bigskip

\noindent Fig. 21. Compositions (upper panels) and chemical potentials
(lower panels) for neutrino-free matter ($Y_\nu=0$) including
hyperons in the field-theoretical model.
\bigskip

\noindent Fig. 22. Absorption mean free paths $\lambda
=(\sigma/V)^{-1}$ for neutrino energy $E_{\nu}=3T$ in matter with no
trapped neutrinos ($Y_\nu=0$) in the field-theoretical model.
Upper panels show both the results for matter without hyperons (dashed
curves) and with hyperons (sold curves).  Lower panels show the
relative contributions to the absorption mean free path for matter
with hyperons.
\bigskip

\noindent Fig. 23. Scattering mean free path for neutrino energy
$E_{\nu}=3T$ in matter with no trapped neutrinos ($Y_\nu=0$) in the
field-theoretical model.  Upper panels show the results for the
total scattering mean free path for matter without hyperons (dashed
curves) and with hyperons (solid curves).  Lower panels show the
relative contributions to the scattering mean free path for matter
with hyperons.
\bigskip

\noindent Fig. 24. Neutrino diffusion coefficient $D_2$ defined in
Eq.~(\ref{dn}) in matter with and with out hyperons for $Y_L=0.4$ in
the field-theoretical model. The left panel shows $D_2$ as a
function of baryon density in matter containing only nucleons and
leptons. The ratio of $D_2$ in matter without hyperons to $D_2$ in
matter with hyperons is shown in the right panel.
\bigskip

\noindent Fig. 25. Neutrino diffusion coefficient $D_4$ defined in
Eq.~(\ref{dn}) in matter with and with out hyperons for $Y_\nu=0$ in
the field-theoretical model. The left panel shows $D_4$ as a
function of baryon density in matter containing only nucleons and
leptons. The ratio of $D_4$ in matter without hyperons to $D_4$ in
matter with hyperons is shown in the right panel.

\newpage
\begin{figure}
\begin{center}
\leavevmode
\epsfxsize=7.0in
\epsfysize=7.0in
\epsffile{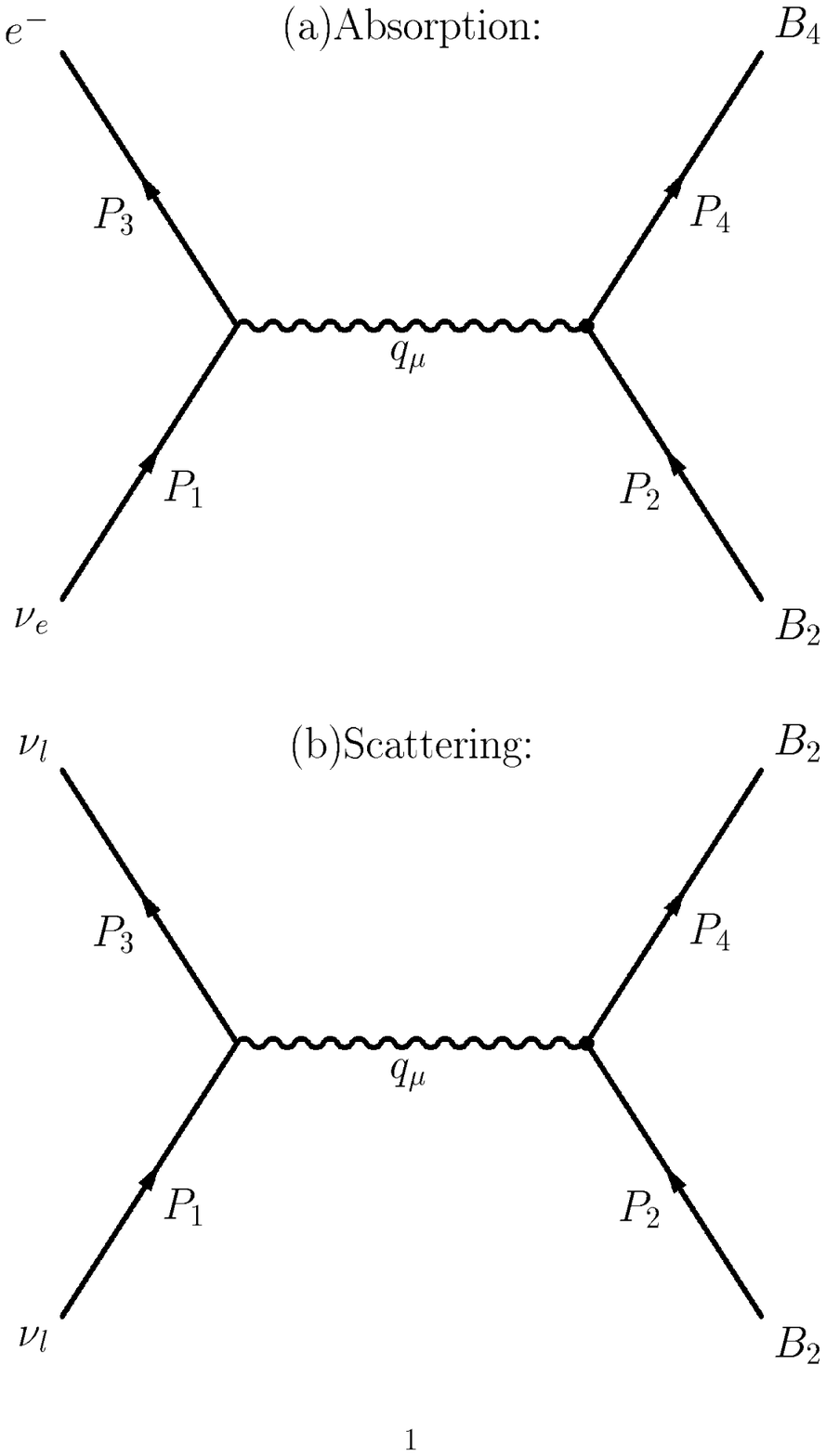}
\end{center}
\caption[]{\footnotesize}
{\label{fig1}}
\end{figure}

\newpage
\begin{figure}
\begin{center}
\leavevmode
\epsfxsize=7.0in
\epsfysize=7.0in
\epsffile{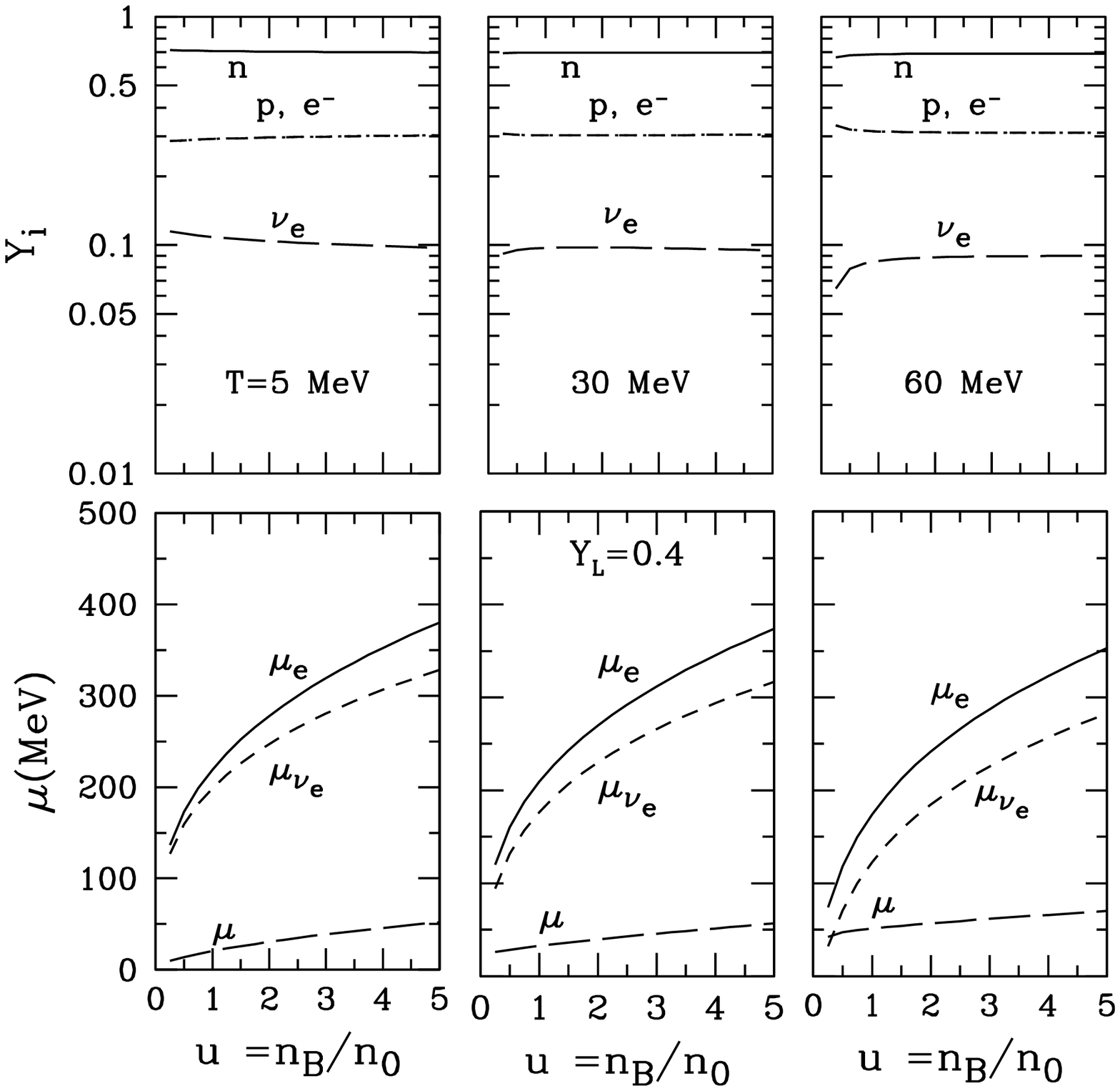}
\end{center}
\caption[]{\footnotesize}
{\label{fig2}}
\end{figure}

\newpage
\begin{figure}
\begin{center}
\leavevmode
\epsfxsize=7.0in
\epsfysize=7.0in
\epsffile{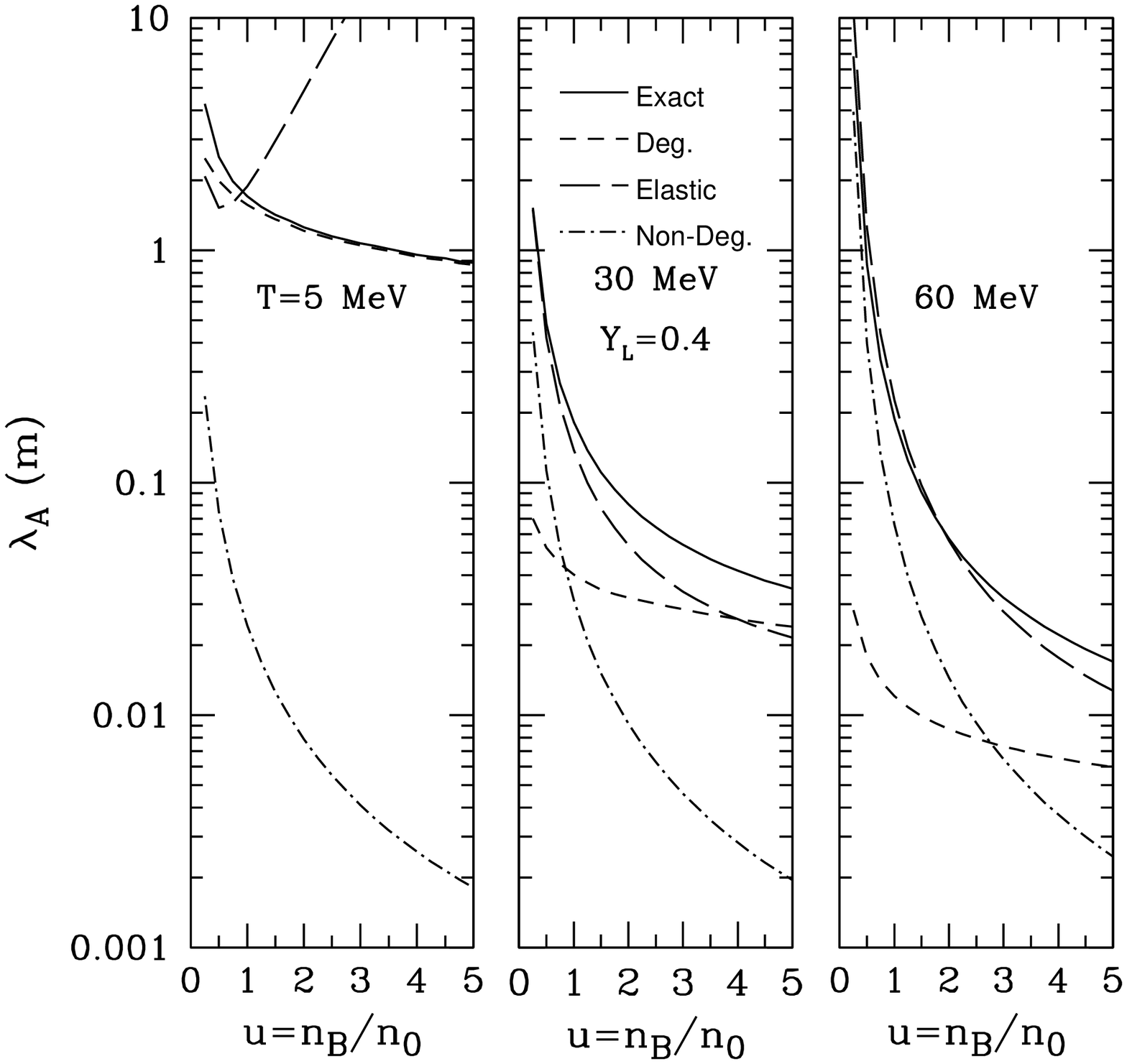}
\end{center}
\caption[]{\footnotesize}
{\label{fig3}}
\end{figure}

\newpage
\begin{figure}
\begin{center}
\leavevmode
\epsfxsize=7.0in
\epsfysize=7.0in
\epsffile{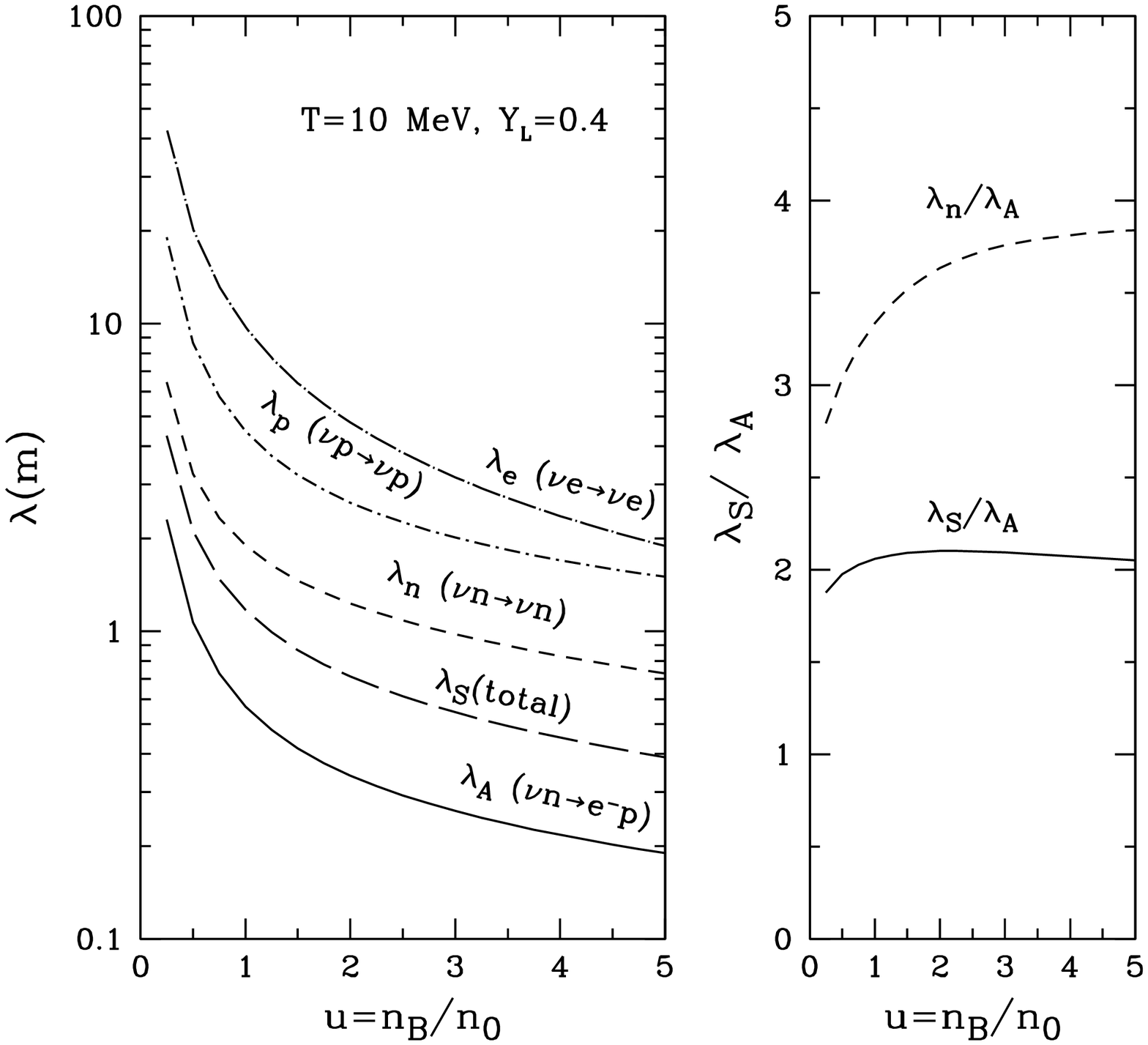}
\end{center}
\caption[]{\footnotesize}
{\label{fig4}}
\end{figure}

\newpage
\begin{figure}
\begin{center}
\leavevmode
\epsfxsize=7.0in
\epsfysize=7.0in
\epsffile{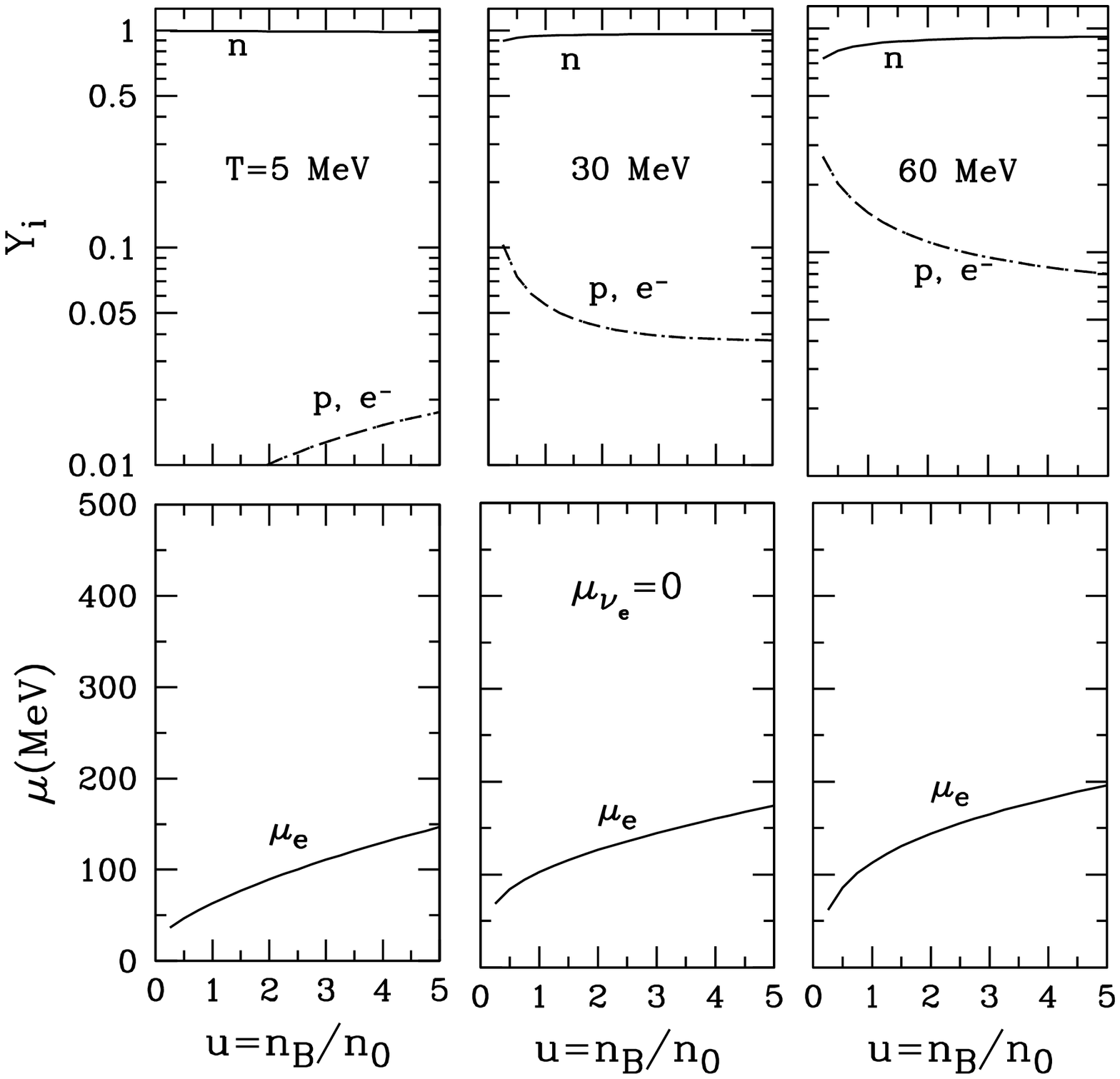}
\end{center}
\caption[]{\footnotesize}
{\label{fig5}}
\end{figure}

\newpage
\begin{figure}
\begin{center}
\leavevmode
\epsfxsize=7.0in
\epsfysize=7.0in
\epsffile{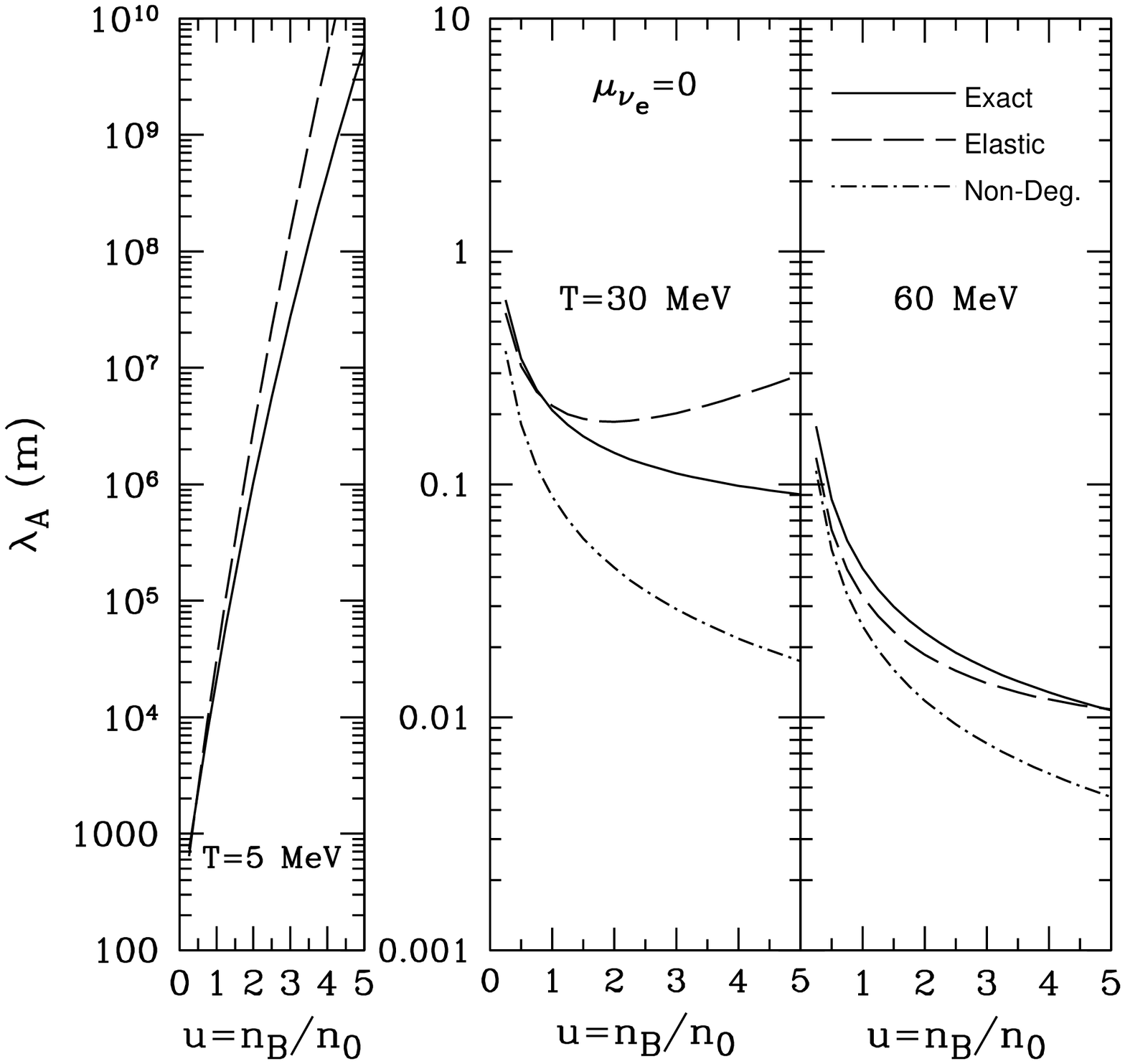}
\end{center}
\caption[]{\footnotesize}
{\label{fig6}}
\end{figure}

\newpage
\begin{figure}
\begin{center}
\leavevmode
\epsfxsize=7.0in
\epsfysize=7.0in
\epsffile{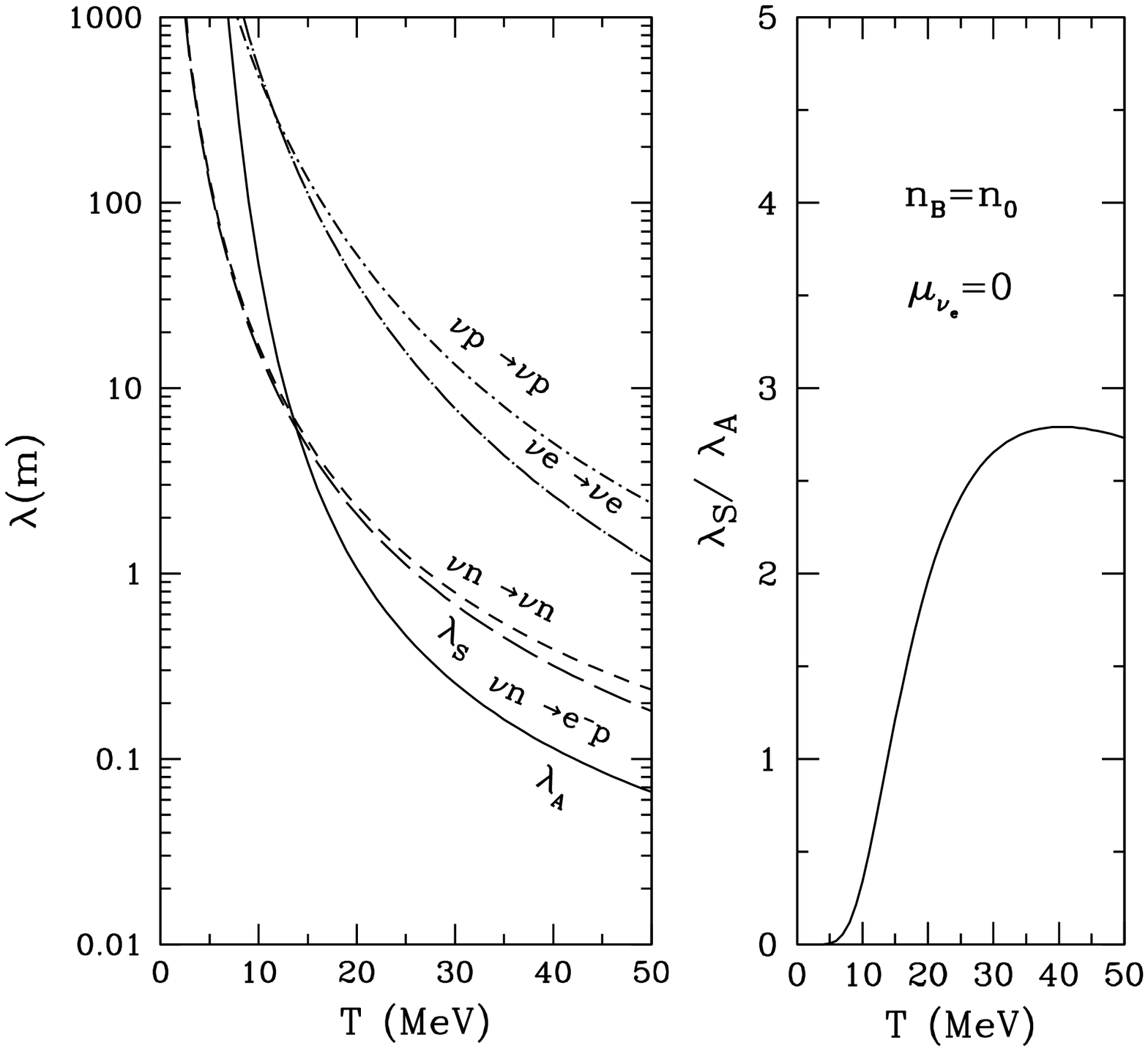}
\end{center}
\caption[]{\footnotesize}
{\label{fig7}}
\end{figure}

\newpage
\begin{figure}
\begin{center}
\leavevmode
\epsfxsize=7.0in
\epsfysize=7.0in
\epsffile{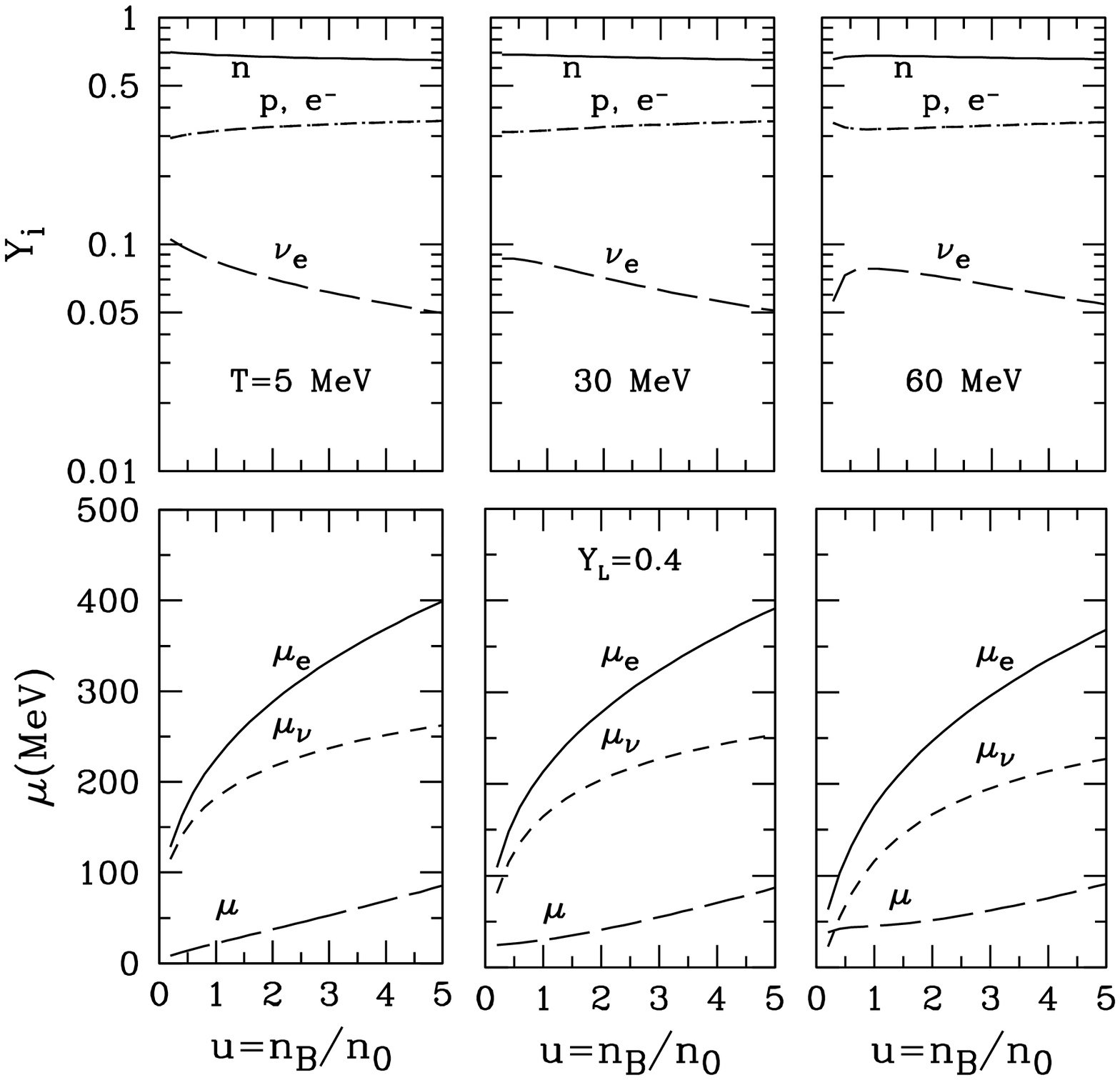}
\end{center}
\caption[]{\footnotesize}
{\label{fig8}}
\end{figure}

\newpage
\begin{figure}
\begin{center}
\leavevmode
\epsfxsize=7.0in
\epsfysize=7.0in
\epsffile{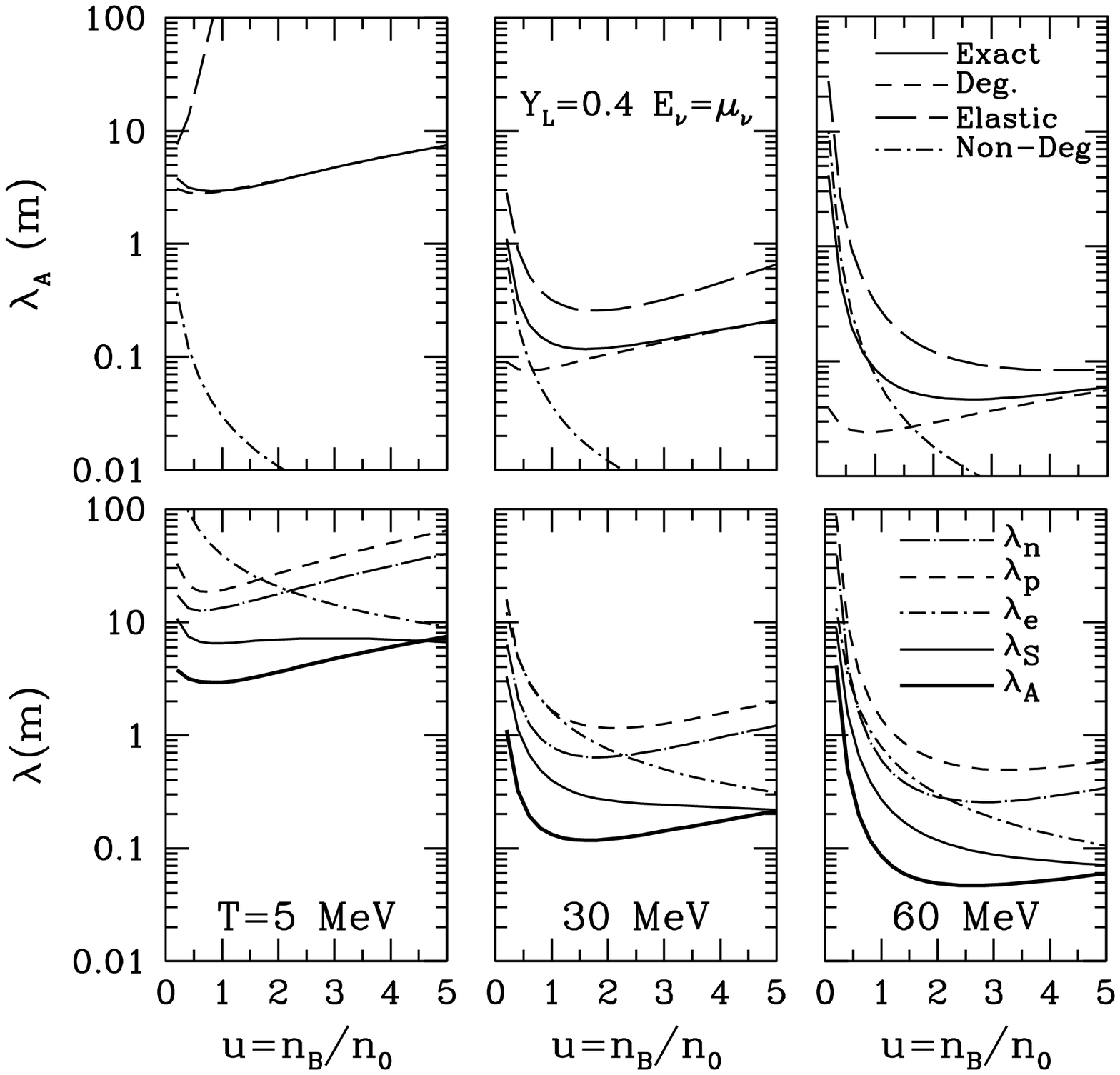}
\end{center}
\caption[]{\footnotesize}
{\label{fig9}}
\end{figure}

\newpage
\begin{figure}
\begin{center}
\leavevmode
\epsfxsize=7.0in
\epsfysize=7.0in
\epsffile{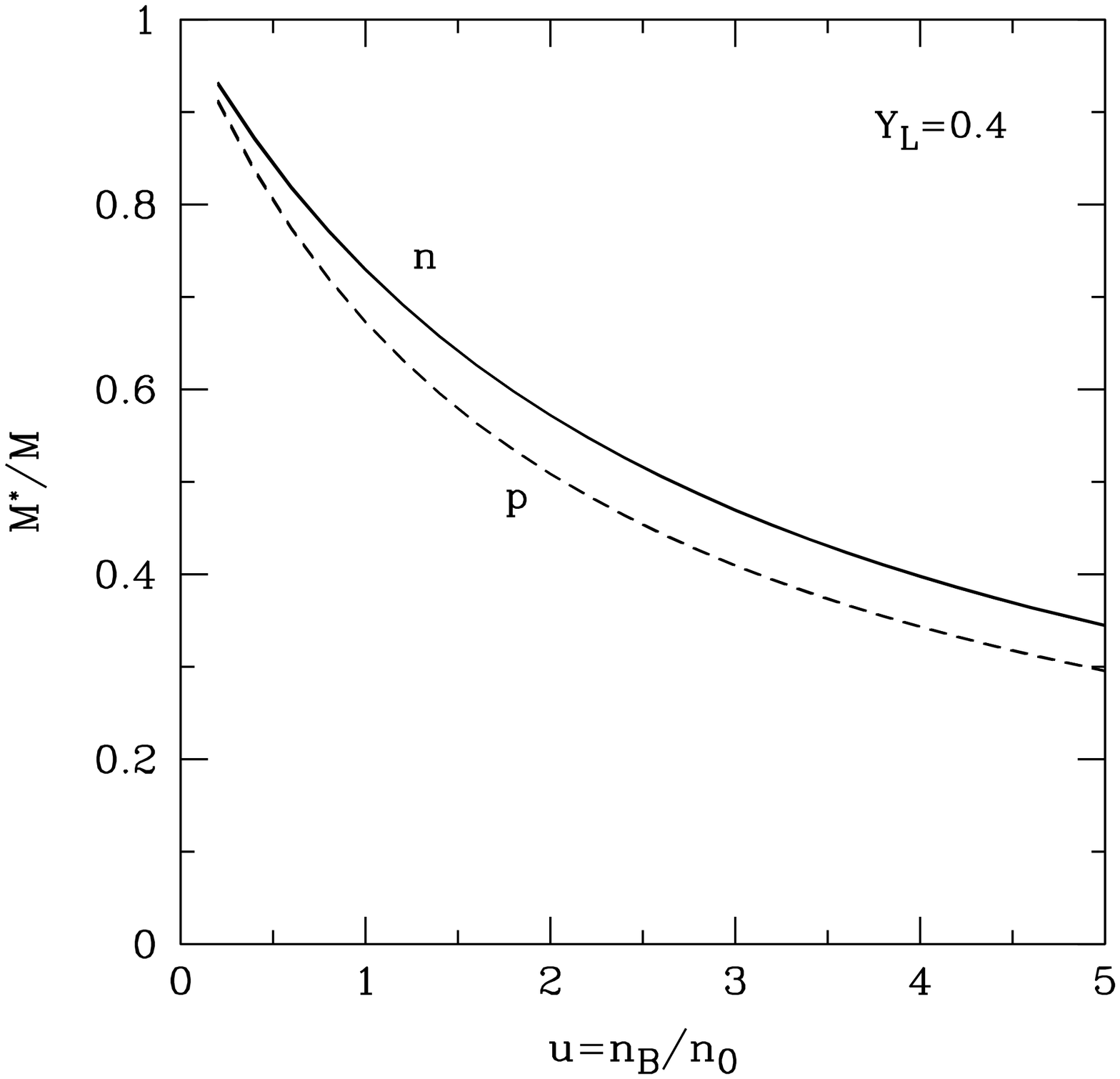}
\end{center}
\caption[]{\footnotesize}
{\label{fig10}}
\end{figure}

\newpage
\begin{figure}
\begin{center}
\leavevmode
\epsfxsize=7.0in
\epsfysize=7.0in
\epsffile{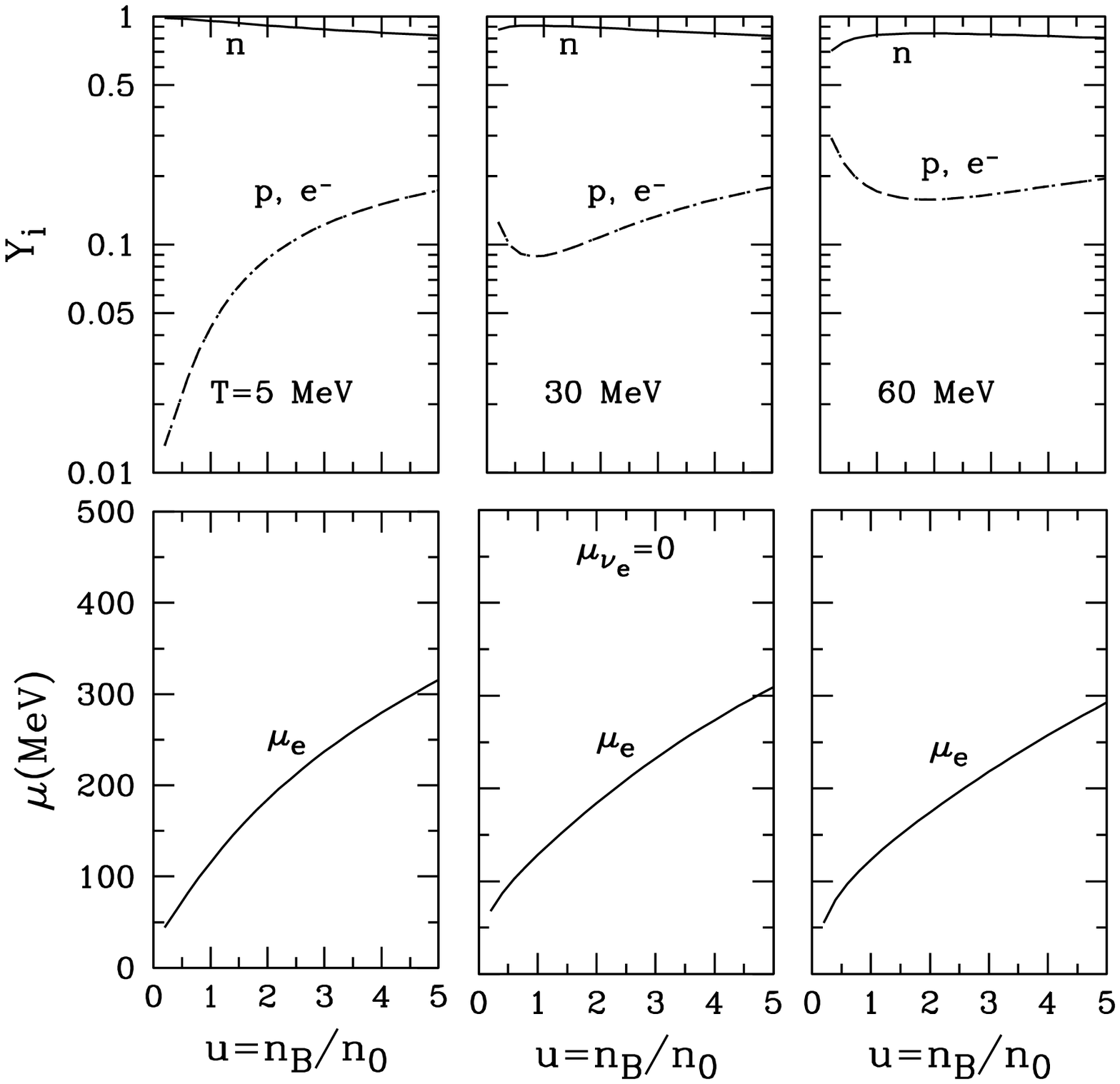}
\end{center}
\caption[]{\footnotesize}
{\label{fig11}}
\end{figure}

\newpage
\begin{figure}
\begin{center}
\leavevmode
\epsfxsize=7.0in
\epsfysize=7.0in
\epsffile{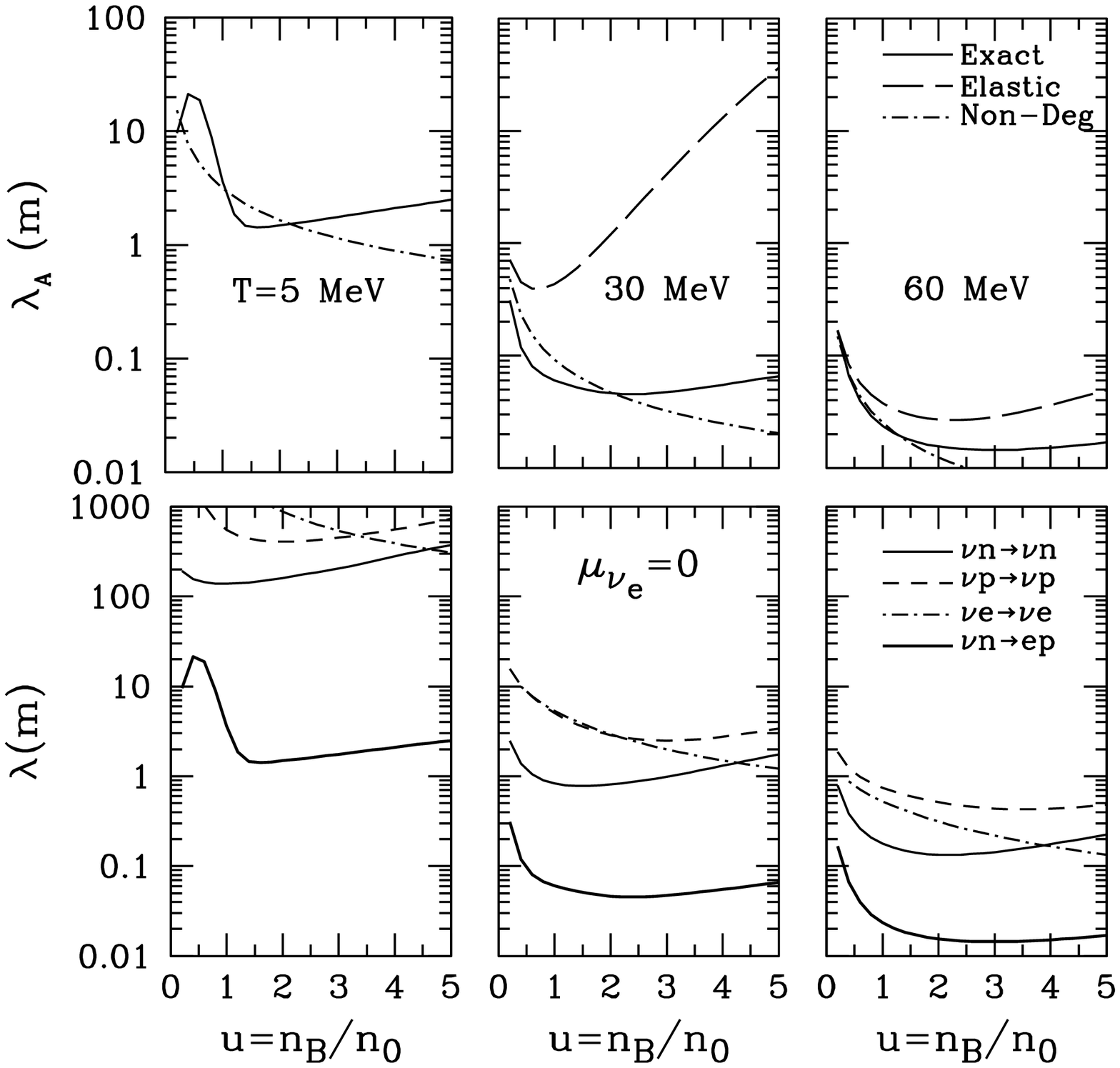}
\end{center}
\caption[]{\footnotesize}
{\label{fig12}}
\end{figure}

\newpage
\begin{figure}
\begin{center}
\leavevmode
\epsfxsize=7.0in
\epsfysize=7.0in
\epsffile{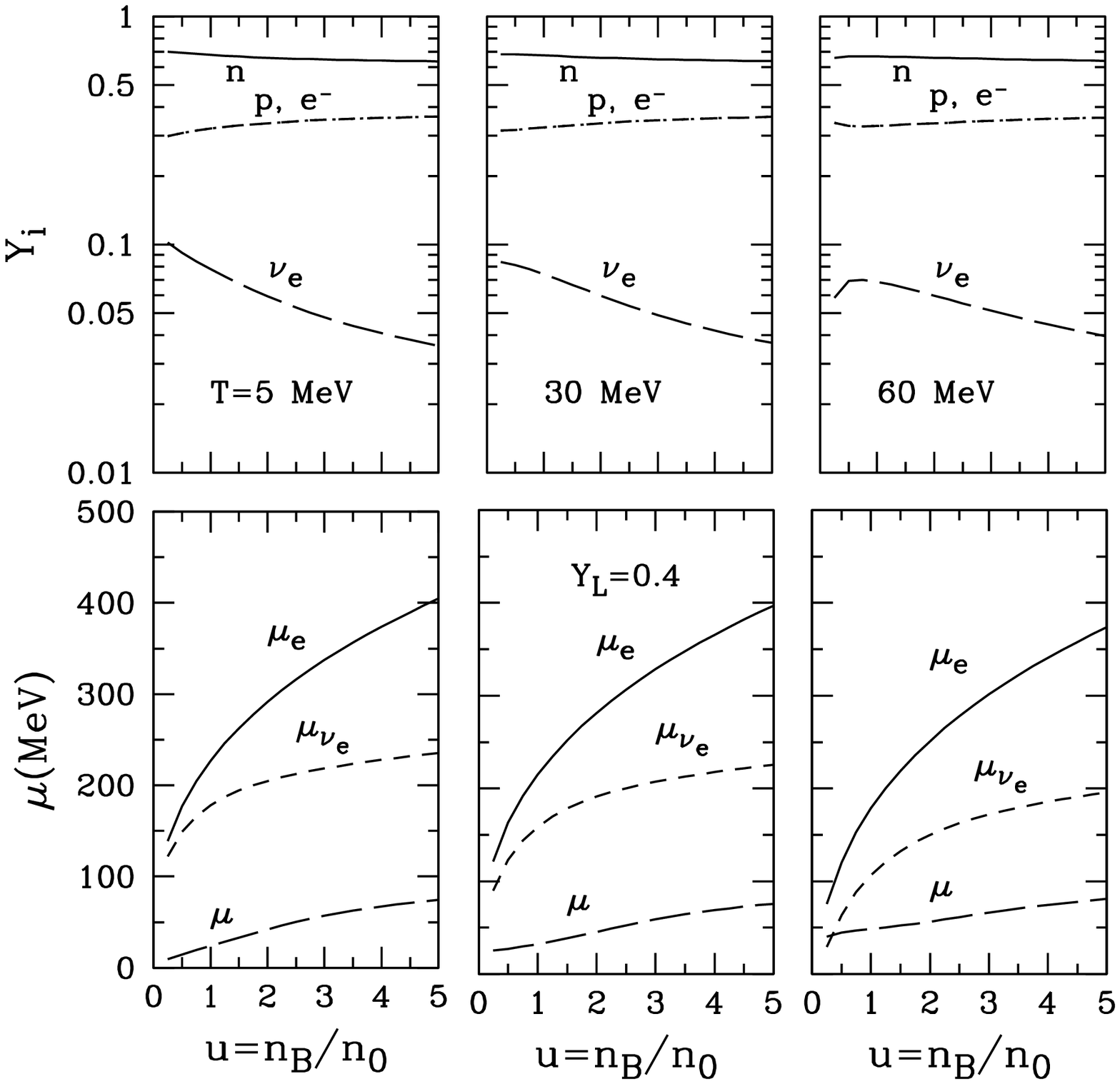}
\end{center}
\caption[]{\footnotesize}
{\label{fig13}}
\end{figure}

\newpage
\begin{figure}
\begin{center}
\leavevmode
\epsfxsize=7.0in
\epsfysize=7.0in
\epsffile{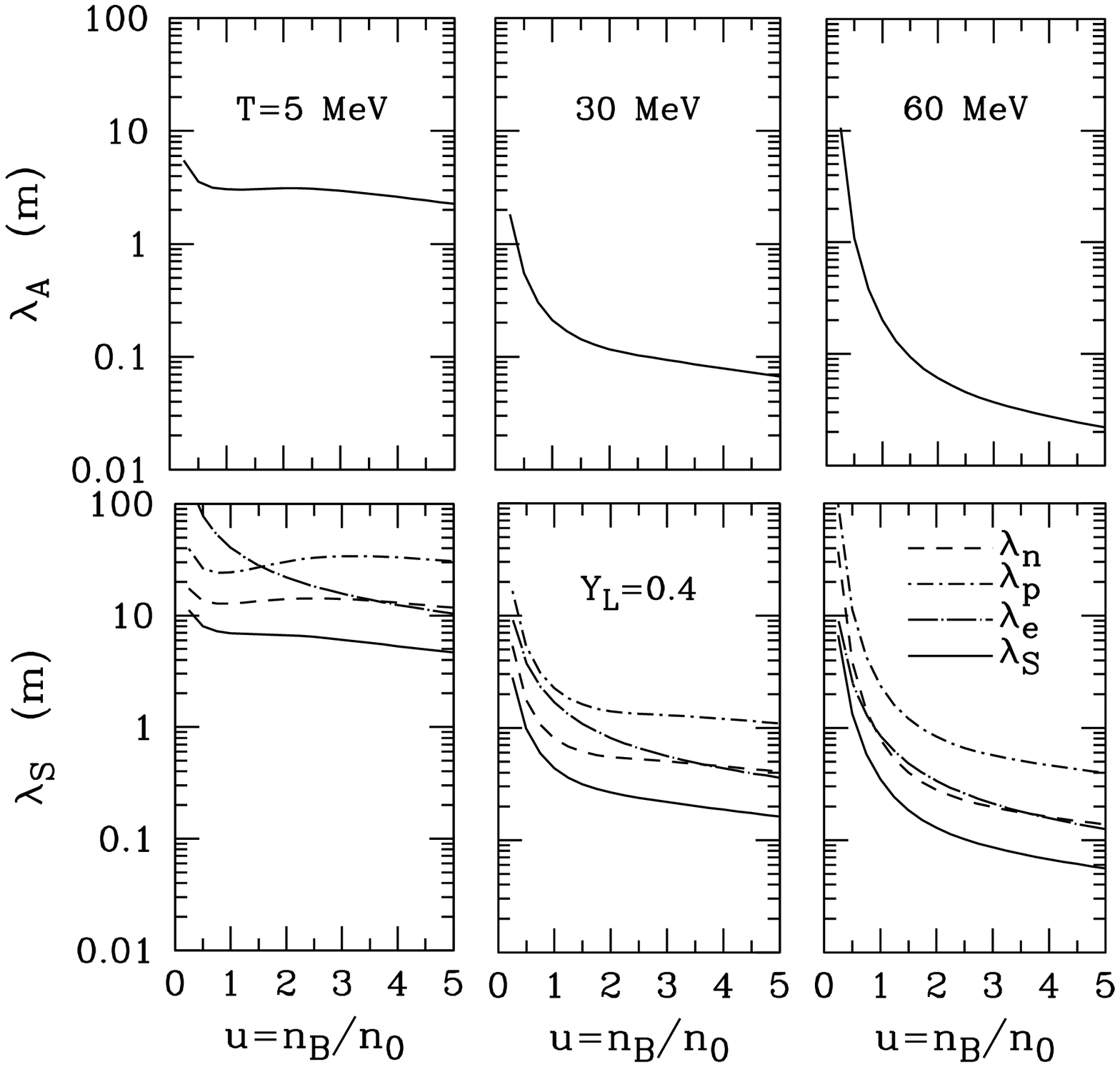}
\end{center}
\caption[]{\footnotesize}
{\label{fig14}}
\end{figure}

\newpage
\begin{figure}
\begin{center}
\leavevmode
\epsfxsize=7.0in
\epsfysize=7.0in
\epsffile{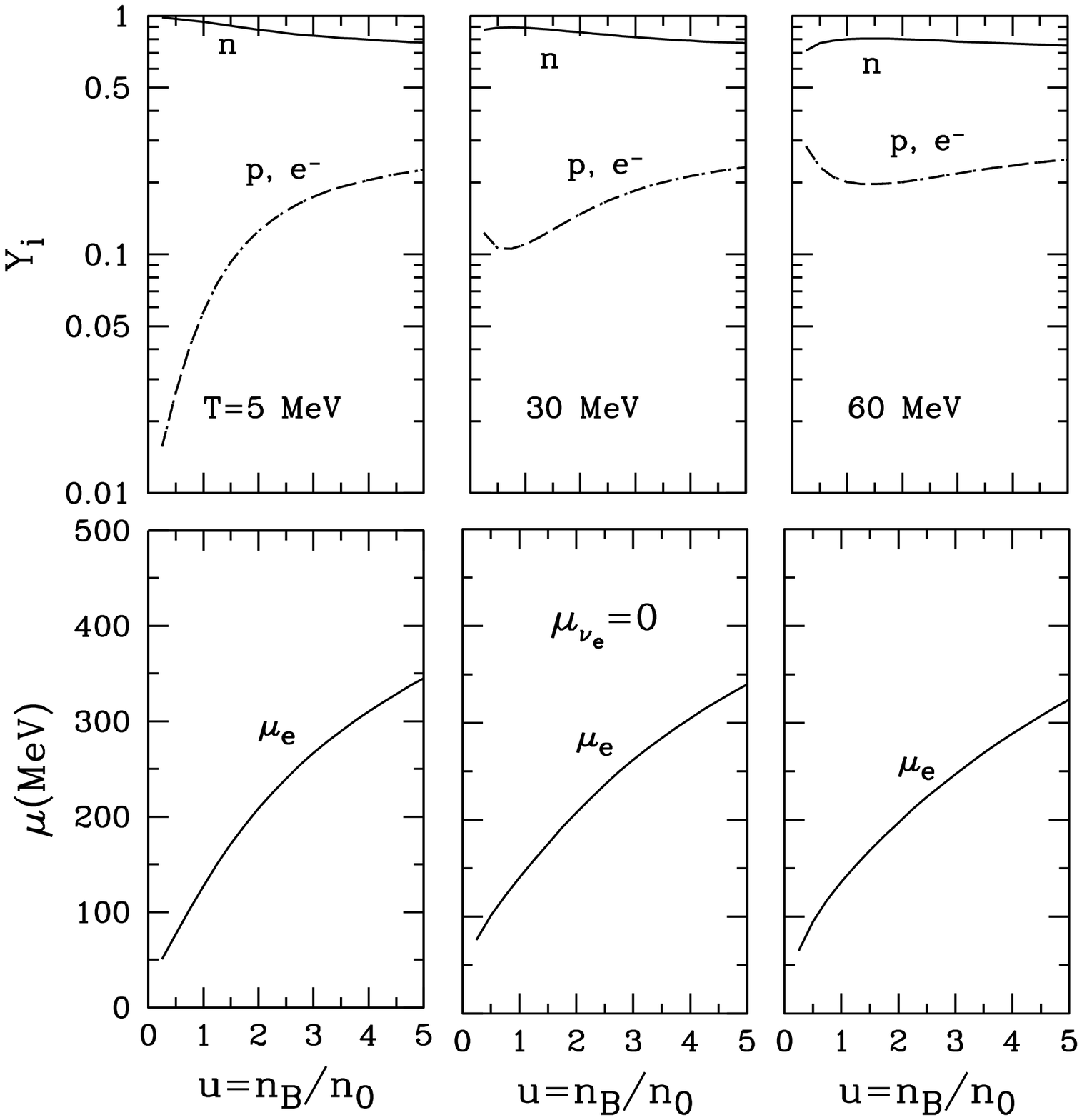}
\end{center}
\caption[]{\footnotesize}
{\label{fig15}}
\end{figure}

\newpage
\begin{figure}
\begin{center}
\leavevmode
\epsfxsize=7.0in
\epsfysize=7.0in
\epsffile{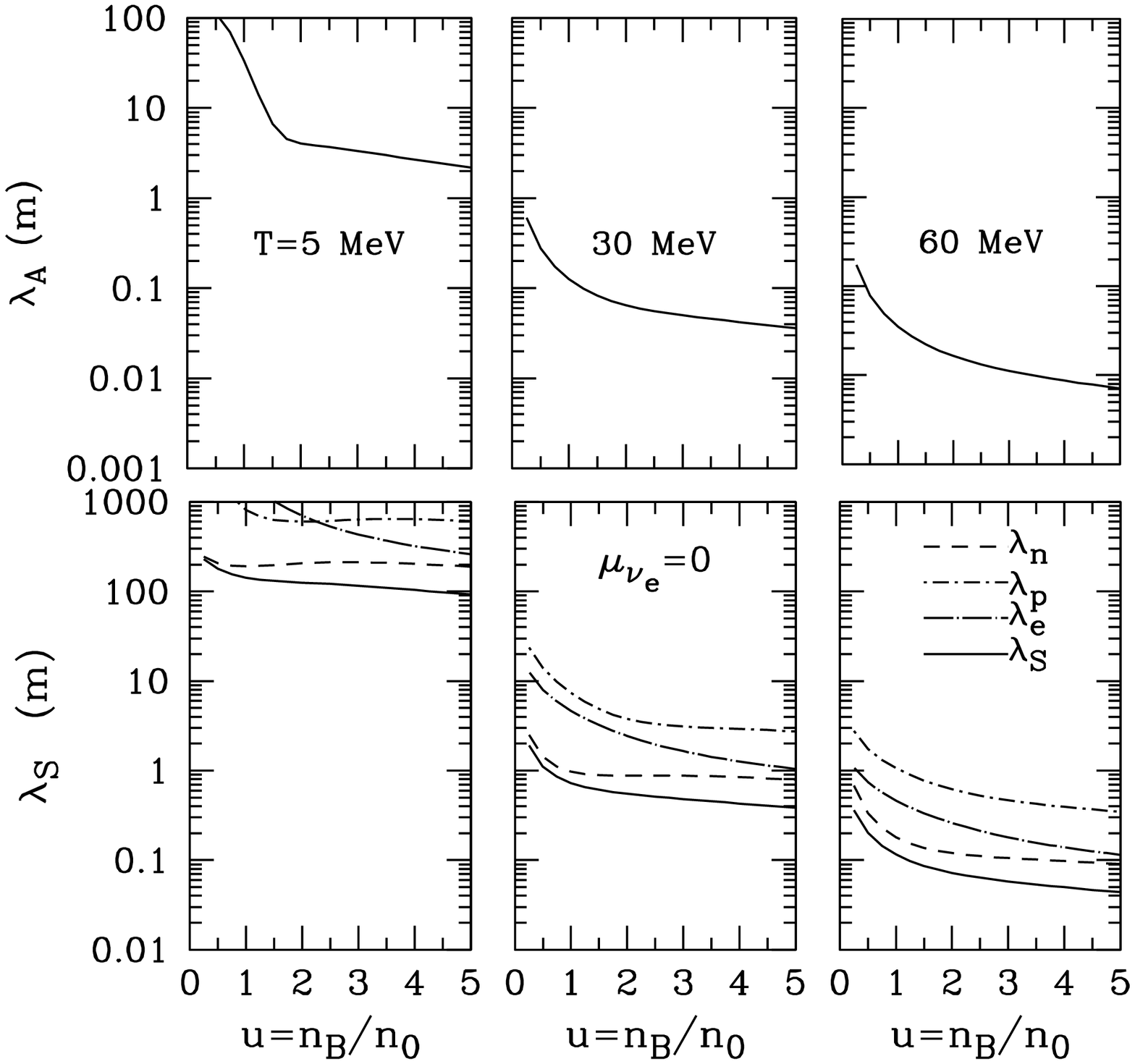}
\end{center}
\caption[]{\footnotesize}
{\label{fig16}}
\end{figure}

\newpage
\begin{figure}
\begin{center}
\leavevmode
\epsfxsize=7.0in
\epsfysize=7.0in
\epsffile{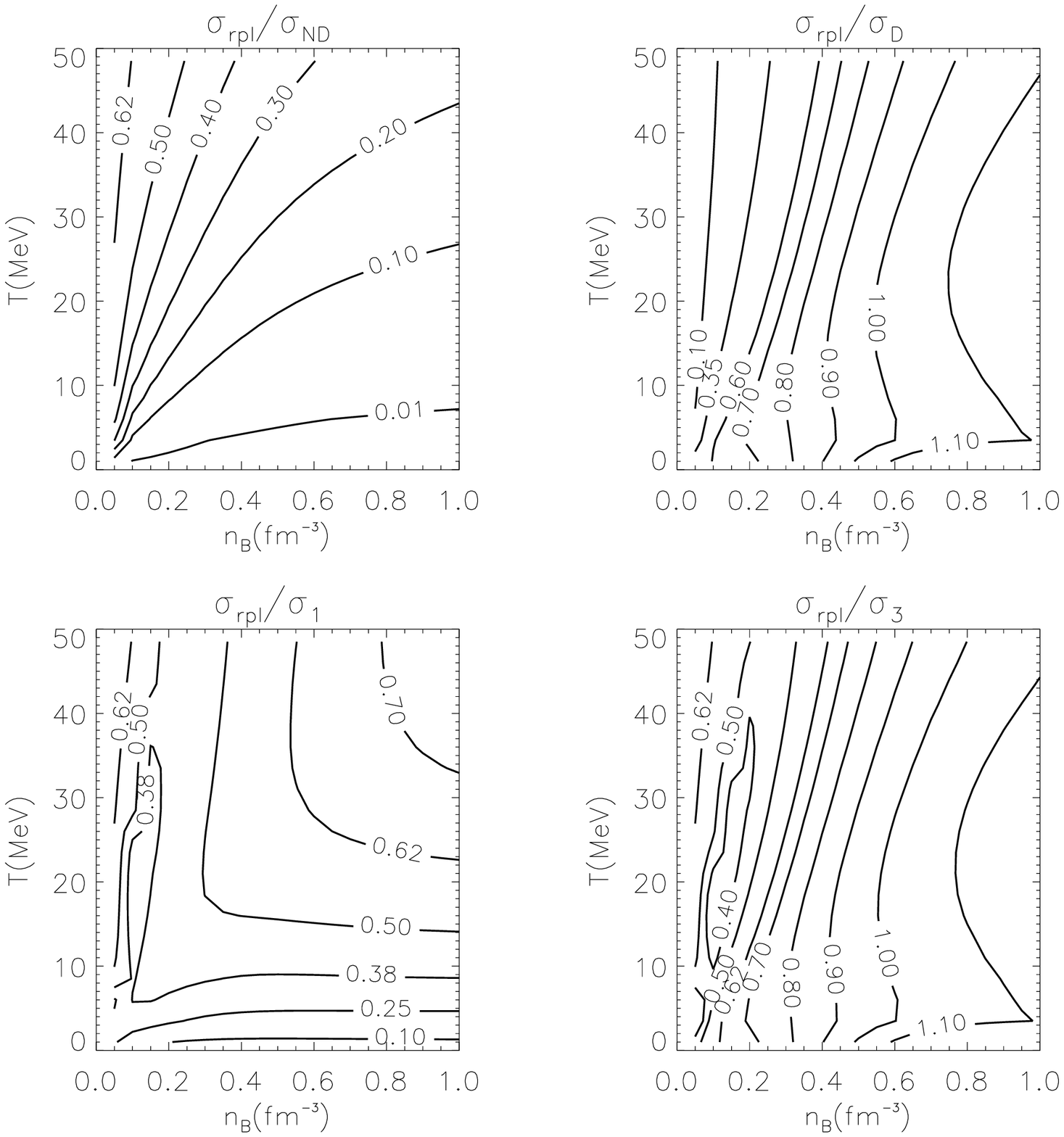}
\end{center}
\caption[]{\footnotesize}
{\label{fig17}}
\end{figure}

\newpage
\begin{figure}
\begin{center}
\leavevmode
\epsfxsize=7.0in
\epsfysize=7.0in
\epsffile{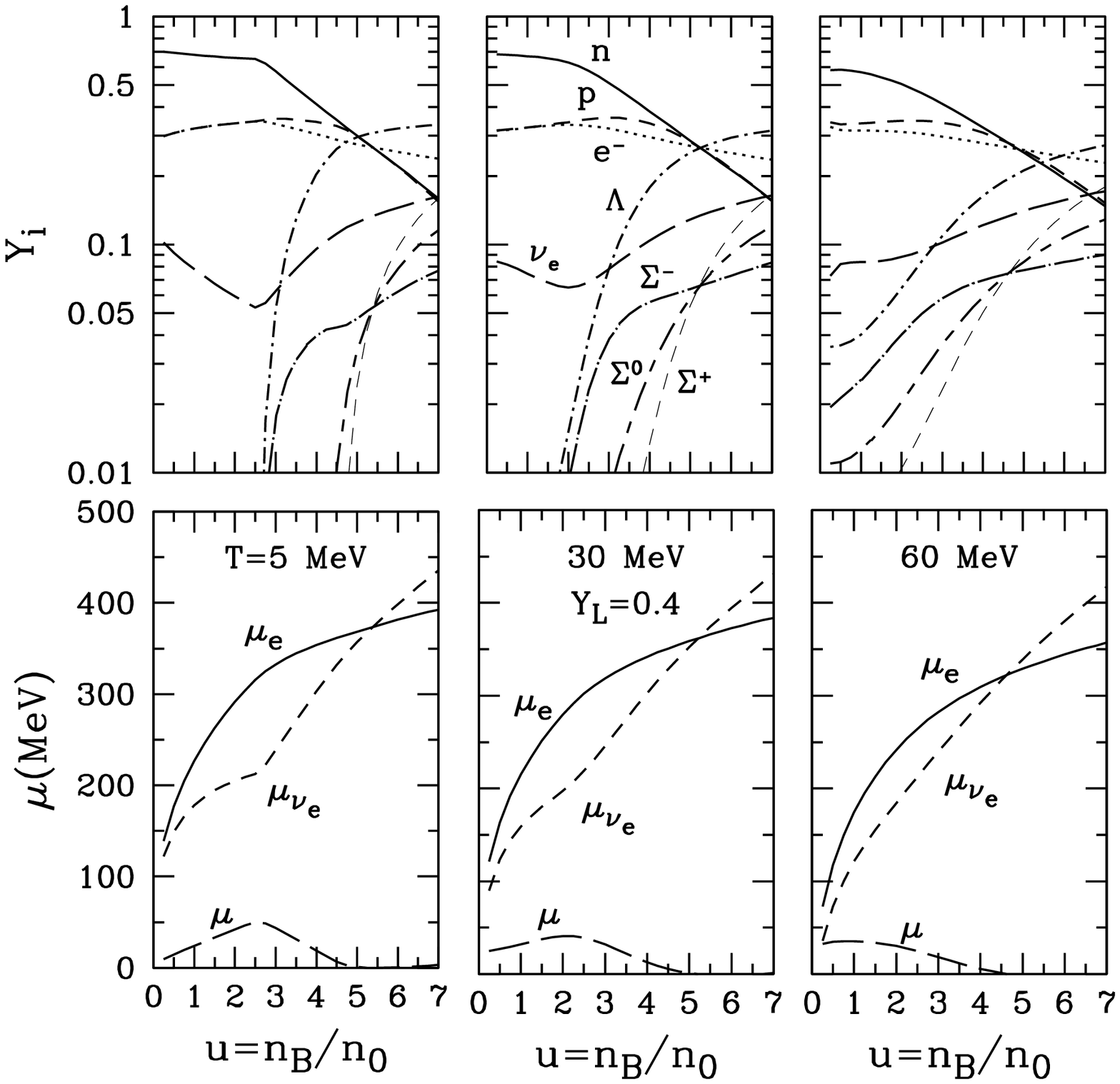}
\end{center}
\caption[]{\footnotesize}
{\label{fig18}}
\end{figure}

\newpage
\begin{figure}
\begin{center}
\leavevmode
\epsfxsize=7.0in
\epsfysize=7.0in
\epsffile{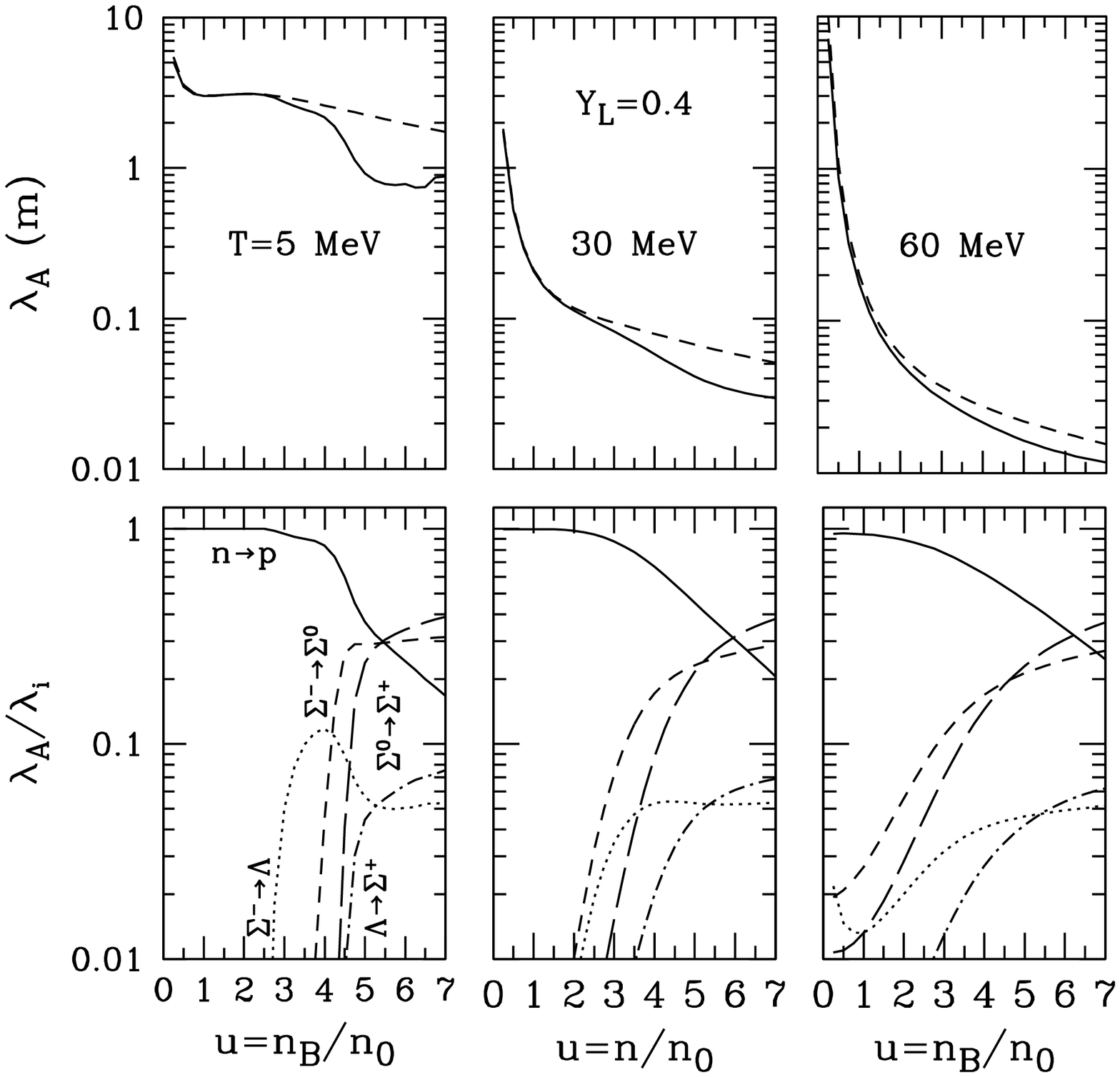}
\end{center}
\caption[]{\footnotesize}
{\label{fig19}}
\end{figure}

\newpage
\begin{figure}
\begin{center}
\leavevmode
\epsfxsize=7.0in
\epsfysize=7.0in
\epsffile{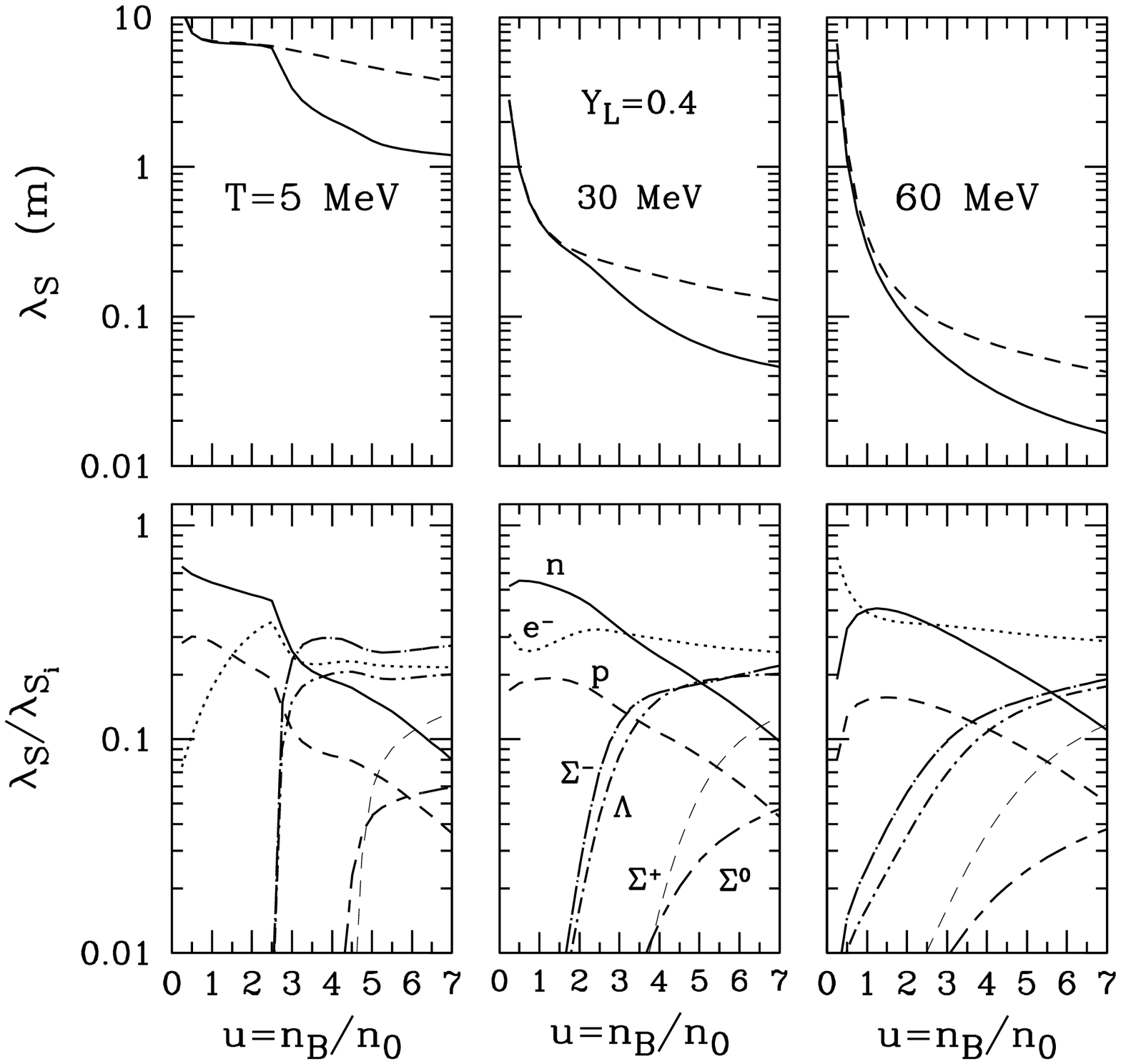}
\end{center}
\caption[]{\footnotesize}
{\label{fig20}}
\end{figure}

\newpage
\begin{figure}
\begin{center}
\leavevmode
\epsfxsize=7.0in
\epsfysize=7.0in
\epsffile{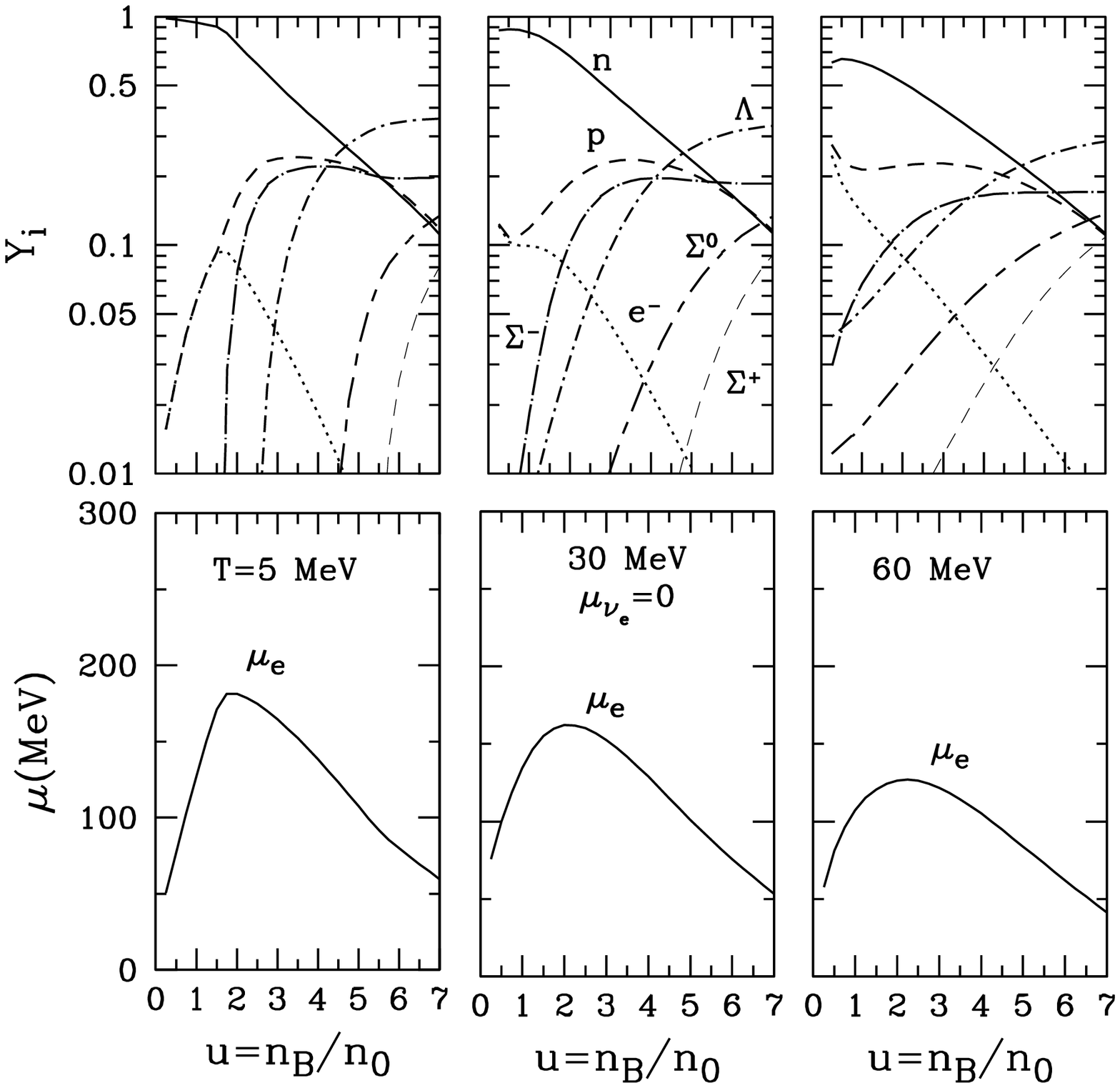}
\end{center}
\caption[]{\footnotesize}
{\label{fig21}}
\end{figure}

\newpage
\begin{figure}
\begin{center}
\leavevmode
\epsfxsize=7.0in
\epsfysize=7.0in
\epsffile{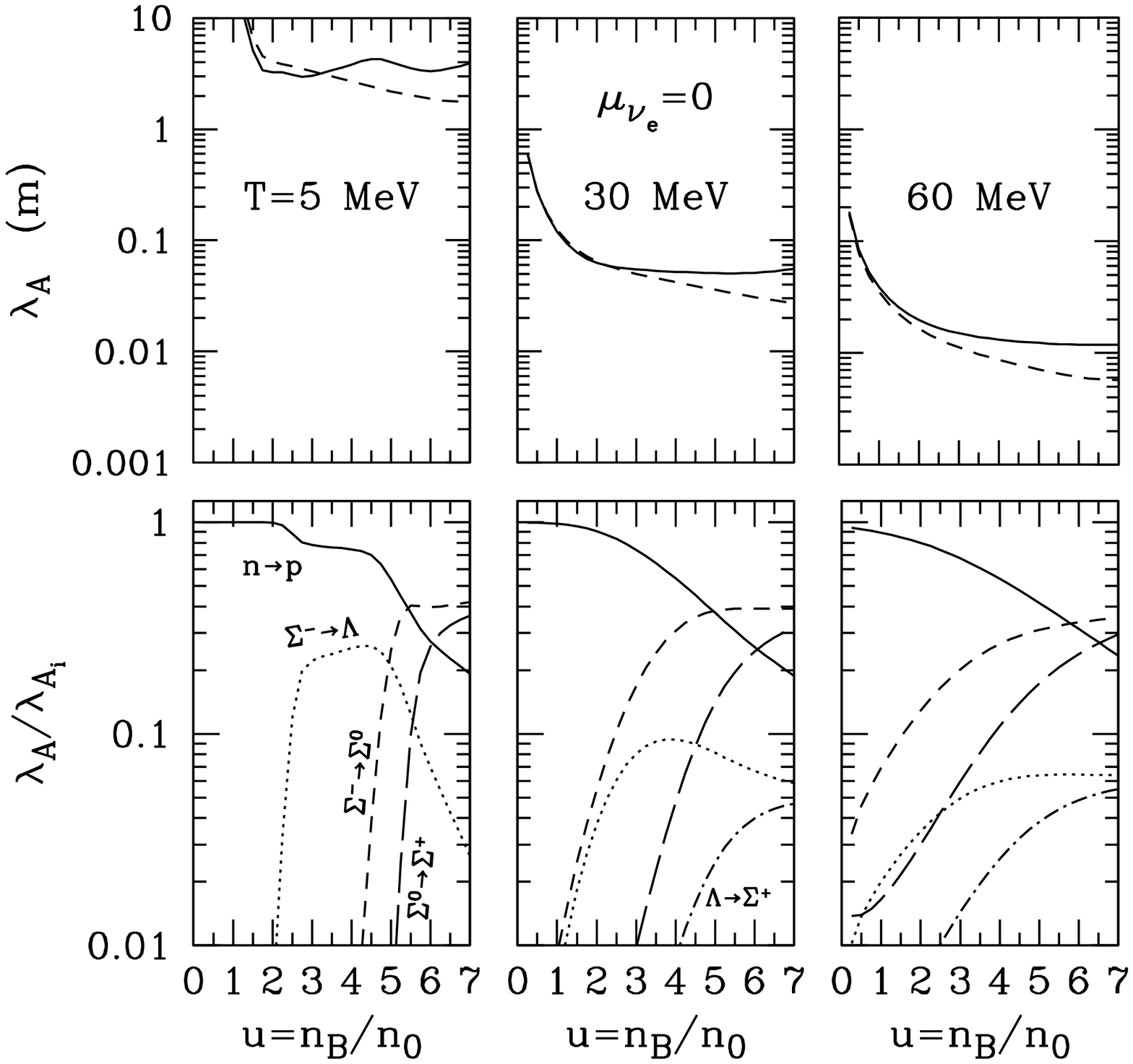}
\end{center}
\caption[]{\footnotesize}
{\label{fig22}}
\end{figure}

\newpage
\begin{figure}
\begin{center}
\leavevmode
\epsfxsize=7.0in
\epsfysize=7.0in
\epsffile{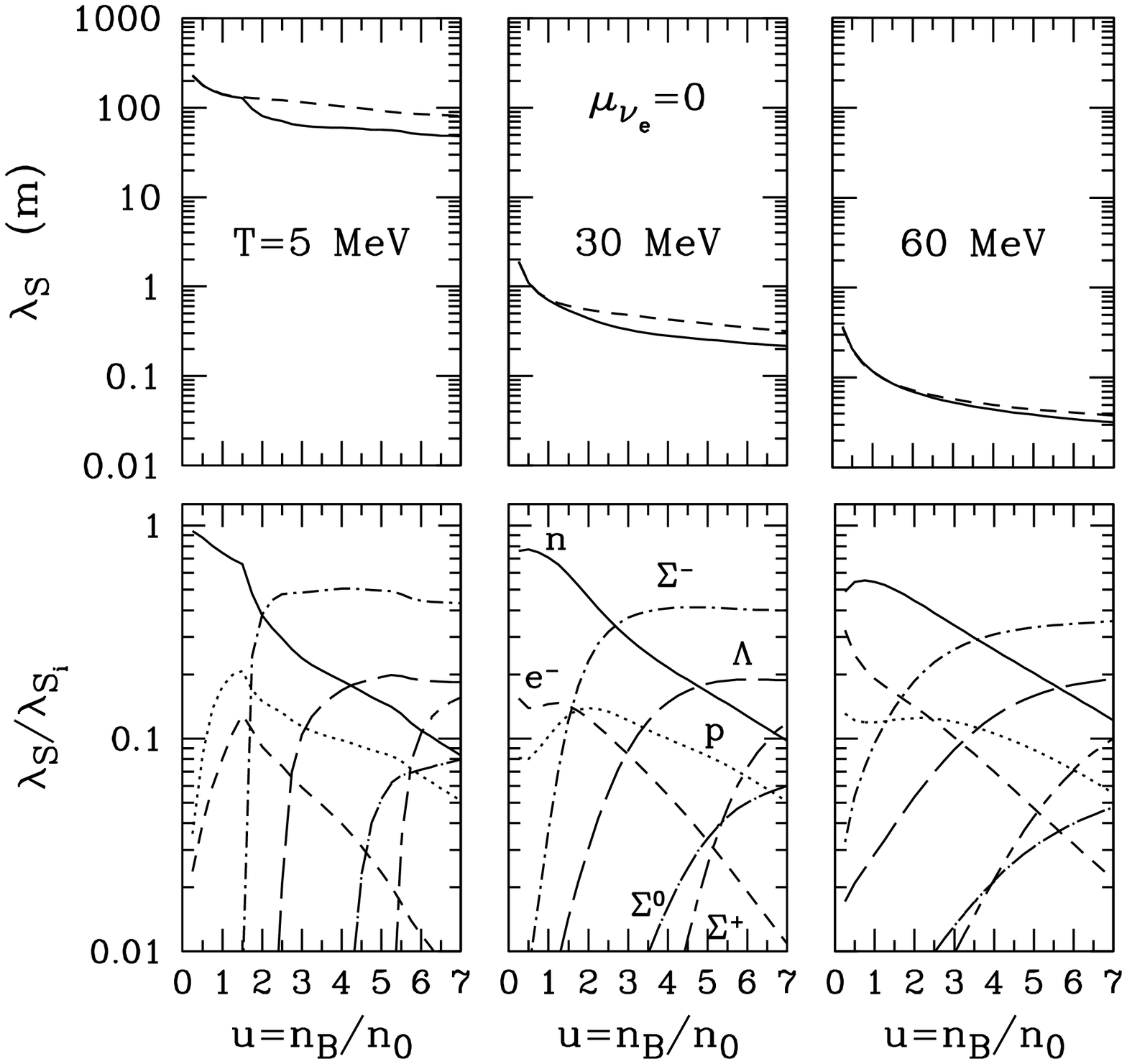}
\end{center}
\caption[]{\footnotesize}
{\label{fig23}}
\end{figure}

\newpage
\begin{figure}
\begin{center}
\leavevmode
\epsfxsize=7.0in
\epsfysize=7.0in
\epsffile{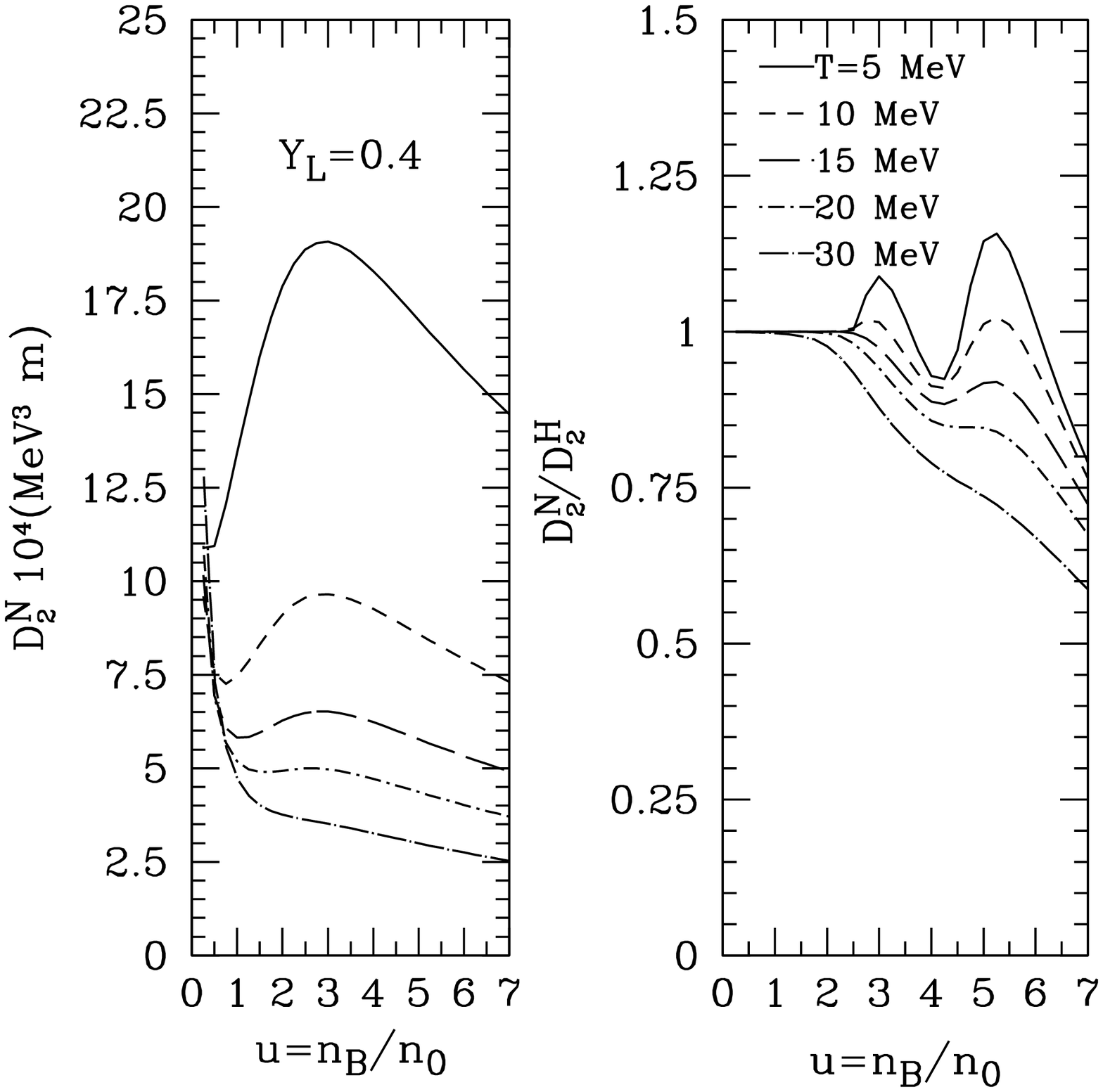}
\end{center}
\caption[]{\footnotesize}
{\label{fig24}}
\end{figure}

\newpage
\begin{figure}
\begin{center}
\leavevmode
\epsfxsize=7.0in
\epsfysize=7.0in
\epsffile{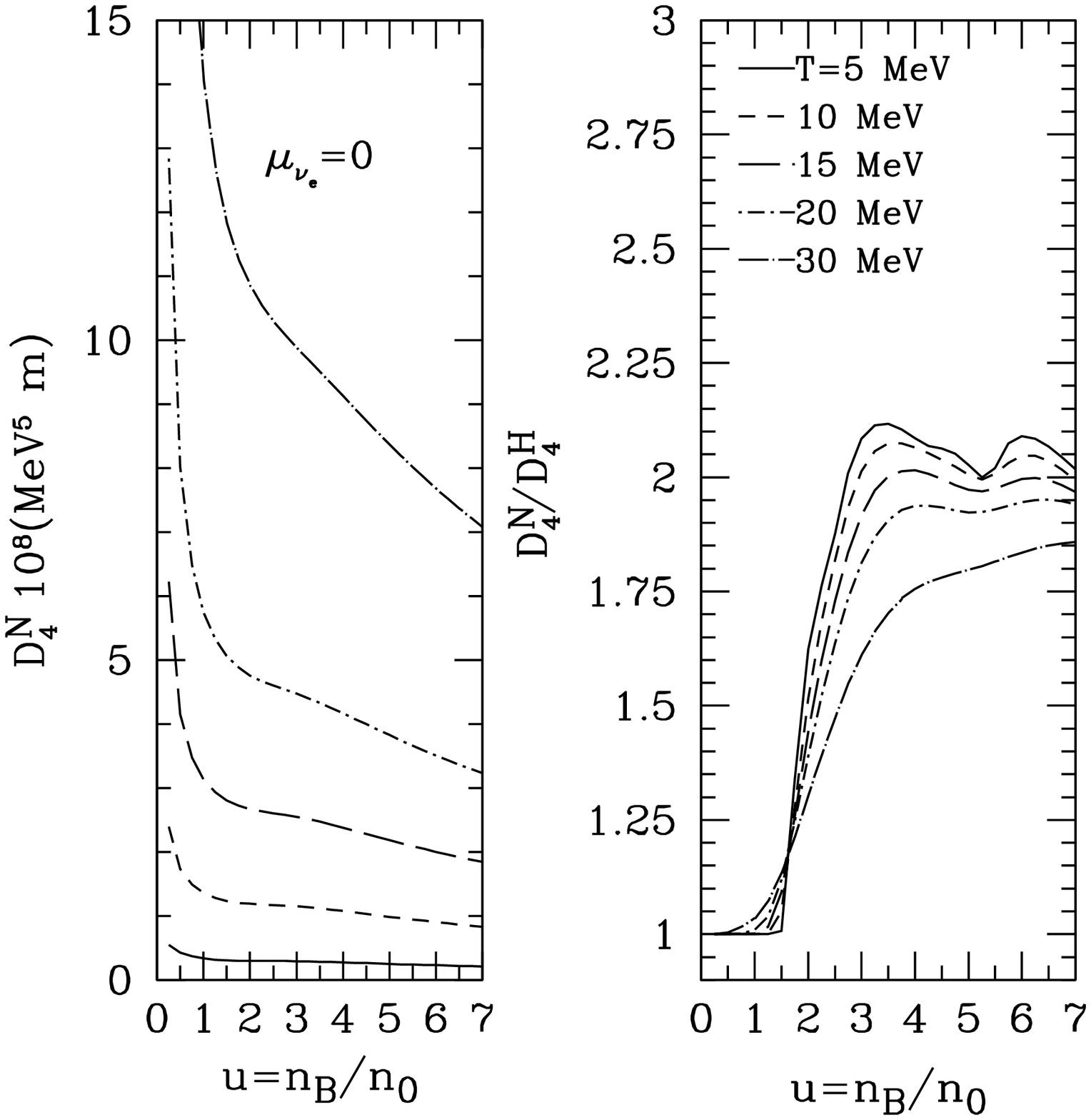}
\end{center}
\caption[]{\footnotesize}
{\label{fig25}}
\end{figure}


\begin{thebibliography}{99}

\bibitem{SN87A} K. Hirata, et. al., {Phys. Rev. Lett.,} {\bf 58} (1987)
1490; R. M. Bionta, et. al., {Phys. Rev. Lett.,} {\bf 58} (1987)
1494.
\bibitem{B1} S. W. Bruenn,
{Astrophys. Jl. Suppl.} {\bf 58} (1985) 771.
\bibitem{MB} A. Mezzacappa and S. W. Bruenn,
{Astrophys. Jl. } {\bf 405} (1993) 637.
\bibitem{BL} A. Burrows and J. M. Lattimer,
{Astrophys. Jl.,} {\bf 307} (1986) 178;
A. Burrows,  {Ann. Rev. Nucl. Sci.,} {\bf 40} (1990) 181; and references
therein.
\bibitem{WM} J. R. Wilson and R. Mayle, in {\em The Nuclear Equation of State},
eds. W. Greiner and H. St{\"o}cker (Plenum Press, New York, 1989) 731.
\bibitem{SuS} H. Suzuki and K. Sato, in {\em The Structure and Evolution of
Neutron Stars}, eds. D. Pines, R. Tamagaki, and S. Tsuruta (New York:
Addison-Wesley, 1992), 276.
\bibitem{K} W. Keil, Prog. part. Nucl. Phys. {\bf 32} (1994) 105.
\bibitem{KJ} W. Keil and H. T. Janka,
{Astronomy and Astrophysics}, {\bf 296} (1995) 145.
\bibitem{TS} D. L. Tubbs and D. N. Schramm,
{Astrophys. Jl.} {\bf 201} (1975) 467.
\bibitem{S1}R. F. Sawyer,
{Phys. Rev.} {\bf D11} (1975) 2740;
{Phys. Rev.} {\bf C40} (1989) 865.
\bibitem{LP} D. Q. Lamb and C. J. Pethick,
{Astrophys. Jl.} {\bf 209} (1976) L77.
\bibitem{L} D. Q. Lamb,
{Phys. Rev. Lett} {\bf 41} (1978) 1623.
\bibitem{T} D. L. Tubbs,
{Astrophys. Jl. Suppl.} {\bf 37} (1978) 287.
\bibitem{BVR} S. Bludman and K. Van Riper,
{Astrophys. Jl.} {\bf 224} (1978) 631.
\bibitem{SS} R. F. Sawyer and A. Soni,
{Astrophys. Jl.} {\bf 230} (1979) 859.
\bibitem{I} N. Iwamoto,
{Ann. Phys.} {\bf 141} (1982) 1.
\bibitem{IP} N. Iwamoto and C. J. Pethick,
{Phys. Rev.} {\bf D25} (1982) 313.
\bibitem{GP} B. T. Goodwin and C. J. Pethick,
{Astrophys. Jl.} {\bf 253} (1982) 816.
\bibitem{BM} A. Burrows and T. J. Mazurek,
{Astrophys. Jl.} {\bf 259} (1982) 330.
\bibitem{G} B. T. Goodwin,
{Astrophys. Jl.} {\bf 261} (1982) 321.
\bibitem{B2} S. W. Bruenn,
{Astrophys. Jl. Suppl.} {\bf 58} (1985) 771.
\bibitem{VC} L. Van Den Horn and J. Cooperstein,
{Astrophys. Jl.} {\bf 300} (1986) 142.
\bibitem{MAX}O. V. Maxwell,
{Astrophys. Jl.} {\bf 316} (1987) 691.
\bibitem{COOP} J. Cooperstein,
{Phys. Rep.} {\bf 163} (1988) 95.
\bibitem{AB} A. Burrows,
{Astrophys. Jl.} {\bf 334} (1988) 891.
\bibitem{PS}P. J. Schinder,
Astrophys. Jl. Suppl. {\bf 74} (1990) 249.
\bibitem{HW} C. J. Horowitz and K. Wehrberger,
Nucl. Phys. {\bf A531} (1991) 665;
Phys. Rev. Lett. {\bf 66} (1991) 272;
Phys. Lett. {\bf B226} (1992) 236.
\bibitem{PPLP} M. Prakash, Manju Prakash, J. M. Lattimer and C. J. Pethick,
{Astrophys. Jl.} {\bf 390} (1992) L80.
\bibitem{RP1}S. Reddy and M. Prakash, in Proc. 11th Winter Workshop on Nuclear
Dynamics, Advances in Nuclear Dynamics, ed. W. Bauer and Migerney (New York:
Plenum) (1995) 237.
\bibitem{KJR} W. Keil, H-T. Janka and G. Raffelt,
Phys. Rev. {\bf D51} (1995) 6635.
\bibitem{JKRS} H-T. Janka, W. Keil, G. Raffelt, and D. Seckel,
Phys. Rev. Lett. {\bf 76} (1996) 2621.
\bibitem{GS} G. Sigl,
Phys. Rev. Lett. {\bf 76} (1996) 2625.
\bibitem{RP2}S. Reddy and M. Prakash,
{Astrophys. Jl.} {\bf 423} (1997) 689.
\bibitem{VB} D. Vautherin and D.M. Brink,
Phys. Rev. {\bf C5} (1972) 626.
\bibitem{SW} B. D. Serot and J. D. Walecka,
Advances in Nuclear Physics, {\bf 16}, eds. J. W. Negele and E. Vogt,
(New York: Plenum); B.D. Serot, Rep. Prog. Phys. {\bf 55} (1992) 1855.

\bibitem{W} S. Weinberg,
Phys. Rev. Lett. {\bf 19} (1967) 1264.
\bibitem{AS} A. Salam, in Proc. Eighth Nobel Symp., Elementary Particle
Theory: Relativistic Groups and Analyticity, ed. N. Svartholm (Stockholm:
Almquist and Wicksell)
\bibitem{SG}S. L. Glashow,
Nucl. Phys. {\bf 22} (1961) 579.

\bibitem{JGGS} J.-M. Gaillard and G. Sauvage,
Ann. Rev. Nucl. Part. Sci., {\bf
34} (1984) 351.
\bibitem{MSJW} M.J. Savage and J. Walden, Phys. Rev. {\bf D55} (1997) 5376.

\bibitem{KS}K. Sato,
{Prog. Theor. Phys.} {\bf 53} (1975) 595.
\bibitem{TM} T. J. Mazurek,
{Astrophys. Space Sci.} {\bf 35} (1975) 117.
\bibitem{BBAL} H. A. Bethe, G. E. Brown, J. Applegate, and J. M. Lattimer,
Nucl. Phys. {\bf A324} (1979) 487.
\bibitem{KG}F. C. Khanna and H. R. Glyde, Can. Jl. Phys. {\bf 54} (1976) 648.
\bibitem{LPPH} J. M. Lattimer, C.J. Pethick, M. Prakash, and P. Haensel,
{Phys. Rev. Lett.} {\bf 66} (1991) 2701.
\bibitem{WFF}R. B. Wiringa, V. Fiks and A. Fabrocine,
Phys. Rev. {\bf C38} (1988) 1010.
\bibitem{PIPELR} M. Prakash, I. Bombaci, Manju Prakash, P. J. Ellis, J. M.
Lattimer and R. Knorren, {Phys. Rep} {\bf 280} (1997) 1.

\bibitem{PAL}M. Prakash, T. L. Ainsworth and J. M. Lattimer,
Phys. Rev. Lett. {\bf 61} (1988) 2518.
\bibitem{FW} A.L. Fetter and J.D. Walecka, {\em Quantum Theory of Many Particle
Systems} (New York: McGraw-Hill), 1971.
\bibitem{DS} S. Doniach and E. H. Sondheimer, {\em Green's Functions for Solid
State Physicists} (Reading, The Benjamin/Cummings Publishing Company, Inc.),
1974.
\bibitem{LH} K. Lim and C. J. Horowitz,
Nucl. Phys. {\bf A501} (1989) 729.
\bibitem{SMS}K. Saito, T. Maruyama and K. Soutame, Phys. Rev. {\bf C40}
(1989) 407.
\bibitem{LL}L. Lewin, {\em Polylogarithms and Associated Functions} (New
York: North-Holland) 1983.
\bibitem{Chandra} S. Chandrasekhar, {\em An Introduction to the Study of
Stellar Structure} (Dover, New York, 1967).

\bibitem{PRPL} J. Pons, S. Reddy, M. Prakash and J. M. Lattimer,
to be published.
\bibitem{WR}D. H. Wilkinson, {Phys. Rev.} {\bf C7} (1973) 930;
M. Rho, {Nucl. Phys.} {\bf A231} (1974) 493.
\bibitem{BRHO}G. E. Brown and M. Rho, Phys. Rev. Lett. {\bf 66} (1991) 2720.
\bibitem{GM}N. K. Glendenning and S. Moszkowski, Phys. Rev. Lett. {\bf 67}
(1991) 2414.

\end{thebibliography}
\end{document}